\documentclass[twocolumn,orientation=90]{aastex63}
\pdfoutput=1
\usepackage{multirow}
\usepackage{amsmath}
\usepackage{epsfig}
\usepackage{epsf}
\input{epsf}
\usepackage{epstopdf}
\usepackage{amssymb}
\usepackage{bbold}
\usepackage{natbib}
\usepackage{graphicx}
\usepackage{color}
\usepackage{hyperref}
\usepackage{subfigure}
\usepackage{threeparttable}
\usepackage{xspace}
\usepackage{float}

\DeclareGraphicsExtensions{.jpg,.pdf,.png,.eps,.ps}

\newcommand{\ltsima}{$\; \buildrel < \over \sim \;$}
\newcommand{\ltsim}{\lower.5ex\hbox{\ltsima}}

\newcommand{\beq}{\begin{equation}}
\newcommand{\eeq}{\end{equation}}

\def\n{{\bf  \hat n}}
\def\q{{\bf q}}

\def \fivefiffi {\textsc{ra5h30dec-55}}
\def \twthreefiffi {\textsc{ra23h30dec-55}}

\def \twonesix {\textsc{ra21hdec-60}}
\def \twonefif {\textsc{ra21hdec-50}}
\def \threesix {\textsc{ra3h30dec-60}}

\def \zerofif {\textsc{ra0h50dec-50}}
\def \fourten {\textsc{ra4h10dec-50}}
\def \twothir {\textsc{ra2h30dec-50}}
\def \onesix {\textsc{ra1hdec-60}}
\def \fourten {\textsc{ra4h10dec-50}}
\def \fivefourfi {\textsc{ra5h30dec-45}}
\def \sixfiffi {\textsc{ra6h30dec-55}}
\def \sixfourfi {\textsc{ra6h30dec-45}}
\def \twthreesix {\textsc{ra23hdec-62.5}}
\def \twonefour {\textsc{ra21hdec-42.5}}
\def \twtwo {\textsc{ra22h30dec-55}}

\def \sixsix {\textsc{ra6hdec-62.5}}
\def \threefour {\textsc{ra3h30dec-42.5}}
\def \onefour {\textsc{ra1hdec-42.5}}
\def \twthreefourfi {\textsc{ra23hdec-45}}

\def\ColoradoAPS{Center for Astrophysics and Space Astronomy and Department of Astrophysical and Planetary Sciences, University of Colorado, Boulder, CO, USA 80309}
\def\ColoradoPhysics{Department of Physics, University of Colorado, Boulder, CO, USA 80309}
\def\SantaBarbara{Department of Physics, University of California, Santa Barbara, CA, 93106}
\def\KICPChicago{Kavli Institute for Cosmological Physics, University of Chicago, Chicago, IL, USA 60637}
\def\AAUChicago{Department of Astronomy and Astrophysics, University of Chicago, Chicago, IL, USA 60637}
\def\IllinoisAstronomy{Astronomy Department, University of Illinois Urbana-Champaign, Urbana, IL 61801}
\def\IllinoisPhysics{Department of Physics, University of Illinois Urbana-Champaign, Urbana, IL 61801}
\def\ESO{European Southern Observatory, Alonso de C\'{o}rdova 3107, Vitacura Santiago, Chile}
\def\Diego{N\'ucleo de Astronom\'{\i}a, Facultad de Ingenier\'{\i}a y Ciencias, Universidad Diego Portales, Av. Ej\'ercito 441, Santiago, Chile}
\def\NIST{NIST Quantum Devices Group, Boulder, CO, USA 80305}
\def\Fermilab{Fermi National Accelerator Laboratory, Batavia, IL 60510}
\def\ArgonneHEP{High Energy Physics Division, Argonne National Laboratory, Argonne, IL, USA 60439}
\def\PhysicsUChicago{Department of Physics, University of Chicago, Chicago, IL, USA 60637}
\def\EFIChicago{Enrico Fermi Institute, University of Chicago, Chicago, IL, USA 60637}
\def\Dalhousie{Department of Physics and Atmospheric Science, Dalhousie University, Halifax, NS B3H 3J5, Canada}
\def\Cambridge{Institute of Astronomy, University of Cambridge, Madingley Road, Cambridge CB3 0HA, UK}
\def\Caltech{California Institute of Technology, Pasadena, CA, USA 91125}
\def\Berkeley{Department of Physics, University of California, Berkeley, CA, USA 94720}
\def\LBNL{Physics Division, Lawrence Berkeley National Laboratory, Berkeley, CA, USA 94720}
\def\McGill{Department of Physics, McGill University, Montreal, Quebec H3A 2T8, Canada}
\def\CIFAR{Canadian Institute for Advanced Research, CIFAR Program in Gravity and the Extreme Universe, Toronto, ON, M5G 1Z8, Canada}
\def\ESO{European Southern Observatory, Munich, Germany}
\def\UChicago{University of Chicago, Chicago, IL, USA 60637}
\def\Davis{Department of Physics, University of California, Davis, CA, USA 95616}
\def\Arizona{Steward Observatory, University of Arizona, 933 North Cherry Avenue, Tucson, AZ, USA 85721}
\def\Michigan{Department of Physics, University of Michigan, Ann  Arbor, MI, USA 48109}
\def\Oslo{Institute of Theoretical Astrophysics, University of Oslo, Oslo, Norway}
\def\Munich{Department of Physics, Ludwig-Maximilians-Universit\"{a}t, 81679 M\"{u}nchen, Germany}
\def\ExcellenceCluster{Excellence Cluster Universe, 85748 Garching, Germany}
\def\MPE{Max-Planck-Institut f\"{u}r extraterrestrische Physik, 85748 Garching, Germany}
\def\Dunlap{Dunlap Institute for Astronomy \& Astrophysics, University of Toronto, 50 St George St, Toronto, ON, M5S 3H4, Canada}
\def\Minnesota{School of Physics and Astronomy, University of Minnesota, Minneapolis, MN, USA 55455}
\def\Melbourne{School of Physics, University of Melbourne, Parkville, VIC 3010, Australia}
\def\CaseWestern{Physics Department, Center for Education and Research in Cosmology and Astrophysics, Case Western Reserve University,Cleveland, OH, USA 44106}
\def\ArtInstChicago{Liberal Arts Department, School of the Art Institute of Chicago, Chicago, IL, USA 60603}
\def\CfA{Harvard-Smithsonian Center for Astrophysics, Cambridge, MA, USA 02138}
\def\KIPACStanford{Kavli Institute for Particle Astrophysics and Cosmology, Stanford University, Stanford, CA 94305}
\def\Stanford{Department of Physics, Stanford University, Stanford, CA 94305}

\def\AAToronto{Department of Astronomy \& Astrophysics, University of Toronto, 50 St George St, Toronto, ON, M5S 3H4, Canada}
\def\MPR{Max-Planck-Institut f\"{u}r Radioastronomie, Auf dem H\"{u}gel 69 D-53121 Bonn, Germany}
\def\TexAust{Department of Astronomy, University of Texas at Austin, Austin, TX, USA 78712}
\def\NASAGoddard{NASA Goddard Space Flight Center, Greenbelt, MD, USA 20771}

\begin{document}

\title{Millimeter-wave Point Sources from the 2500-square-degree SPT-SZ Survey: Catalog and Population Statistics}

\author [0000-0002-5370-6651] {W.~B.~Everett}\affiliation{\ColoradoAPS}
\author{L.~Zhang}\affiliation{\SantaBarbara}\affiliation{\IllinoisPhysics}
\author[0000-0001-9000-5013]{T.~M.~Crawford}\affiliation{\KICPChicago}\affiliation{\AAUChicago}
\author[0000-0001-7192-3871]{J.~D.~Vieira}\affiliation{\IllinoisPhysics}\affiliation{\IllinoisAstronomy}
\author[0000-0002-6290-3198]{M.~Aravena}\affiliation{\Diego}
\author{M.~A.~Archipley}\affiliation{\IllinoisAstronomy}
\author[0000-0002-6338-0069]{J.~E.~Austermann}\affiliation{\NIST}\affiliation{\ColoradoPhysics}
\author[0000-0002-5108-6823]{B.~A.~Benson}\affiliation{\KICPChicago}\affiliation{\AAUChicago}\affiliation{\Fermilab}
\author[0000-0001-7665-5079]{L.~E.~Bleem}\affiliation{\KICPChicago}\affiliation{\ArgonneHEP}
\author[0000-0002-2044-7665]{J.~E.~Carlstrom}\affiliation{\KICPChicago}\affiliation{\AAUChicago}\affiliation{\ArgonneHEP}\affiliation{\PhysicsUChicago}\affiliation{\EFIChicago}
\author[0000-0002-6311-0448]{C.~L.~Chang}\affiliation{\KICPChicago}\affiliation{\AAUChicago}\affiliation{\ArgonneHEP}
\author{S.~Chapman}\affiliation{\Dalhousie}\affiliation{\Cambridge}
\author{A.~T.~Crites}\affiliation{\KICPChicago}\affiliation{\AAUChicago}\affiliation{\Caltech}
\author{T.~de~Haan}\affiliation{\Berkeley}\affiliation{\LBNL}
\author{M.~A.~Dobbs}\affiliation{\McGill}\affiliation{\CIFAR}
\author[0000-0001-7874-0445]{E.~M.~George}\affiliation{\Berkeley}\affiliation{\ESO}
\author[0000-0003-2606-9340]{N.~W.~Halverson}\affiliation{\ColoradoAPS}\affiliation{\ColoradoPhysics}
\author{N.~Harrington}\affiliation{\Berkeley}
\author{G.~P.~Holder}\affiliation{\IllinoisPhysics}\affiliation{\IllinoisAstronomy}\affiliation{\CIFAR}
\author{W.~L.~Holzapfel}\affiliation{\Berkeley}
\author{J.~D.~Hrubes}\affiliation{\UChicago}
\author{L.~Knox}\affiliation{\Davis}
\author[0000-0003-3106-3218]{A.~T.~Lee}\affiliation{\Berkeley}\affiliation{\LBNL}
\author{D.~Luong-Van}\affiliation{\UChicago}
\author[0000-0003-2385-6904]{A.~C.~Mangian}\affiliation{\IllinoisAstronomy}
\author[0000-0002-2367-1080]{D.~P.~Marrone}\affiliation{\Arizona}
\author{J.~J.~McMahon}\affiliation{\Michigan}
\author{S.~S.~Meyer}\affiliation{\KICPChicago}\affiliation{\AAUChicago}\affiliation{\PhysicsUChicago}\affiliation{\EFIChicago}
\author{L.~M.~Mocanu}\affiliation{\KICPChicago}\affiliation{\AAUChicago}\affiliation{\Oslo}
\author[0000-0002-6875-2087]{J.~J.~Mohr}\affiliation{\Munich}\affiliation{\ExcellenceCluster}\affiliation{\MPE}
\author{T.~Natoli}\affiliation{\KICPChicago}\affiliation{\AAUChicago}
\author{S.~Padin}\affiliation{\KICPChicago}\affiliation{\AAUChicago}\affiliation{\Caltech}
\author{C.~Pryke}\affiliation{\Minnesota}
\author[0000-0003-2226-9169]{C.~L.~Reichardt}\affiliation{\Melbourne}
\author[0000-0001-7477-1586]{C.~A.~Reuter}\affiliation{\IllinoisAstronomy}
\author{J.~E.~Ruhl}\affiliation{\CaseWestern}
\author[0000-0002-1062-1842]{J.~T.~Sayre}\affiliation{\ColoradoAPS}\affiliation{\ColoradoPhysics}
\author{K.~K.~Schaffer}\affiliation{\KICPChicago}\affiliation{\EFIChicago}\affiliation{\ArtInstChicago}
\author[0000-0002-2757-1423]{E.~Shirokoff}\affiliation{\KICPChicago}\affiliation{\AAUChicago}\affiliation{\Berkeley}
\author[0000-0003-3256-5615]{J.~S.~Spilker}\affiliation{\TexAust}
\author{B.~Stalder}\affiliation{\CfA}
\author{Z.~Staniszewski}\affiliation{\CaseWestern}
\author[0000-0002-2718-9996]{A.~A.~Stark}\affiliation{\CfA}
\author{K.~T.~Story}\affiliation{\KIPACStanford}\affiliation{\Stanford}
\author{E.~R.~Switzer}\affiliation{\NASAGoddard}
\author{K.~Vanderlinde}\affiliation{\Dunlap}\affiliation{\AAToronto} 
\author{A.~Wei{\ss}}\affiliation{\MPR}
\author{R.~Williamson}\affiliation{\KICPChicago}\affiliation{\AAUChicago}
 
\email{wendeline.everett@colorado.edu}
 
\begin{abstract}
We present a catalog of emissive point sources detected in the SPT-SZ survey, a contiguous 2530-square-degree area surveyed with the South Pole Telescope (SPT) from 2008 - 2011 in three bands centered at 95, 150, and 220\,GHz.  
The catalog contains 4845 sources measured at a significance of 4.5\,$\sigma$ or greater in at least one band, corresponding to detections above approximately 9.8, 5.8, and 20.4\,mJy in 95, 150, and 220\,GHz, respectively.  
Spectral behavior in the SPT bands is used for source classification into two populations based on the underlying physical mechanisms of compact, emissive sources that are bright at millimeter wavelengths: synchrotron radiation from active galactic nuclei and thermal emission from dust.
The latter population includes a component of high-redshift sources often referred to as submillimeter galaxies (SMGs).
In the relatively bright flux ranges probed by the survey, these sources are expected to be magnified by strong gravitational lensing.
The survey also contains sources consistent with protoclusters, groups of dusty galaxies at high redshift undergoing collapse.
We cross-match the SPT-SZ catalog with external catalogs at radio, infrared, and X-ray wavelengths and identify available redshift information.
The catalog splits into 3980 synchrotron-dominated and 865 dust-dominated sources and we determine a list of 506 SMGs.  
Ten sources in the catalog are identified as stars.
We calculate number counts for the full catalog, and synchrotron and dusty components, using a bootstrap method and compare our measured counts with models.  
This paper represents the third and final catalog of point sources in the SPT-SZ survey.
\end{abstract}



\section{Introduction} 
\label{sec:intro}
The South Pole Telescope (SPT, \citealt{carlstrom11}) is a 10-m millimeter-wavelength telescope which has provided an immensely rich set of survey data.  
From 2008 -- 2011, the SPT was used to conduct a 2500-square-degree survey of the southern sky in three bands centered at 95, 150, and 220\,GHz with arcminute resolution.
While the primary science goal of this survey, the South Pole Telescope Sunyaev-Zel'dovich (SPT-SZ) Survey, was a search for galaxy clusters using the Sunyaev-Zel'dovich effect~\citep{bleem15b}, the dataset is also ideal for finding compact, extragalactic sources of emission~\citep{vieira10}.  
The large area, high resolution, and comparatively low noise of the full SPT-SZ survey provide an extensive catalog of new sources selected at millimeter (mm) wavelengths spanning flux densities of a few mJy to many Jy.
The multi-frequency nature of the dataset further provides the opportunity for population separation based on spectral characteristics of different types of sources.

Broadly speaking, extragalactic sources that are bright at mm wavelengths fall into two categories: sources whose flux increases with frequency, and sources whose flux is either nearly constant or decreasing with frequency.  
Flat- or falling-spectrum sources are generally associated with active galactic nuclei (AGN), where the source of the mm flux is from acceleration of relativistic charged particles producing synchrotron radiation.  
Rising-spectrum sources are predominantly dusty star-forming galaxies (DSFGs). The high dust content of these sources makes them difficult to detect at optical wavelengths, but the mm and submillimeter (submm) flux from these sources is thermal emission from the dust itself, making the mm/submm wavebands particularly useful for identifying and observing this population.

Historically, the synchrotron population has been well-studied at radio wavelengths (a review of the current understanding of radio source populations from millimeter and radio surveys can be found in \citealt{dezotti10}).  
The spectra of radio sources are generally characterized by a power law relating source flux density, $S$, to frequency, $\nu$: $S \propto \nu^\alpha$.  
AGN-fueled radio sources can be roughly separated into two populations: flat-spectrum sources, generally defined to have  $\alpha > -0.5$, and steep-spectrum sources with $\alpha < -0.5$.  
In the currently accepted ``unified model"~\citep[e.g.,][]{urry95,netzer15}, these two populations are actually the same type of physical object whose spectral appearance depends on the orientation of the observer relative to the axis of the characteristic jets emerging from the central black hole.  
In side-on observations relative to the typically-extended jets, the optically thin lobes create a steep component of the spectrum at radio frequencies, and the central black hole engine is obscured by the dusty accretion torus.  
For sight lines along the axis of the jet, the object appears as a compact flat-spectrum source also referred to as a blazar.  

The characterization of dusty sources has progressed significantly as mm- and submm-wave surveys have grown in size and resolving power in the last several decades.  
In the 1980s, the all-sky infrared satellite IRAS discovered a population of 18,351 extra-galactic sources~\citep{saunders00b}.  
Most of these were at relatively low redshifts, $z \ltsim 0.3$, with emission dominated by dust, and were classified as luminous infrared (IR) galaxies (LIRGs) (10$^{11} < L_{\textrm{IR}} < 10^{12}$ L$_\odot$) and ultraluminous IR galaxies (ULIRGs) (10$^{12} < L_{\textrm{IR}} < 10^{13}$ L$_\odot$), compared with typical spiral galaxies with luminosities around 10$^{10}$ L$_\odot$ \citep{blain02}.  
Beginning in the late 1990s, observations at 450 and 850\,$\mu$m with the Submillimeter Common-User Bolometer Array (SCUBA) instrument on the James Clerk Maxwell Telescope (JCMT)~\citep{holland99} discovered a high-redshift component of the DSFG population, which were termed submillimeter galaxies (SMGs), due to the wavelength at which they were identified.  
These early surveys of SMGs covered relatively small areas, only a few square degrees at most, and as a result traced out populations of relatively dim sources~\citep[e.g.,][]{smail97,hughes98c,barger98,eales00,cowie02,scott02b,bertoldi07,weiss09}.  

The advent of large-area and multi-band surveys allowed detections probing the brightest and rarest SMGs.  
This included surveys conducted using the SPT at 1.4, 2.0, and 3.2\,mm \citep[e.g.,][]{vieira10}, the Spectral and Photometric Imaging Receiver (SPIRE) at 250, 350, and 500\,$\mu$m on the {\it Herschel Space Observatory}\citep[e.g.,][]{eales10}, the Planck satellite~\citep[e.g.,][]{planck15-26}, and the Atacama Cosmology Telescope~\citep[e.g.,][]{gralla19}.  

The first released compact-source sample from the SPT, \citet{vieira10}, included a population of extremely bright ($\sim$30\,mJy at 1.4\,mm), rising-spectrum sources that did not have counterparts in IRAS catalogues (indicating they were most likely at high redshift).
Follow-up observations of these sources and a similarly bright population of sources detected in early {\it Herschel} surveys using telescopes such as the Atacama Large Millimeter/Submillimeter Array (ALMA) and the Submillimeter Array (SMA) have demonstrated that these objects are indeed at high redshift and most of them are magnified by strong gravitational lensing by a massive object along the line of sight~\citep{negrello10,vieira13,spilker16,hezaveh16}.

Thermal dust emission at high redshift is probed almost uniquely by moderate-to-high-resolution, mm/submm observatories, including the SPT and {\it Herschel}. Where high-redshift observations of other emission mechanisms at other wavelengths suffer from cosmological dimming, mm/submm observations benefit from a strong negative K-correction~\citep{blain96} that results in nearly constant observed flux density for a source with a dust-like spectrum, out to approximately $z=10$ for mm wavelengths. The combination of this effect and the phenomenon of gravitational lensing makes large-area mm/submm surveys uniquely powerful in studying the nature of star formation at the highest redshifts possible.

In this work, we present results from the full 2500 square degrees of the SPT-SZ survey; this analysis is an extension on the work of two previous papers: \citet{vieira10} (hereafter V10), and \citet{mocanu13} (hereafter M13), and builds on the same analysis pipeline.  
V10 developed the source-finding pipeline and applied it to a single field covering 87 square degrees observed in 2008 in two frequencies.  
M13 expanded that analysis to 5 fields, two observed in 2008 and three in 2009 (771 square degrees in total), and added a third frequency.  
In this current paper, we add 1759 square degrees of previously unanalyzed data and include additional data for two fields which were re-observed in 2010 and 2011.  
We adjust the previous pipeline to be compatible with the goals of the full survey (full area coverage) and work to optimize elements in the pipeline chain.  
Sections~\ref{sec:observ} and \ref{sec:reduction} present an overview description of the data and analysis pipeline.  
Section~\ref{sec:catalog} provides a description and characterization of the catalog, including source population separation, and Section~\ref{sec:number_counts} presents the number counts.  
Section~\ref{sec:discussion} presents a discussion of the results and conclusions can be found in Section~\ref{sec:conclusion}.
Throughout the work, we assume a standard $\Lambda$CDM cosmology with $H_0 = 70$, $\Omega_m = 0.3$, and $\Omega_\Lambda = 0.7$. 


\section{Observations} \label{sec:observ}
The South Pole Telescope (SPT, \citealt{carlstrom11}), a 10-meter telescope designed for observations in millimeter wavelengths, is located at the geographic South Pole and was designed to measure low-contrast sources such as CMB anisotropies with high sensitivity.
The first camera for the SPT, SPT-SZ, contained a 960-pixel array of transition-edge-sensor bolometers, with sensitivity in three bands centered at 95, 150, and 220\,GHz (3.2, 2.0, and 1.4\,mm, respectively).  
The SPT-SZ receiver had an angular resolution of roughly 1.7, 1.2, and 1.0\,arcmin at 95, 150, and 220\,GHz, respectively, with a 1-degree diffraction-limited field-of-view.  
The pixels on the focal plane were arranged into 6 triangular wedges forming a hexagon, with each wedge sensitive in a single band. 

The SPT-SZ survey represents the culmination of four years of observations, 2008--2011, of roughly 2500 square degrees on the sky.  
The sky area covered spans the region in the southern hemisphere from roughly declination (decl.) -65 to -40 degrees and from right ascension (R.A.) $20^\text{h}$ to $7^\text{h}$, avoiding sky area contaminated by emission from the Galaxy.   
Over the duration of the survey, the composition of the receiver changed slightly.  
In 2008, the focal plane was composed of three 150\,GHz wedges, two 220\,GHz wedges, and a single 95\,GHz wedge, but the 95\,GHz wedge failed to produce science-quality data.  
For 2009, one 220\,GHz wedge was swapped for another 150 GHz wedge, and the 95\,GHz wedge was upgraded to an improved-quality wedge, resulting in four 150\,GHz wedges, and one each of 220\,GHz and 95\,GHz.  
The composition of the focal plane then remained the same for 2009, 2010, and 2011.

\begin{figure*}[!ht] 
\begin{center} 
\subfigure{\includegraphics[width=17cm]{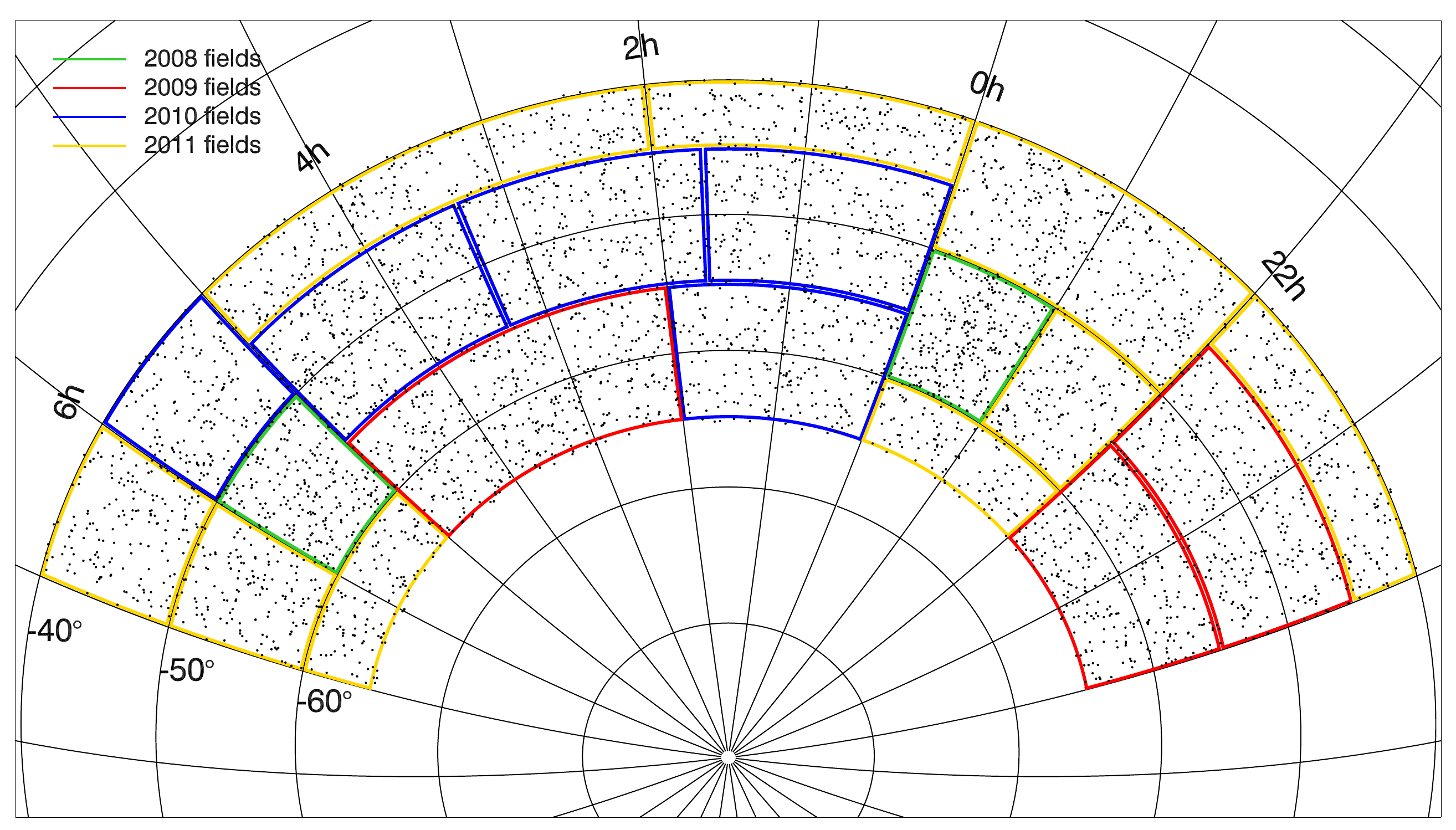}}
\caption{The 2500-square-degree SPT-SZ survey was observed in 19 separate fields shown in outlines.  Field outlines are only illustrative of field locations and areas and are not the masks used in the analysis.  The two fields observed in 2008 were re-observed in 2010 and 2011, which is not indicated in this figure.  Black dots indicate the locations of all sources reported in the catalog of this work.
\label{fig:surveyfields}}
\end{center}
\end{figure*}

The full 2500-square-degree area was split into 19 contiguous fields which were observed independently.  
The characteristics of each field are presented in Table~\ref{tab:fieldstable}, and Figure~\ref{fig:surveyfields} shows the location of each field on the sky.  
In observing a given field, the telescope started in one corner, scanned back and forth across the sky in constant elevation and then took a step in elevation and repeated until it had covered the desired area in that field.  
Scan speeds varied between 0.25 and 0.42\,deg/sec.  
Between observations, the telescope initial starting position was dithered to achieve uniform coverage of each field.  
Only data from the constant-speed portion of each scan is used in the map for that particular observation.  
The three 2009 fields, {\sc ra21hdec-50}, {\sc ra3h30dec-60}, and {\sc ra21hdec-60}, and one 2008 field, {\sc ra23h30dec-55}, were observed using a lead-trail scan strategy, in which the field is split into two halves, left and right.  
The two halves were observed independently, delayed such that due to sky rotation, the second half had drifted so that the two halves were observed over the same azimuth range.  
This allows for the possibility of the removal of ground-synchronous contamination.  
However, ground contamination in those fields was measured to be negligible, so the lead and trail portions are simply coadded in this analysis.  
The rest of the fields were observed using a simple scan in azimuth, except for the {\sc ra21hdec-50} field, for which a portion of the observations used an elevation scan, where the telescope scans up and down in elevation while allowing the field to drift through the field-of-view in azimuth.  
Techniques for analyzing this field are discussed in detail in M13.  
The observation strategy for each field was designed to produce as close as possible a uniform-depth survey across the full area, except for two fields, {\sc ra5h30dec-55} and {\sc ra23h30dec-55}, both of which were observed originally in 2008 and then re-observed in either 2010 or 2011, to add data at 95\,GHz which was unavailable in 2008 and nominally to observe to twice the depth of the 2008 survey in 150\,GHz.


 \begin{deluxetable*}{lcccccccc}
\centering
\tablecaption{SPT fields used in this work}
\tablehead{
\colhead{Name} & \colhead{Year} & \colhead{R.A.} & \colhead{Decl.} & \colhead{$\Delta$R.A.} & \colhead{$\Delta$Decl.} &\colhead{Eff. Area} & \colhead{No. sectors}\\
&& [$^\circ$] & [$^\circ$] & [$^\circ$] & [$^\circ$] & [deg$^2$] & \\
&&&&&&&}
\startdata
\fivefiffi & 		2008/2011		&	82.5 &	-55.0 &	15 &	10 &		89 &		3x3\\
\twthreefiffi & 	2008/2010	&	352.5 &	-55.0 &	15 &	10 &		108 &	3x3\\
\twonesix &	2009 		&	315.0 &	-60.0 &	30 &	10 &		150 &	6x3\\
\threesix &	2009			& 	52.5 &	-60.0 &	45 &	10 &		225 &	8x3\\
\twonefif &		2009 		&	315.0 &	-50.0 &	30 &	10 &		193 &	6x3\\
\fourten &		2010			& 	62.5 &	-50.0 &	25 &	10 &		166 &	5x3\\
\zerofif &	 	2010			& 	12.5 &	-50.0 &	25 &	10 &		152 &	5x3\\
\twothir &	 	2010			&	37.5 &	-50.0 &	25 &	10 &		155 &	5x3\\
\onesix & 		2010	  		&	15.0 &	-60.0 &	30 &	10 &		140 &	6x3\\
\fivefourfi &	2010			&	82.5 &	-45.0 &	15 &	10 &		105 &	3x3\\
\sixfiffi & 		2011			&	97.5 &	-55.0 &	15 &	10 &		82 &		3x3\\
\twthreesix & 	2011			&	345.0 &	-62.5 &	30 &	5 &		65 &		6x2\\
\twonefour & 	2011			&	315.0 &	-42.5 &	30 &	5 &		118 &	6x2\\
\twtwo & 		2011			&	337.5 &	-55.0 &	15 &	10 &		73 &		3x3\\
\twthreefourfi &	2011 		& 	345.0 &	-45.0 &	30 &	10 &		221 &	6x3\\
\sixsix &		2011 		&	90.0 &	-62.5 &	30 &	5 &		65 &		6x2\\
\threefour & 	2011			&	52.5 &	-42.5 &	45 &	5 &		185 &	8x2\\
\onefour &		2011 		&	15.0 &	-42.5 &	30 &	5 &		126 &	6x2\\
\sixfourfi &	 	2011			&	97.5 &	-45.0 &	15 &	10 &		112 &	3x3\\
\hline
Total: & & & & & & 2530
\enddata
\label{tab:fieldstable}
\tablecomments{Locations and sizes of the fields included in this work.  For each field we give the center of the field in Right Ascension (R.A.) and Declination (Decl.), the extent of the field in Right Ascension and Declination, the number of sectors the field is divided into (see Section~\ref{sec:optfiltsection}), the effective field area as defined by the apodization mask.}
\end{deluxetable*}


\section{Data reduction and analysis} 
\label{sec:reduction}
The following section describes the steps in the analysis pipeline from timestream data for individual bolometers to source catalogs.  
These steps include: filtering of each bolometer's timestream data for each scan; forming a single-observation map by coadding each bolometer's contribution to map pixels, and then forming a single map for each field by coadding all single-observation maps; constructing masks to define the high-weight regions of the fields for source-finding; developing an optimal filter to amplify the signal-to-noise for detecting compact sources; extracting sources separately for each band using a CLEAN algorithm; and finally, forming a single multi-band catalog, taking into account the effect of flux biases and overlap regions between fields.

\subsection{Timestream filtering and mapmaking}
\label{sec:filtering}
The response of each detector is recorded at 100\,Hz as time-ordered data (TOD) as the telescope scans across the sky.  
We apply a set of filters to the TOD to suppress noise above the temporal frequency corresponding to the map pixel size and low-frequency noise due to atmosphere.  
The filtering we apply in this work is very similar to that in M13.  
The data are low-pass filtered above a temporal frequency corresponding to $\ell = 37500$ in the scan direction to remove noise on scales smaller than the chosen map pixel size, 0.25\,arcmin.
To mitigate atmospheric noise, we apply a first-order polynomial subtraction and a high-pass filter below $\ell = 246$.  
Since atmospheric noise will be spatially coherent on the size scale of the detector wedges, we also remove a mean across each wedge of the receiver from all well-performing bolometers at each time step.

The filtered TOD for each bolometer are then coadded into 0.25 by 0.25\,arcmin pixels by inverse-variance weighting, adding contributions from bolometers to each pixel to form a single map per observation.  
The weights for each bolometer are calculated from the power spectral density of each detector's TOD in the range from 1 to 3\,Hz.  
We pixelize each field using an oblique Lambert equal-area projection.  
This choice of projection is important for source-finding because it preserves the source shape across the full area of the map.  
However, it also produces complications in the analysis, since the scan direction rotates with pixel location in the map.  
The ramifications of this are discussed in Section~\ref{sec:optfiltsection}.

To make final coadded maps from all the observations, we apply several cuts (which have been previously shown to be useful for SPT data--see, e.g., \citealt{schaffer11}), based on the mean weights and mean RMS of the uniform-weight region of each single-observation map.  
We cut on excessively high median weights which occurs when a bolometer's TOD has anomalously low noise.  
In the past, this has been shown to correspond to poor bolometer behavior, such as when a detector changes operating point due to shifts in the amount of loading~\citep{schaffer11}.  
For maps with reasonably good weather conditions, the weights scale well with the RMS of the map; however, for poor weather days, $1/f$ noise dominates the RMS and the 1--3\,Hz range is no longer a good estimate of  the weight that should be assigned to that bolometer.  
Therefore, we also perform a cut on observations where the map RMS does not scale properly with the median weight in the map.  
The single-observation maps that survive the cuts are coadded by inverse-variance-weighting each pixel to form a single coadded map per field and per band.

To calibrate the maps, we use both a relative and absolute calibration.  
The relative calibration of the TOD from one observation to the next is done through repeated observations of the galactic HII region RCW38 and reference from a thermal calibrator source installed in the bolometer optical path~\citep{schaffer11}.  
The absolute calibration is determined from comparisons of the SPT power spectrum to that from \citet{planck15-11} in the $\ell$ range from 682 to 1178.  
This results in fractional errors in temperature of 1.05\%, 1.15\%, and 2.24\% in 95, 150, and 220\,GHz, respectively.

The pointing model used for constructing maps is based on regular measurements of galactic HII regions in addition to data recorded by sensors on the telescope measuring temperature, linear displacement, and tilt.  
To check the absolute astrometry, we correct the global pointing of each field by cross-matching the positions of the brightest 40 SPT-SZ sources in each of the three bands per field to source locations in the Australia Telescope 20 GHz (AT20G) Survey catalog~\citep{murphy10}, which has an RMS positional accuracy of 1 arcsec.  
We then fit for a global pointing correction using the cross-matched locations. 
We iterate on this process of cross-matching and fitting for a correction until the calculated offset is smaller than the residual scatter, and we find the RMS residual pointing scatter for the on-average 26 brightest sources with cross-matches in each field to be 4.3\,arcsec in declination and 4.6\,arcsec in R.A.$\cdot \cos$(Decl.).

\subsection{Mask construction}\label{masks}
Each field is analyzed separately in our pipeline for extracting sources.  
We then cross-match the single-field catalogs at the end, accounting for places where fields overlap, to form a single catalog for the survey.  
Field masking is needed to exclude low signal-to-noise edges due to turn-around regions of the scan strategy and non-uniform array coverage between bands on the focal plane.  
However, we also want to define masks such that we have continuous coverage of the full 2500-square-degree survey area.  
This requires that we define separate masks per band for each field, because different bands occupy physically offset locations in different wedges on the focal plane and therefore observe slightly offset regions on the sky.  
In principle, this choice only adds slightly more complicated bookkeeping for cataloging sources, since now a source detected on the edge of one field in one band could be detected in an adjacent field in a different band.  
To achieve continuous coverage, we also need to extend the field masks to lower signal-to-noise regions compared with M13, making the noise level within each field slightly less uniform.

\subsection{Optimal filtering for source extraction}
\label{sec:optfiltsection}
As the sources we detect are expected to be unresolved by the telescope (except for nearby sources), a source in our maps should manifest in the maps as an SPT beam with the time-stream filtering applied.  
We can improve our signal-to-noise for detecting objects with an expected source shape using an appropriate optimal filter.  
The filter takes advantage of knowledge of the source shape and the noise in the region of the map where the source is located, which includes residual atmosphere, instrument noise, and the primary anisotropies of the CMB, which acts as a source of noise for the detection of compact, extragalactic sources.  
The first component needed in constructing the source profile is the beam, which is measured using a combination of observations of Jupiter and Venus, as well as the brightest point sources in the fields.  
The main lobes of the beams are measured to be well-described by Gaussian functions with FWHM of 1.7, 1.2, and 1.0\,arcmin for 95, 150, and 220\,GHz, respectively.  
The sidelobes of the beams are downweighted in the filter, and therefore are unimportant for the point source analysis pipeline.  
To model the source profile, we insert a beam into a noiseless map, and then reobserve the source once for each single-observation map using the characteristics of the telescope's performance for that particular observation and the time-stream filtering.  
The Fourier-domain version of this source profile is used as the transfer function for our maps.  
All the single-observation transfer functions are then coadded into a single transfer function for the coadded map.  

Following formalism set up in \citet{tegmark98} and \citet{haehnelt96}, to maximize the signal-to-noise of sources in the map, we filter the map using an appropriately normalized version of the signal-to-noise of the source.  
We apply the optimal filter, $\psi$, in the Fourier domain given by:
\begin{equation}
\psi = \frac{\tau^TN^{-1}}{\tau^TN^{-1}\tau},
\end{equation}
where $\tau$ is the transfer function and $N$ is the 2D noise power spectral density (PSD), resulting in a filtered map still in units of temperature.  
In addition to the source profile, we also need to characterize the noise around each source.  
To do this, we find the PSD of the noise of the coadded map by averaging 100 versions of difference maps.  
Each difference map is constructed by multiplying a randomly chosen half of the individual observation maps by -1 and adding them.  
The 2D power spectrum of the Fourier transform of each difference map are then averaged to generate a single 2D noise PSD for the coadded map.  
Because differencing two individual observation maps cancels out the contribution to the noise from the CMB anisotropies, we add back in a Gaussian realization of the best-fit CMB power spectrum from \citet{keisler11}.
Smaller contributions to the noise, such as from secondary CMB anisotropies, thermal and kinetic SZ effects, are neglected in the filter construction.

The construction of the optimal filter is complicated by two characteristics of the SPT data.  
Because of the telescope's location at the South Pole, the scan direction is always along constant declination.  
This means that the effect of time-stream filtering is anisotropic in the maps, and we essentially have an anisotropic beam.  
We account for the smearing of the beam in the scan direction by calculating the transfer function and applying the transfer function during source extraction.  
But, for point source work, we use an area-preserving projection, which causes the scan direction to rotate with respect to the axes of the pixel orientation of the maps.  
Therefore, our anisotropic beam in the map rotates with respect to the pixel $x$-$y$ location.  
The second characteristic of the SPT data is that the noise in the maps varies with declination.  
Because the telescope scans the same distance in azimuth in the same amount of time regardless of elevation, but this distance corresponds to less physical distance farther from the equator, the result is a noise level with a gradient in declination through our maps, with slightly less noise at higher declination.  
To account for these two position-dependent complications, we divide up each field in a number of sectors which are small enough that an assumption of zero source rotation and noise uniformity is reasonable.  
We then calculate separate transfer functions and noise PSDs for each sector independently, construct a single optimal filter for each sector, and extract sources separately per sector.  
Further description of the process for creating these data products and their salient features can be found in the SPT 2008 data release paper~\citep{schaffer11}.  
The number of sectors per field is shown in Table~\ref{tab:fieldstable}.

Essentially, splitting up each field into sectors is a compromise between computation time and accuracy.  
We test that the sizes of our sectors are appropriate, i.e. that the measured flux density of sources is unaffected by the size of the sector we choose, by applying the transfer function and noise PSD from adjacent sectors to a sector where the effects of noise and scan rotation angle are the most severe, and check that the resultant change in the flux densities of the sources in that sector are below the noise level of the sector.  
We also test that the noise in different sectors does not differ by more than 5\%.

We found in creating and testing our optimal filter that there was residual noise due to incomplete averaging in the creation of the PSD.  
This resulted in excess noise in the source extraction template, resulting in excess noise in the optimally filtered maps.  
To mitigate this effect, we apply a smoothing kernel to the optimal filter in the Fourier domain.  
To test that the strength of the filtering is optimal, we sweep through a range of kernel size while monitoring the noise.  
As we apply a stronger and stronger smoothing to the filter, we see that the noise level in the optimally filtered map is reduced, indicating that the excess noise being introduced by the filter is being diminished.  
But, applying stronger smoothing past a certain point eventually causes the noise level to once again rise, as real noise information in the PSD will begin to be cut, and the filter becomes a less realistic description of the actual signal-to-noise in the map and therefore less optimal.  
We take the minimum noise level as our optimized smoothing kernel size.  

\subsection{Source extraction algorithm}
\label{sec:clean}
After optimally filtering the map, we locate and extract source flux densities using a CLEAN algorithm~\citep{hogbom74}.  
CLEANing was developed originally for radio interferometry, where uneven baseline sampling and a finite number of antennae produce incomplete sampling of the Fourier domain.  
In turn, this effect produces sidelobes on the beam (a so-called ``dirty beam"), which is analogous to the wings on the SPT beam due to the total applied optimal filter.  
The CLEAN algorithm detects and removes sources iteratively using a template source profile, which allows for the detection of fainter sources hidden underneath the dirty-beam wings of brighter sources.  
The source template we employ for CLEANing takes into account that we have optimally filtered the map, however, technically this optimal filter is only optimal for a source located at the center of a sector (which is where the simulated beam was placed when calculating the transfer functions for each sector).  
For sources off center in the sector, this optimal filter is at a slightly incorrect rotation angle.  In order to form a template for each source, $\tau'$, we rotate the source profile (which is the map space version of the transfer function, $\tau$) to the correct rotation angle for the $x$-$y$ pixel location in the sector, and then convolve it with the optimal filter for that sector (which is not rotated).  
Effectively, our source template (in the Fourier domain) is given by, 
\begin{equation}
\tau' = \psi \tau,
\end{equation}
where $\psi$ is the optimal filter function.  
Each sector of a field has been filtered separately, so we also perform the cleaning separately per sector and then unite the catalogs of detected sources from all fields.  
Since sources have long wings in the scan direction due to time-stream filtering, we need to account for the possibility of false detections from the wings of sources bleeding into a sector from sources just outside the sector.  
We do this by defining a sector pixel mask to outline the source-finding area for each sector and a second mask which covers a larger area than this sector mask.  
We define the larger masks such that the extra space on the left and right sides relative to the sector mask edges will be wider than the wings on all but the very obviously brightest sources, which we check by hand if they occur at the edge of a sector.  
The CLEAN algorithm is applied to the area of the larger mask for each sector, but only the sources that are within the smaller sector pixel mask are saved into the catalog.  

To better account for non-uniformity in noise level across each sector, we construct a scaled noise map using the weight map for each field's coadded map.  
We apply the optimal filter for each sector to the inverse of the weight map, and then scale each sector's RMS noise by the square-root of the ratio of each sector's median weight to its filtered weight map.  
In essence, we construct a local scaled-noise map, which can be used to construct a local signal-to-noise map when combined with the optimally filtered map.  
Thus, rather than assume a single noise value per sector when CLEANing, we take into account any local noise non-uniformity and CLEAN down to a locally-determined signal-to-noise threshold.  
The most noticeable differences resulting from the implementation of this method arise along the edges of the map, which are noisier than the RMS noise of the sectors which include these regions, and fields which were observed with a lead-trail observing strategy and have low-noise strips where the lead and trail observations overlap.

The steps of the CLEANing are as follows:

\begin{enumerate}
\item Find the location of the brightest pixel in a given sector in the optimally filtered map.
\item Rotate the source profile for that sector by the appropriate rotation angle for that $x$-$y$ pixel location, and convolve it with the optimal filter for that sector.  This is the source template.
\item Subtract the source template, scaled to the flux of the pixel and multiplied by a loop gain coefficient.  The loop gain is a multiplicative factor between 0 and 1 to account for non-ideal characteristics of the CLEANing pipeline, such as imperfections in the source model, extended sources, and finite pixelization in the map.  We choose a loop gain of 0.1.
\item Find the next brightest pixel in the map and repeat the process until all pixels in the map have significance below the chosen signal-to-noise detection threshold, in this case 4.5 times the scaled RMS noise of that pixel location.
\end{enumerate}
We extract negative sources as well as positive sources during the CLEANing process.
Because the CLEANing is performed with a loop gain, bright sources will be broken up into multiple brightest pixels during the CLEANing.  
Once the CLEANing is finished (i.e. no pixels in the map remain above the chosen significance threshold), the pixels found by the CLEAN are associated into sources using a radius of association that is brightness-dependent, scaling from roughly 38\,arcsec for detections of 4.5\,$\sigma$ up to 2\,arcmin for detections of 200\,$\sigma$ or larger.
All of the pixels associated with a single source are used in a centroiding process to find the source's position.  
The post-CLEAN map, with all sources removed, is called a residual map, and will be used in later steps of the analysis.

After locating sources in the map, we convert from units of CMB temperature fluctuations to units of flux density.  
Optimally filtering the map is equivalent to fitting the map with a source shape, and the value of the brightest pixel of each source can be used to calculate the integrated source flux.  
Specifically, we calculate the flux of each source by stacking all of the CLEAN components removed for a given source onto the residual map and taking the maximum in a cutout region.
The maps are calibrated in units of CMB temperature fluctuations, so we convert to flux density units using 
\begin{equation}
S[\text{Jy}] = T_\text{peak}\cdot \Delta\Omega_\text{f} \cdot 10^{26} \cdot \frac{2k_\text{B}}{c^2}\left(\frac{k_\text{B}T_\text{CMB}}{h}\right)^2\frac{x^4e^x}{(e^x-1)^2}, 
\end{equation}
where $x = h\nu/(k_\text{B}T_\text{CMB})$, and $\Delta \Omega_\text{f}$ is the effective solid angle under a filtered source template, given by
\begin{equation}
\Delta \Omega_\text{f} = \left[\int d^2 k\ \psi(k_x,k_y)\ \tau(k_x,k_y)\right]^{-1}.
\end{equation}

We inspect detected sources for obvious spurious detections, such as false sources created by the effect of the timestream filtering on bright galaxy clusters and spurious detections very close to extremely bright sources.  
We also inspect for extended sources, discussed in more detail in Section \ref{sec:extsrcsection}.  
Obvious false detections are trimmed; for the sake of completeness in the catalog, information on extendedness  is not used to remove any sources, but is retained as a flag in the catalog.


\subsection{Flux biases and three-band flux deboosting}
The raw fluxes in our catalogs are subject to several biases, which must be carefully considered before the fluxes can be used for population statistics.  
The first is due to the fact that the underlying source number count populations are steep functions of flux.  
We expect the noise in the map to be Gaussian, but since there are many more dim sources than bright sources, it is much more likely that a detection at a given significance is a dim source on top of a positive noise fluctuation than a bright source on top of a negative noise fluctuation.  
Therefore, more sources below a significance threshold will be bumped above the threshold and detected as sources due to noise than will be bumped below, resulting in a positive flux bias which most strongly affects low signal-to-noise sources.  
This bias is closely related to Eddington bias, although that term is generally applied to counts as a function of flux rather than the fluxes of individual objects~\citep{mocanu13}.  
When applied to individual sources, we refer to this bias as ``flux boosting" and its correction as ``flux de-boosting."

A second bias is due to the fact that we estimate source flux based on peak pixel brightness.  
A positive noise fluctuation near a source will pull the detected peak position away from the true position and also return a higher flux, whereas a nearby negative noise fluctuation will not have nearly as as strong a corresponding opposite effect on either the returned position or flux~\citep{austermann10}.  
For a significance threshold of $S/N_\text{meas} = 4.5$, this is a roughly 5\% effect and will be less important for all higher-significance detections (see e.g. \citealt{vanderlinde10}).  
We therefore neglect this bias in this work.

Finally, a third bias arises from the fact that for sources that we detect only in one or two bands but not all three, the flux(es) for the source in the non-detected band(s) will be subject to a slight negative bias.  
This is due to the fact that we measure source flux in the non-detection band(s) using a source position determined from a band where the source is detected, and positional uncertainty biases the flux low.  
This bias is expected to be small given the small positional uncertainty for a 4.5-$\sigma$ detection.  
We calculate that a 1-$\sigma$ positional offset would result in a flux underestimate of 5\%, and therefore neglect this bias.  

\subsubsection{Bayesian flux deboosting}
One standard method for dealing with flux deboosting in mm and submm surveys is the application of a Bayesian approach, where a posterior probability distribution is calculated given prior knowledge about the underlying source populations~\citep{coppin05}.  
The usual Bayesian posterior distribution can be expressed as
\begin{equation}
P(S_\text{true} | S_{\text{meas}}) \propto P(S_{\text{meas}} | S_\text{true})\ P(S_\text{true}),
\end{equation}
where $P(S_\text{true} | S_{\text{meas}})$ is the posterior probability, expressing the probability of the true source flux $S_\text{true}$ given the measured value $S_{\text{meas}}$.  
$P(S_\text{meas} | S_\text{true})$ is the likelihood, expressing the probability of measuring a flux $S_{\text{meas}}$ given that the true flux of the source is $S_\text{true}$.  
Most simply, the likelihood is taken to be a Gaussian with width given by the map noise.  
$P(S_\text{true})$ is the prior, which expresses previous knowledge about the population of sources being detected, which in our case is proportional to the differential number counts as a function of flux, $dN/dS$.

\citet{crawford10} present an argument for slightly altering the expressions above to account for the fact that we expect the number of sources to rise steeply with decreasing flux and the reality that the telescope observes the sky with some finite resolution (which we further pixelate when creating a map).  
Therefore, there is a confusion limit due to faint sources coexistent in a single pixel which contributes to the noise of each detection.  
The standard Bayesian approach can be modified slightly to account for this:

\begin{equation} 
P(S_{\text{max}} | S_{\text{meas}}) \propto P(S_{\text{meas}} | S_{\text{max}})\ P(S_{\text{max}}),
\end{equation}
where now the posterior, $P(S_{\text{max}} | S_{\text{meas}})$, gives the probability that the highest flux source contributing to the pixel brightness is $S_{\text{max}}$ given the measured flux of $S_{\text{meas}}$ in that pixel.  
Similarly, $P(S_{\text{meas}} | S_{\text{max}})$ expresses the likelihood that $S_{\text{meas}}$ will be measured given that the brightest source contributing to that pixel brightness has flux $S_{\text{max}}$.  
The likelihood includes the uncertainty in the flux due to the presence of fainter sources.  
The prior, $P(S_{\text{max}})$, is still expressed by the differential number counts, $dN/dS$, but now multiplied by an exponential suppression at low flux representing the probability that no other sources brighter than $S_{\text{max}}$ exist in that pixel.

\subsubsection{Simultaneous three-band deboosting}\label{dbdb}
Also presented in \citet{crawford10} is the framework for expanding the single-band deboosting presented above to a deboosting of fluxes for sources detected in multiple bands simultaneously.  
\citet{crawford10} expands the analysis from one to two bands, and M13 presents the extension to three bands.  
We use the same method for deboosting as in M13 and present an overview of the methodology below, see M13 for further details.  

The goal of multi-band deboosting is to estimate the posterior probability for the flux of the source in multiple bands using its measured flux in one or more bands and any prior information known.  
The simplest way to write this would be
\begin{multline}
P\left(S^{\text{max}}_{95}, S^{\text{max}}_{150}, S^{\text{max}}_{220} | S^{\text{meas}}_{95}, S^{\text{meas}}_{150}, S^{\text{meas}}_{220}\right) \propto\\ P \left(S^{\text{meas}}_{95}, S^{\text{meas}}_{150}, S^{\text{meas}}_{220} | S^{\text{max}}_{95}, S^{\text{max}}_{150}, S^{\text{max}}_{220}\right) \centerdot \\ P\left(S^{\text{max}}_{95}, S^{\text{max}}_{150}, S^{\text{max}}_{220}\right),
\end{multline}
which would express the 3-dimensional posterior probability distribution for the true flux for the detected source in the three bands, given the measured fluxes for that source in three bands.  
For the multi-band prior, one could assume that the priors for each band are independent and therefore could be separated as
\begin{equation}
P\left(S^{\text{max}}_{95}, S^{\text{max}}_{150}, S^{\text{max}}_{220}\right) = P\left(S^{\text{max}}_{95}\right)\ P\left(S^{\text{max}}_{150}\right)\ P\left(S^{\text{max}}_{220}\right).
\end{equation}
However, in general this assumption would only be accurate if the three bands probed completely separate populations of sources with no overlap.  
In general, while more synchrotron sources are detected in 95 GHz, and 220 GHz is a stronger probe of dusty sources, there is certainly population overlap between the bands.  

To accommodate this issue, we can express the prior as the combination of a prior on flux for one band (for example 150 GHz), and two priors describing the power law behavior connecting two fluxes as a function of frequency:
\begin{equation}
\begin{split}
S_{95} = S_{150}\left(\frac{\nu_{95}}{\nu_{150}}\right)^{\alpha_{95-150}} \\
S_{220} = S_{150}\left(\frac{\nu_{220}}{\nu_{150}}\right)^{\alpha_{150-220}}.
\label{alphadef}
\end{split}
\end{equation}
Note that the effective band centers for SPT depend slightly on the assumed spectral index of the source.  
We assume a flat spectral index of zero, which gives effective band centers of 97.6, 152.9, and 218.1\,GHz.  
M13 found that source fluxes are not affected significantly by making this assumption.
Through a change of variables, then, we can express the three-flux prior in terms of one flux and two spectral indices ($\alpha$):
\begin{multline}
P\left(S^{\text{max}}_{95}, S^{\text{max}}_{150}, S^{\text{max}}_{220}\right) =\\ P\left( S^{\text{max}}_{150}, \alpha^{\text{max}}_{95-150},\alpha^{\text{max}}_{150-220}\right)\ \frac{d \alpha^{\text{max}}_{95-150}}{d S^{\text{max}}_{95}}\ \frac{d \alpha^{\text{max}}_{150-220}}{d S^{\text{max}}_{220}}
\end{multline}
where the $\frac{d\alpha}{dS^{\text{max}}}$ can be found from Eqn. \ref{alphadef}.

We then make the assumption that the prior written in this way is made up of three independent components:
\begin{multline}
P\left( S^{\text{max}}_{150}, \alpha^{\text{max}}_{95-150},\alpha^{\text{max}}_{150-220}\right) = \\P\left(S^{\text{max}}_{150}\right) P\left(\alpha^{\text{max}}_{95-150}\right) P\left(\alpha^{\text{max}}_{150-220}\right).
\label{priorindependence}
\end{multline}
By separating them, we are assuming that the spectral indices are independent of flux and the two spectral indices are not correlated with each other.  
Strictly speaking, we know that this assumption of independence is also incorrect -- fainter sources tend to have more dust-like spectral indices.  
More fundamentally, simply changing variables does not change the issue of the priors being correlated, since the amount of information contained in the priors has stayed the same.  
However, since we are interested in measuring $\alpha$ in this analysis, and allow for the possibility of sources with non-typical spectral indices, expressing the priors in this way allows us to place weak flat priors on both spectral indices between the physically motivated range of $-3 \le \alpha \le 5$, allowing the intrinsic population characteristics to emerge.

We now have for the 3-dimensional posterior on fluxes:
\begin{multline}
P\left(S^{\text{max}}_{95}, S^{\text{max}}_{150}, S^{\text{max}}_{220} | S^{\text{meas}}_{95}, S^{\text{meas}}_{150}, S^{\text{meas}}_{220}\right) \propto \\ P \left(S^{\text{meas}}_{95}, S^{\text{meas}}_{150}, S^{\text{meas}}_{220} | S^{\text{max}}_{95}, S^{\text{max}}_{150}, S^{\text{max}}_{220}\right) \centerdot \\P\left( S^{\text{max}}_{150}\right)P\left(\alpha^{\text{max}}_{95-150}\right)P\left(\alpha^{\text{max}}_{150-220}\right)\ \frac{d \alpha^{\text{max}}_{95-150}}{d S^{\text{max}}_{95}}\ \frac{d \alpha^{\text{max}}_{150-220}}{d S^{\text{max}}_{220}}
\end{multline}
The likelihood, $P \left(S^{\text{meas}}_{95}, S^{\text{meas}}_{150}, S^{\text{meas}}_{220} | S^{\text{max}}_{95}, S^{\text{max}}_{150}, S^{\text{max}}_{220}\right)$, is given by a multivariate Gaussian
\begin{multline}
P\left(S^{\text{\text{meas}}}_{95}, S^{\text{meas}}_{150}, S^{\text{meas}}_{220} | S^{\text{max}}_{150}, S^{\text{max}}_{95}, S^{\text{max}}_{220} \right) =\\ \frac{1}{\sqrt{(2\pi)^3 det(\textbf{C})}}\text{exp}(-\frac{1}{2}\textbf{r}^T\textbf{C}^{-1}\textbf{r}).
\end{multline}
The noise covariance $\textbf{C}$ represents the flux uncertainty due to instrument noise, atmosphere, and uncertainties in beam and absolute calibration. 

The residual vector, $\textbf{r}$, is given by:
\begin{equation}
\textbf{r} = \left[ S^{\text{meas}}_{95} - S^{\text{max}}_{95}, S^{\text{meas}}_{150} - S^{\text{max}}_{150}, S^{\text{meas}}_{220} - S^{\text{max}}_{220}\right]
\end{equation}

For the flux prior, we estimate the number counts $dN/dS$ based on a sum of synchrotron and dusty population models.  
Synchrotron populations are calculated using the \citet{dezotti05} model at 150\,GHz and extrapolated to the other two bands.    
Dusty populations at 150 and 220\,GHz are estimated by use of updated \citet{negrello07} models (M. Negrello, private communication).
The population at 95\,GHz is estimated using an extrapolation of the \citet{negrello07} prediction at 850\,$\mu$m using spectral indices of 3.1 for high-redshift sources (calculated from the spectral energy distribution of the ULIRG Arp 220 shifted to $z \sim 3$) and 2.0 for low-redshift sources (from IRAS observations). 
This is the same method as employed in M13.   

There is an asymmetry introduced in our current deboosting algorithm, namely, that one band is chosen to have much stricter prior information applied to it through the flux prior, and the other two bands have much less restrictive priors applied through loose $\alpha$ priors.  
Therefore, for any given source, with flux information in three bands, the amount of deboosting each band's flux receives depends on the choice made in selecting which band the flux prior is applied to.  
In \citet{crawford10} and M13, this band is termed the ``detection band" but this is slightly confusing terminology, since a given source could in fact be detected simultaneously in all three bands or some combination of bands.  
To avoid this confusion, here we employ the term ``flux-prior band" to refer to the band which has the flux prior applied as opposed to a prior on $\alpha$.  
In practice, the deboosted fluxes reported in the catalog are calculated using the band with the highest significance detection in raw flux as the flux-prior band.  
For number counts, we use the band for which we are calculating number counts as the flux-prior band and then restrict to only sources with a detection in that band.

Since we are interested in calculating posterior distributions for spectral indices in addition to fluxes, we calculate in parallel the posteriors for one flux and two $\alpha$'s:
\begin{multline}
P\left(S^{\text{max}}_{150}, \alpha^{\text{max}}_{95-150},\alpha^{\text{max}}_{150-220} | S^{\text{meas}}_{95}, S^{\text{meas}}_{150}, S^{\text{meas}}_{220}\right) \propto \\P\left(S^{\text{meas}}_{95}, S^{\text{meas}}_{150}, S^{\text{meas}}_{220} | S^{\text{max}}_{150}, \alpha^{\text{max}}_{95-150},\alpha^{\text{max}}_{150-220}\right) \centerdot \\ P\left( S^{\text{max}}_{150}, \alpha^{\text{max}}_{95-150},\alpha^{\text{max}}_{150-220}\right).
\end{multline}

The prior is identical to that used for three fluxes, and the likelihood is very similar:
\begin{multline}
P\left(S^{\text{meas}}_{95}, S^{\text{meas}}_{150}, S^{\text{meas}}_{220} | S^{\text{max}}_{150}, \alpha^{\text{max}}_{95-150},\alpha^{\text{max}}_{150-220}\right) = \\\frac{1}{\sqrt{(2\pi)^3 det(\textbf{C})}}\text{exp}(-\frac{1}{2}\textbf{r}^T\textbf{C}^{-1}\textbf{r}),
\end{multline}
the same as before, but where the residual vector is now:
\begin{multline}
\textbf{r} = \left[ S^{\text{meas}}_{95} - S^{\text{max}}_{150}\left(\frac{\nu_{95}}{\nu_{150}}\right)^{\alpha^{\text{max}}_{95-150}},\ S^{\text{meas}}_{150} - S^{\text{max}}_{150}\right.,\\ \left. S^{\text{meas}}_{220} - S^{\text{max}}_{150}\left(\frac{\nu_{220}}{\nu_{150}}\right)^{\alpha^{\text{max}}_{150-220}}\right].
\end{multline}
The likelihood values are identical for the corresponding locations in the different parameter spaces.

From our 3-dimensional posterior probability distributions, we marginalize over two of the three parameters in the posterior to find the corresponding 1-dimensional posteriors for a parameter of interest.  
We then integrate the PDFs to the 16\%, 50\%, and 84\% levels in the cumulative distribution to calculate the best-fit values and 1-$\sigma$ error bars.

\subsection{Radial cross-match method}
\label{sec:radxmatchmethod}
There are several instances in the analysis pipeline where a cross-match method is employed: cross-matching between the 19 SPT fields within a given band, cross-matching between SPT bands, and cross-matching between the SPT catalog and external catalogs.  
The same general principle is applied in each case, while the details that differ will be discussed topically in following sections.  
A cross-match criterion involving only a radial offset is appropriate when the source densities of the two groups of sources under comparison are comparable or, in the case of cross-matching with external information, the source density of the external catalog is similar or lower than the SPT catalog, and the positional uncertainty is small relative to the typical distance between sources.  
An appropriate cross-matching radius can then be chosen either analytically or empirically using the measured source density.
Depending on the application, either all of the sources within the radial distance are considered associated (in the case of cross-matching between SPT fields for the same band), or the closest candidate within the radial criterion, if one exists, is considered associated (in the case of cross-matching between SPT bands and between the SPT catalog and external catalogs).  
Selecting a radial threshold that is excessively large will result in falsely associating physically unrelated objects, whereas a radial threshold that is too small risks missing true associations that are shifted in position due to map noise or residual pointing error.  
Further details of the cross-matching between SPT detections to form a single three-band full-survey-area catalog can be found in the following subsection, and details of cross-matches with external catalogs and redshift information are further detailed in Sections~\ref{sec:xchecksection} and \ref{sec:zassoc}.

\subsection{Catalog generation}
To be included in the source catalog, we require that a source must exceed the detection threshold of 4.5\,$\sigma$ in raw flux signal-to-noise in at least one band.  
The threshold of 4.5\,$\sigma$ was chosen to align with V10 and M13, which calculated purity levels of roughly 90\% at 150\,GHz and a 4.5\,$\sigma$ threshold.  
We note that the purity level in V10 for 220\,GHz was also roughly 90\%, whereas it is somewhat lower in the current work (see Section~\ref{sec:purity} below), which is due to the 2008-observed fields having deeper 220\,GHz data than the rest of the survey.
To form a united single SPT catalog including all fields and bands, we first cross-match across all 19 fields for detections in a single band.  
About 10\% of the full survey area falls in overlap regions covered by multiple fields, and sources that lie in overlap regions will have repeat detections in different fields.
We remove repeat detections by concatenating all fields' detections in each band and employing a radial cross-match as discussed in Section~\ref{sec:radxmatchmethod}.  
We then keep the detection that comes from the map with the lowest noise at that location and throw out the others.  
To determine the cross-match radius, we use the analytic formalism in Appendix B of \citet{ivison07} which takes into account the measured beam FWHM for that band and the source signal-to-noise to yield a positional error calculated analytically.  
Assuming there is equal and uncorrelated error in both positional directions, we use a 3-$\sigma$ positional error for a 4.5-$\sigma$ detection to cross-match, corresponding to 57.8, 40.3, and 34.0\,arcsec for 95, 150, and 220\,GHz, respectively.  
(Note: in reality the positional error is correlated in the two orthogonal directions in the map when cross-matching source positions, so assuming errors are uncorrelated is not technically correct).  
We test that this is an appropriate radius by comparing to the density of source detections within each single field and also check that we don't associate (and therefore remove) sources detected within the same field.  
We note that the radial criterion used to associate CLEAN components into sources (as discussed in Section~\ref{sec:clean}) corresponds to just under 2\,$\sigma$ for a detection of 4.5\,$\sigma$ at 95\,GHz, the band with the largest beam; therefore, sources within this radius of another source within the same field and band very likely would have been considered a component detection of that source.  
We remove six sources flagged within the crossmatch radius of a source in the same field, five of these are sources detected at 95\,GHz, one at 150\,GHz, and none at 220\,GHz.
These six appear to either be multiple detections of the same source or component detections of extended sources.

We additionally remove all sources that lie in regions with overlapping coverage from multiple fields where the source is detected in a field with a higher noise level but not detected in an overlapping field with lower noise, as it's expected that these detections would be false.  
This step removes 51 sources at 95\,GHz, 44 sources at 150\,GHz, and 47 sources at 220\,GHz, which is roughly 3\% of sources in 220\,GHz, and a smaller percentage for 95 and 150\,GHz.  
We check that the distribution in signal-to-noise of sources trimmed is sensible, i.e. that almost all trimmed sources are near the detection threshold of 4.5\,$\sigma$, and therefore likely to be false detections due to map noise.  
The one notable exception is a 12.3-$\sigma$ detection in 220\,GHz that is removed from field \onesix\, due to overlapping coverage by \threesix\, which has lower noise at that source location but a non-detection of the source in that band.  
This source appears to be a flaring radio source that became brighter over the course of observing the 2500-square-degree area of the SPT-SZ survey, such that it was brighter in \onesix, observed in 2010, compared with \threesix, observed in 2009.
We note that due to detections of this source above 4.5\,$\sigma$ in 95 and 150\,GHz, this source does survive to the final catalog as SPT-S J015917-6055.9, but with recorded fluxes in the three bands that were not measured contemporaneously.

The next step in creating a multi-band catalog is to cross-match across the SPT bands.  
We employ a radial cross-match method and use a 30\,arcsec radius of association, which is chosen similarly to above using the analytical positional uncertainty of SPT sources calculated from the formalism in~\citet{ivison07}.  
For the band with the widest beam (1.7\,arcmin at 95\,GHz), 30\,arcsec is roughly a 1.5-$\sigma$ positional error for a 4.5-$\sigma$ detection.  
Since the source densities in all three SPT bands are quite low relative to the 30\,arcsec association radius, the expected rate of random association of two unrelated sources between bands is also very low.  
Using just the full-survey average source density, the probability of random association is $0.024\%$, $0.034\%$, and $0.012\%$ for 95, 150, and 220\,GHz, respectively.  

At this step, we also remove any sources with incomplete coverage across the three bands.  
Because the masks of usable area for each field cover physically offset regions of the sky for the three bands, some area of the survey at the edges will be covered only by one or two bands, and for the sake of consistency, we trim any sources detected in these areas.  
This removes 115 sources from the final catalog, or about 2\% of the catalog.

\begin{figure*}[!ht] 
\begin{center} 
\subfigure{\includegraphics[width=17.5cm]{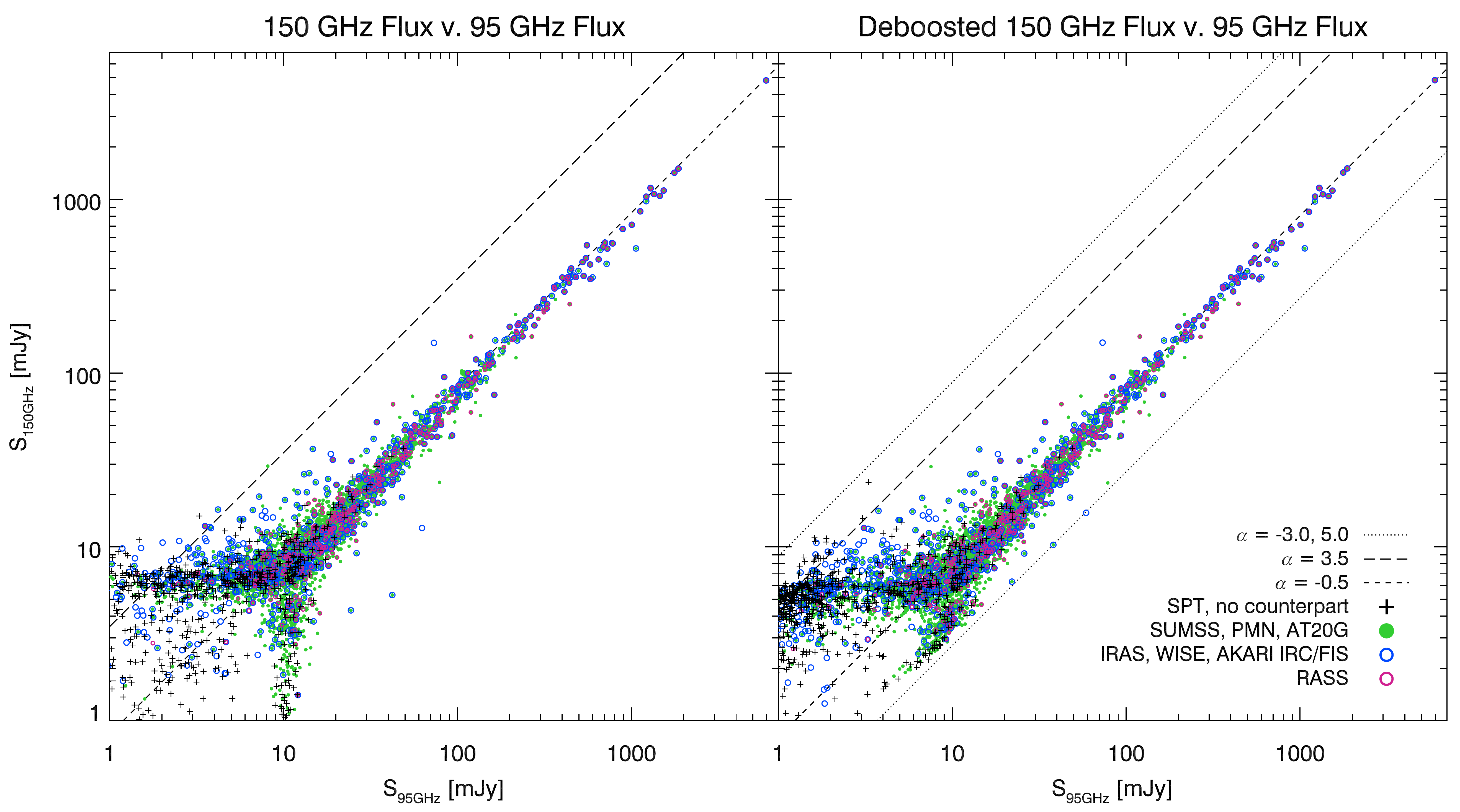}}
\subfigure{\includegraphics[width=17.5cm]{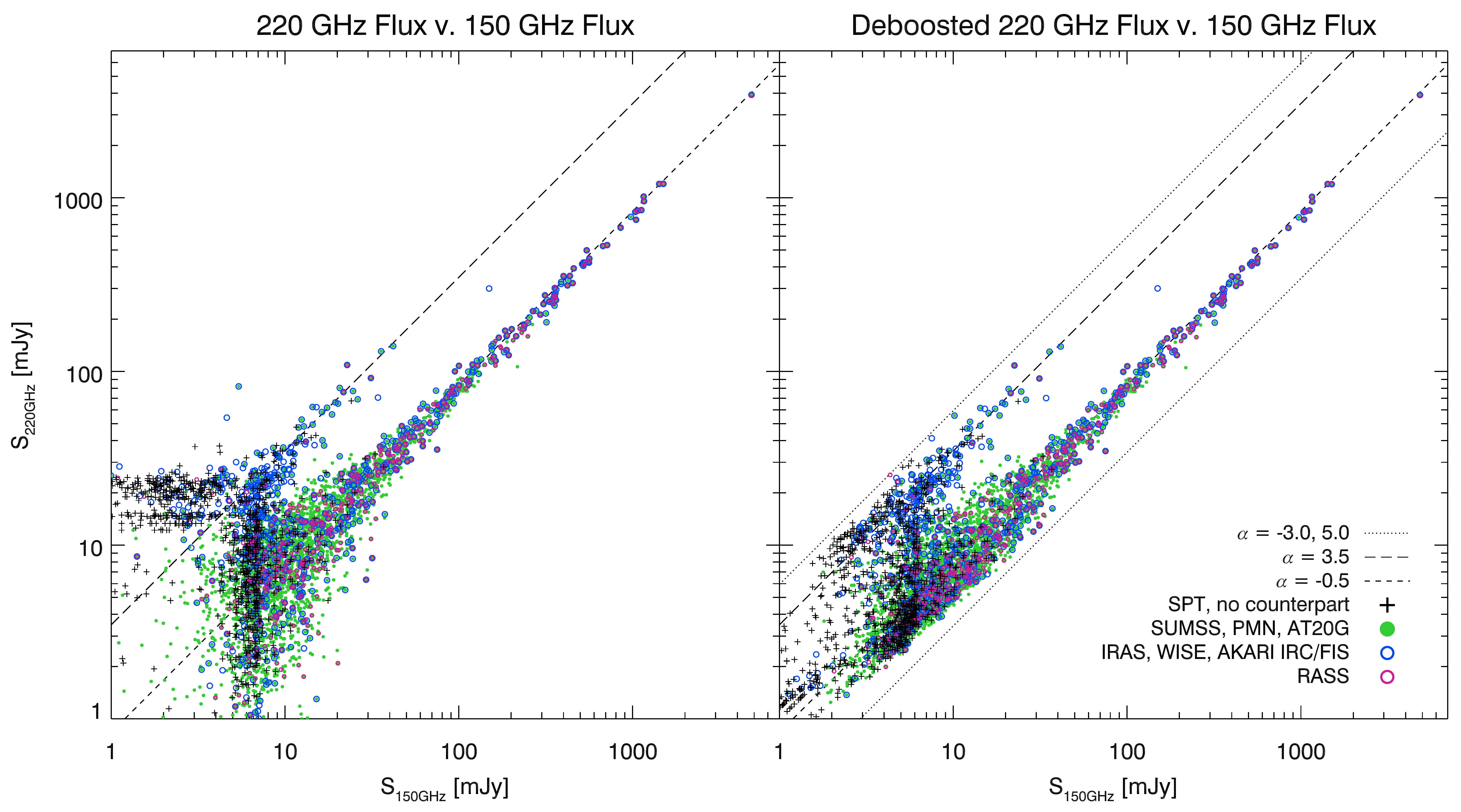}}
\caption{{\it Upper panels:} raw ({\it left}) and deboosted ({\it right}) fluxes for 150\,GHz vs. 95\,GHz for all sources in the catalog.  Lower panels: raw ({\it left}) and deboosted ({\it right}) fluxes for 220\,GHz vs. 150\,GHz.  Colors and symbols show cross-matches with external catalogs and black crosses indicate sources detected by the SPT-SZ survey with no counterparts in external catalogs.  Dashed lines show expected spectral indices for synchrotron and dusty populations, and dotted lines in the right panels show bounds applied as priors on spectral index for the deboosting.  The vertical and horizontal clusters of sources at relatively low flux in the left panels show the detection thresholds of $4.5\,\sigma$, which translate to a different flux threshold for each field.  For example, the clusters at clearly different flux levels in 220\,GHz represent the $4.5\,\sigma$ thresholds in the double-depth fields and in the rest of the catalog.
\label{fig:rawf}}
\end{center}
\vspace{0.4in}
\end{figure*}

\begin{figure*}[!ht] 
\begin{center} 
\subfigure{\includegraphics[width=17.5cm]{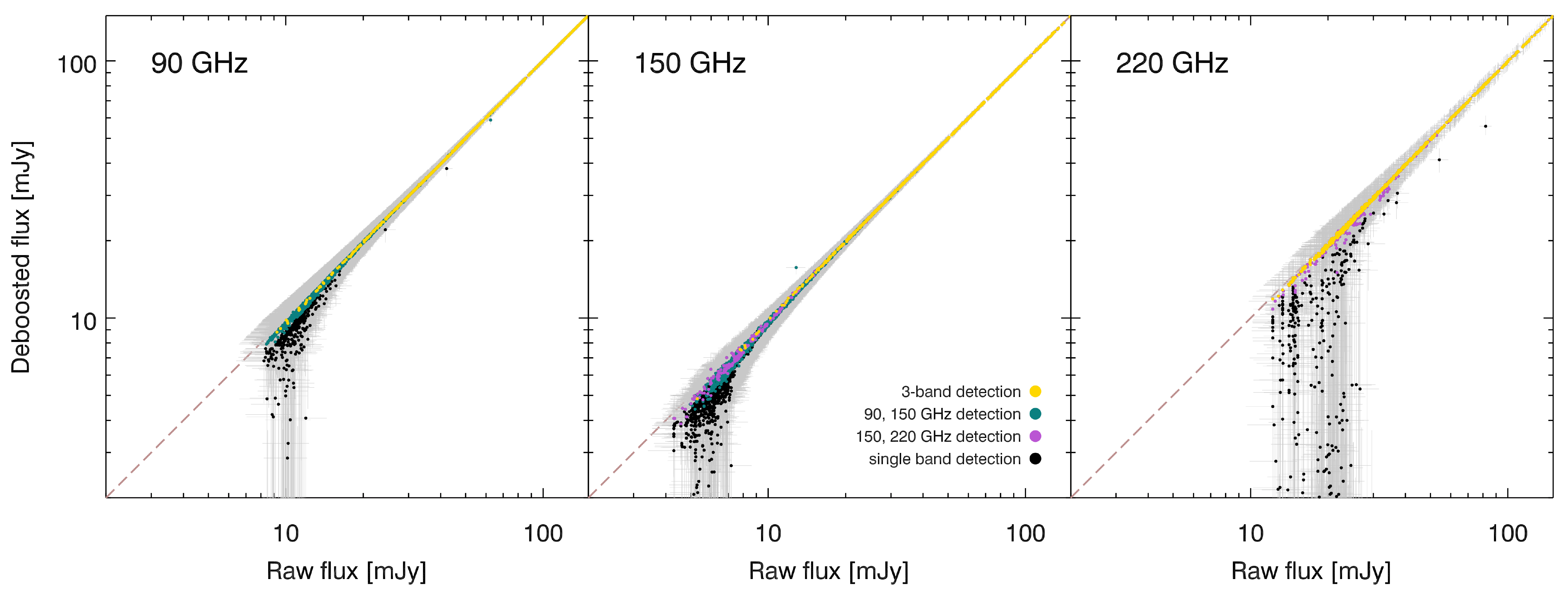}}
\caption{Deboosted flux compared with raw flux for all sources in the catalog, for 95 ({\it left}), 150 ({\it middle}), and 220\,GHz ({\it right}), focusing on the lower flux range of the catalog where the deboosting has the largest effect.
\label{fig:dbVsrawFlux}}
\end{center}
\end{figure*}

\begin{figure*}[!ht]
\begin{center} 
\subfigure{\includegraphics[width=8.9cm]{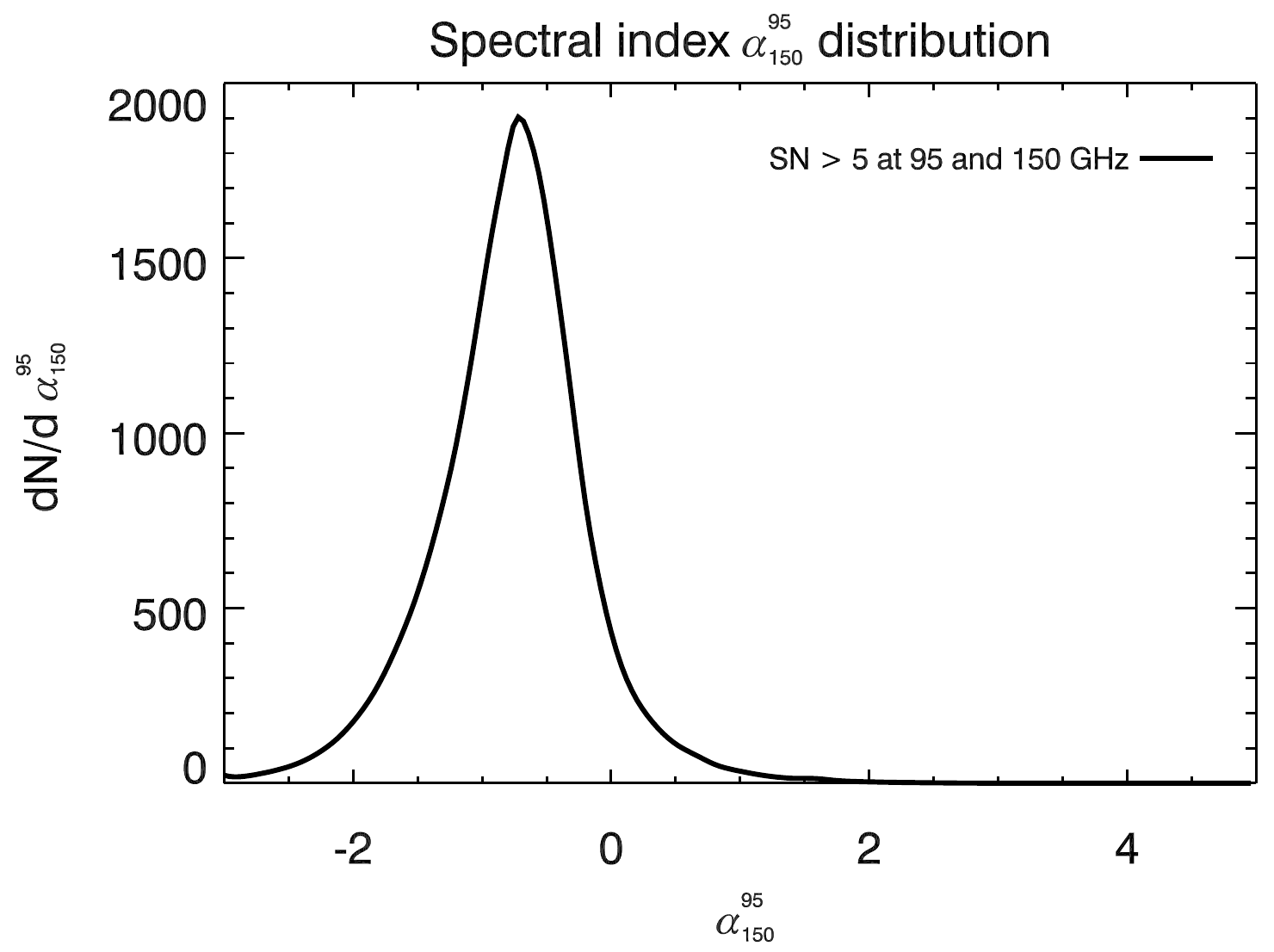}}
\subfigure{\includegraphics[width=8.9cm]{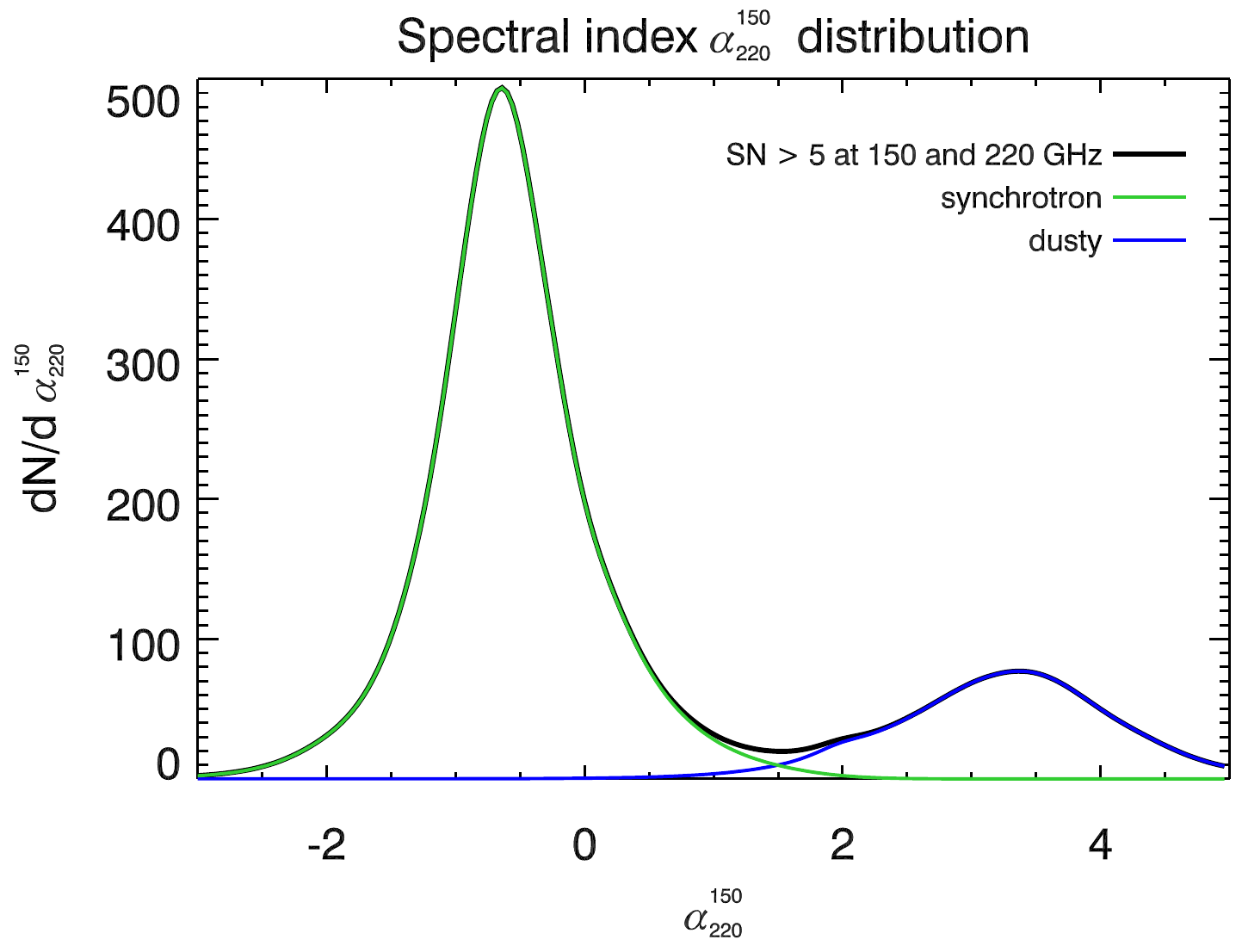}}
\end{center}
\caption{Summed normalized posterior distributions for the 95-150\,GHz spectral index ({\it left}) and the 150-220\,GHz spectral index ({\it right}), choosing for each spectral index only sources with detections above 5\,$\sigma$ in both bands spanned by that index.  As expected, the distribution for $\alpha^{\text{max}}_{95-150}$ shows a single peak at $\alpha^{\text{max}}_{95-150} \sim -0.7\pm0.6$, since synchrotron sources are expected to dominate at 95\,GHz, especially for bright, and therefore high-signal-to-noise, sources.  The distribution for $\alpha^{\text{max}}_{150-220}$ shows distinct peaks for synchrotron and dusty populations, with peaks at $\alpha^{\text{max}}_{150-220} \sim -0.6\pm0.6$ and $\alpha^{\text{max}}_{150-220} \sim 3.4\pm0.8$, respectively, and we select the minimum at $\alpha^{\text{max}}_{150-220} = 1.51$ as the threshold for applying a categorization for each source in the catalog.
\label{fig:alphaposts}}
\end{figure*}


\begin{figure*}[!ht]
\begin{center}
\subfigure{\includegraphics[width=8.9cm]{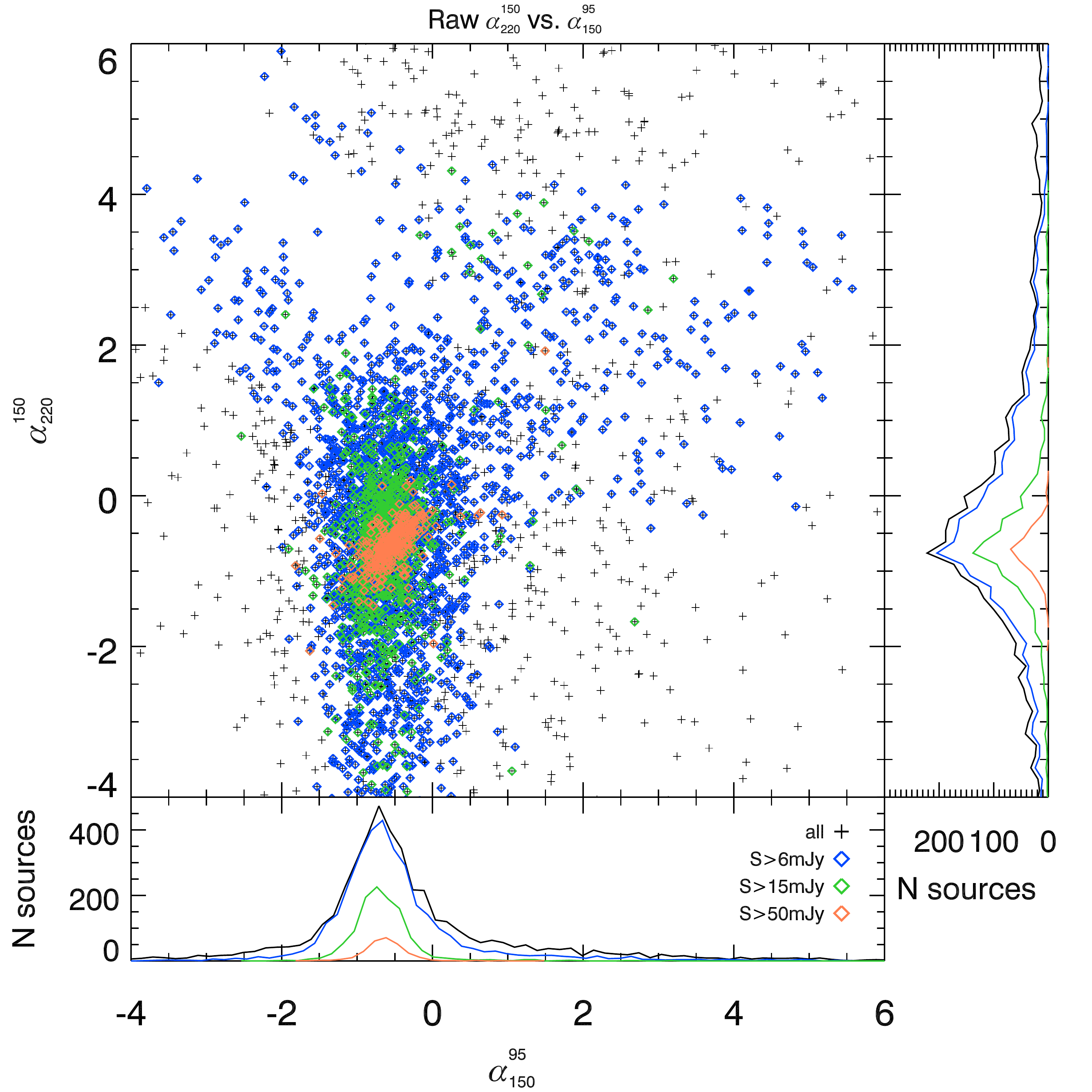}}
\subfigure{\includegraphics[width=8.9cm]{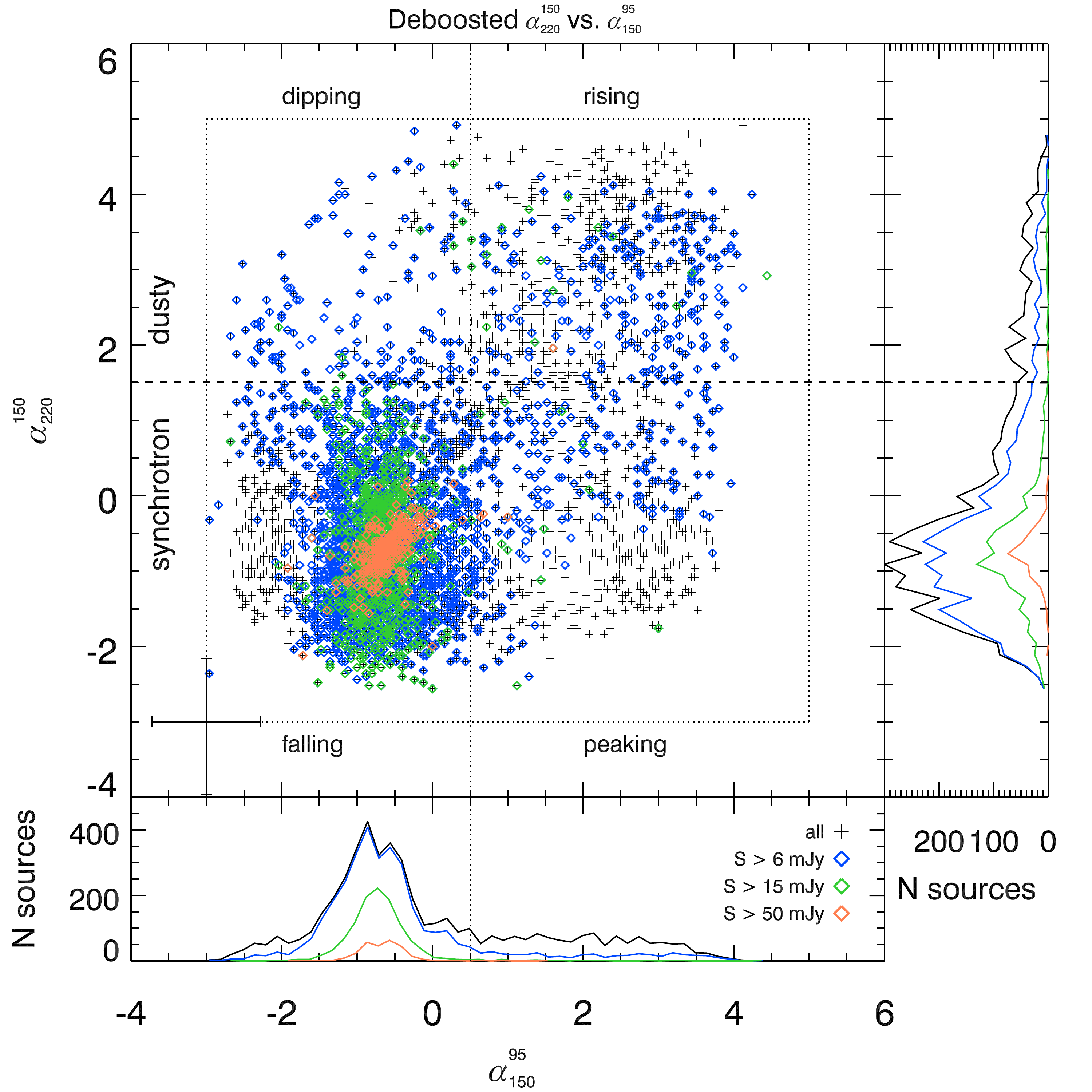}}
\subfigure{\includegraphics[width=8.9cm]{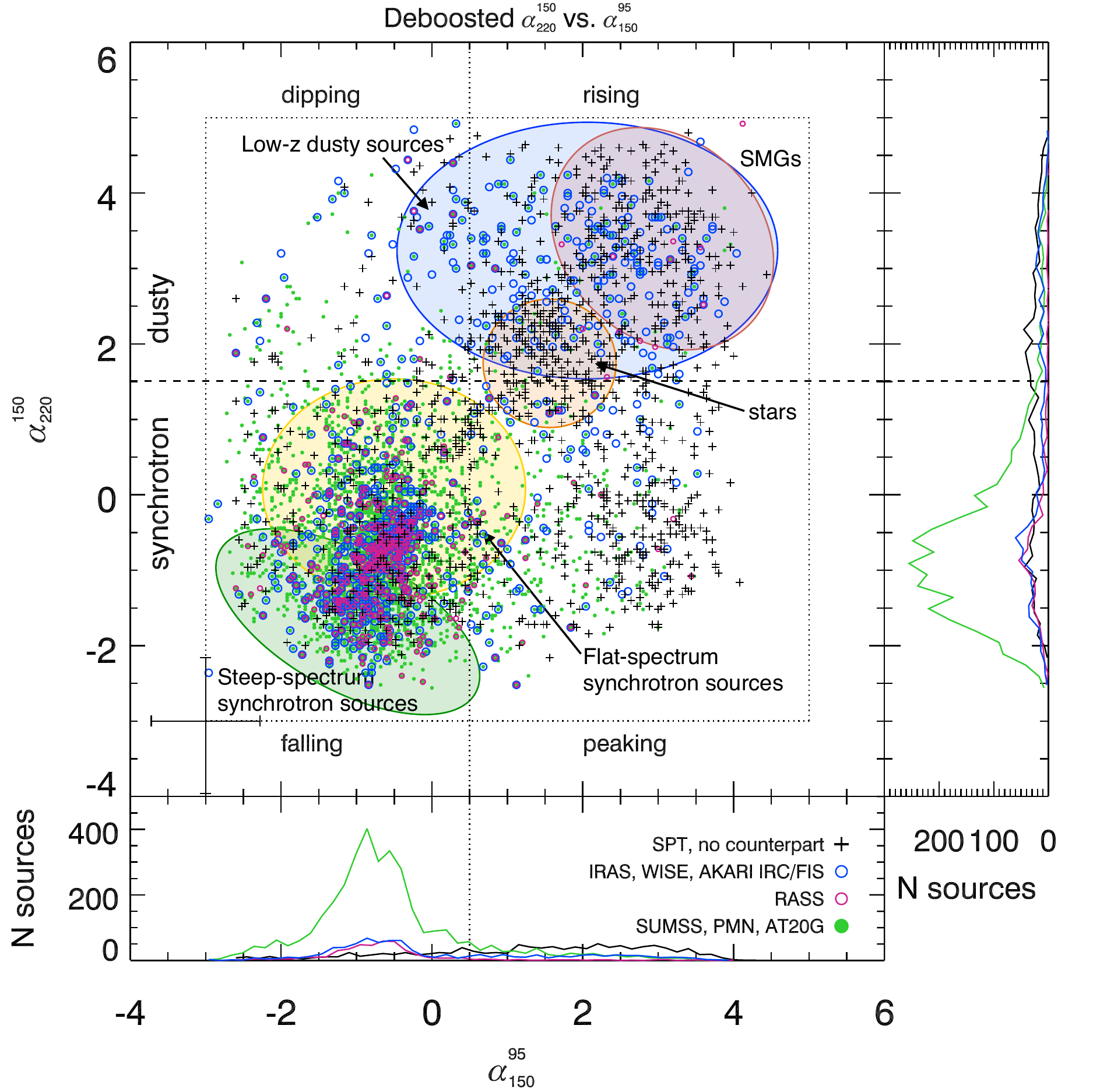}}
\subfigure{\includegraphics[width=8.9cm]{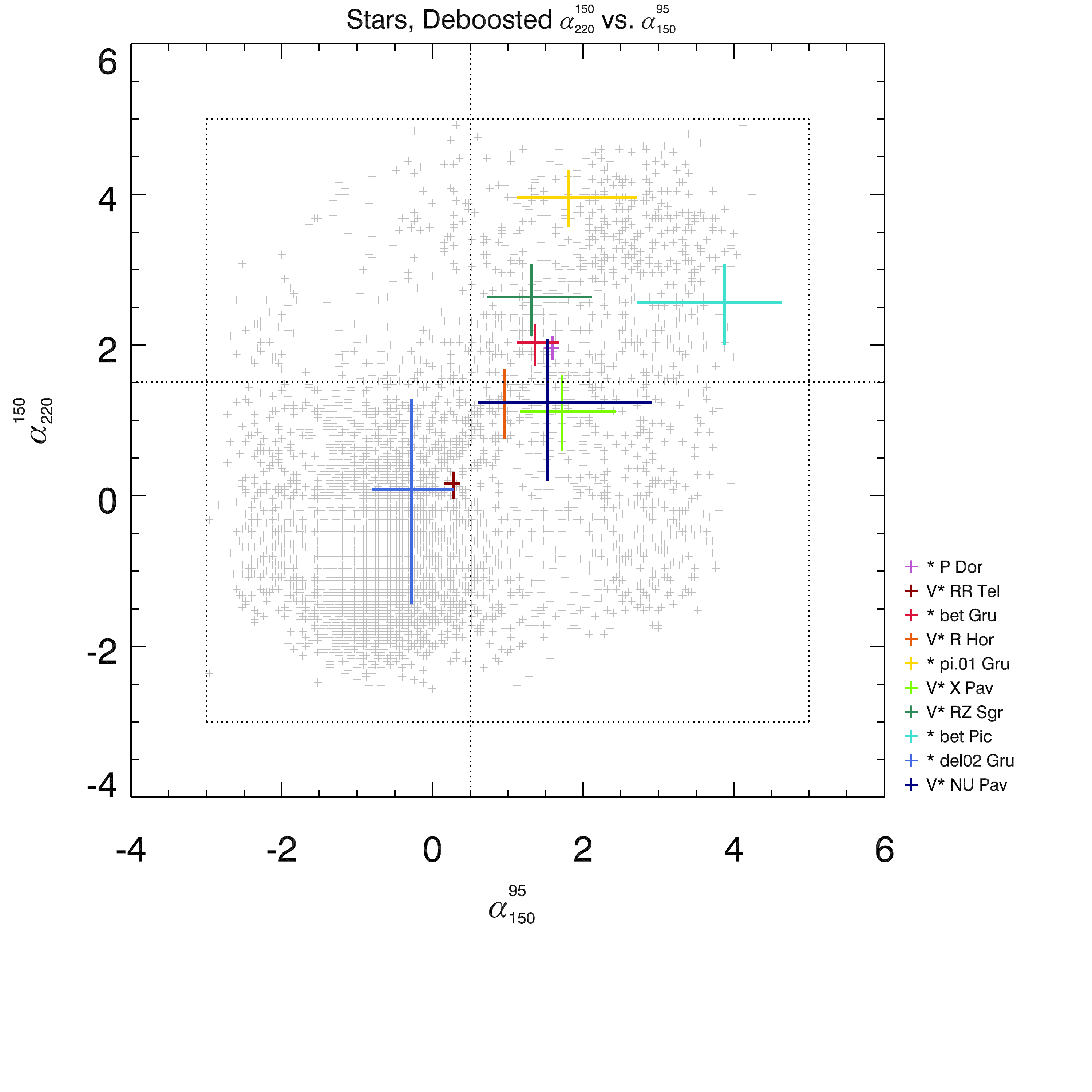}}
\end{center}
\caption{{\it Upper panels:} raw ({\it left}) and deboosted ({\it right}) $\alpha_{150-220}$ vs. $\alpha_{95-150}$, where colors and symbols indicate flux level.  Dotted box indicates the bounds of the prior on spectral index applied during deboosting, and dotted horizontal and vertical dashed lines indicate $\alpha_{150-220} = 1.51$, the minimum of the 150-220\,GHz summed posterior and the corresponding separation index between dusty and synchrotron in 95-150\,GHz from reexamining the distributions of summed spectral index posteriors after population separation.  These separation lines are used to categorize sources into four quadrants of ``falling," ``rising," ``peaking," and ``dipping."  The typical error for a source at the 4.5-$\sigma$ detection threshold is shown in the lower left corner of the deboosted spectral index plots.  {\it Lower panels: Left:} deboosted $\alpha_{150-220}$ vs. $\alpha_{95-150}$, where colors and symbols indicate cross-matches with external catalogs, and black crosses indicate SPT sources with no cross-matches in external catalogs.  {\it Right:} Measured spectral indices for ten stars detected in the catalog, overplotted on the rest of the catalog, shown by grey crosses.
\label{fig:alpha1v2}}
\vspace{0.4in}
\end{figure*}


\section{Catalog: description and characterization}
\label{sec:catalog}
\subsection{Single-band and multi-band catalogs}
Our 3-band integrated catalog for the full 2530 square degrees of survey area contains 2774 sources detected above 4.5\,$\sigma$ at 95\,GHz, 3909 at 150\,GHz, and 1435 at 220\,GHz.  
The median purity at this detection threshold across all fields in the survey is 94.4\%, 94.8\%, and  83.4\% at 95, 150, and 220\,GHz, respectively; see Section~\ref{sec:purity} for further detail.
Cross-matching across SPT bands, this yields a multi-band catalog with 4845 total sources detected at a minimum of 4.5\,$\sigma$ in at least one band.  
The noise levels for individual matched-filtered maps are shown in Table~\ref{tab:completeness}; taking the median noise level across all fields, 4.5\,$\sigma$ corresponds to detections above 9.8, 5.8, and 20.4\,mJy in 95, 150, and 220\,GHz, respectively.  
Of the 4845 sources in the catalog, 722 sources are detected at $\ge 4.5\,\sigma$ in all three bands.  
1662 are detected only in 95 GHz and 150\,GHz, and 167 are detected only in 150\,GHz and 220\,GHz.  
390 are detected only in 95\,GHz, 1358 are detected only in 150\,GHz, and 546 are detected only in 220\,GHz.  
Of all the detections in the catalog, roughly 8\% have fluxes in different bands drawn from multiple different fields, which is consistent with about 10\% of the area of the survey falling in overlap regions covered by multiple fields.  
Similarly, of all the sources detected above 4.5\,$\sigma$ in all three bands, about 9\% have fluxes drawn from multiple fields.  
We compare raw fluxes and deboosted fluxes in the combined catalog in Figure~\ref{fig:rawf}.  
Overplotted are expected values for spectral indices between the bands, and we see that for the most part, sources follow the characteristic lines for dusty and synchrotron sources.  
Similarly, we plot $\alpha_{95-150}$ vs. $\alpha_{150-220}$ for both raw spectral indices and deboosted values in Figure~\ref{fig:alpha1v2}.  
We note in these plots, that spectral index does seem to correlate with source brightness, as expected, where the brightest sources are synchrotron-dominated.  
We also note that while there are sources where $\alpha_{95-150}$ correlates with $\alpha_{150-220}$, there are also numerous sources with spectral indices which are not correlated, indicating sources with a spectral break, which will be discussed further in Section~\ref{sec:discussion}.
To show the effect of the deboosting, Figure~\ref{fig:dbVsrawFlux} plots deboosted flux as a function of raw flux for each of the three SPT-SZ bands.
An overview of the number of sources above 4.5 and $5.0\,\sigma$ are shown in Table~\ref{tab:dettable}.

\subsection{Population separation}
\label{sec:popsep}
To explore the distributions in spectral indices that we find from deboosting and to separate sources into populations based on spectral index, we normalize each source's posterior probability distribution for $\alpha$, such that the integral of the marginalized posterior over all possible values of $\alpha$ is unity, and then sum all the posteriors from different sources.  
In Figure \ref{fig:alphaposts}, we show these distributions for sources with signal-to-noise greater than or equal to 5.0 in both of the bands that a particular spectral index spans.  
We restrict to higher signal-to-noise sources for this part of the analysis to provide a cleaner population separation.

From Figure~\ref{fig:alphaposts}, we see that the posteriors for $\alpha^{\text{max}}_{95-150}$ show only the presence of a synchrotron population peaking at $\alpha^{\text{max}}_{95-150} \sim -0.7\pm0.6$.  
As shown in Figure \ref{fig:alpha1v2}, synchrotron sources do dominate the high signal-to-noise sources in general, and dusty sources, with a positive spectral index, are much more likely to be below the detection threshold at 95\,GHz.  
In contrast, the posteriors for $\alpha^{\text{max}}_{150-220}$ show two peaks in the distribution, representing contributions from both synchrotron and dusty populations, peaking at $\alpha^{\text{max}}_{150-220} \sim -0.6\pm0.6$ and $\alpha^{\text{max}}_{150-220} \sim 3.4\pm0.8$, respectively.
Once again, the synchrotron peak is stronger since we are restricting to relatively high signal-to-noise detections, which are synchrotron-dominated.

We take the minimum of our summed posterior distribution on $\alpha^{\text{max}}_{150-220}$ as the dividing criterion to produce separate catalogs of synchrotron and dusty sources.  
From Figure~\ref{fig:alphaposts}, this produces a population separation at $\alpha^{\text{max}}_{150-220} = 1.51$.  
To classify each source as either dusty or synchrotron, we find the probability for each source that $\alpha^{\text{max}}_{150-220} > 1.51$ from each source's marginalized posterior.  
If the probability that a source has $\alpha^{\text{max}}_{150-220} > 1.51$ is less than 50\%, we classify the source as synchrotron, and conversely, if the the probability that a source has $\alpha^{\text{max}}_{150-220} > 1.51$ is greater than or equal to 50\%, the source is classified as dusty.

\subsection{Catalog description}
The columns in our catalog are described as follows; sources in the catalog are listed in order of detection significance, using the highest-significance detection across all bands.  
The catalog for the full survey will be available online\footnote{https://pole.uchicago.edu/public/data/everett20/}.
\begin{enumerate}
\item Source I.D.: Source IAU identification 
\item RA: Right ascension (J2000) in degrees 
\item DEC: Declination (J2000) in degrees 
\item $S^{\text{meas}}_{95}/N_{95}$: Raw signal-to-noise in 95\,GHz
\item $S^{\text{meas}}_{95}$: Raw flux in 95\,GHz, [mJy]
\item $S^{\text{max}}_{95}$: Deboosted flux in 95\,GHz taken from integrating 50\% of the posterior PDF, with 16\% and 84\% taken as 1-$\sigma$ error bars, [mJy]
\item $S^{\text{meas}}_{150}/N_{150}$: Raw signal-to-noise in 150\,GHz, [mJy]
\item $S^{\text{meas}}_{150}$: Raw flux in 150\,GHz, [mJy]
\item $S^{\text{max}}_{150}$: Deboosted flux in 150\,GHz taken from integrating 50\% of the posterior PDF, with 16\% and 84\% taken as 1-$\sigma$ error bars,  [mJy]
\item $S^{\text{meas}}_{220}/N_{220}$: Raw signal-to-noise in 220\,GHz, [mJy]
\item $S^{\text{meas}}_{220}$: Raw flux in 220\,GHz, [mJy]
\item $S^{\text{max}}_{220}$: Deboosted flux in 220\,GHz taken from integrating 50\% of the posterior PDF, with 16\% and 84\% taken as 1-$\sigma$ error bars, [mJy]
\item $\alpha^{\text{meas}}_{95-150}$: Spectral index between 95\,GHz and 150\,GHz as calculated from the raw 95 and 150\,GHz fluxes.
\item $\alpha^{\text{max}}_{95-150}$: Spectral index between 95 and 150\,GHz taken from integrating 50\% of the posterior PDF from the deboosting algorithm.  1-$\sigma$ error bars from integrating 16\% and 84\% of the posterior PDF.
\item $\alpha^{\text{meas}}_{150-220}$: Spectral index between 150\,GHz and 220\,GHz as calculated from the raw 150 and 220\,GHz fluxes.
\item $\alpha^{\text{max}}_{150-220}$: Spectral index between 150 and 220\,GHz taken from integrating 50\% of the posterior PDF from the deboosting algorithm.  1-$\sigma$ error bars from integrating 16\% and 84\% of the posterior PDF.
\item Type: Classification of a source as either synchrotron or dusty depending on the fraction of the integrated 150-220\,GHz spectral index posterior above the threshold of $\alpha^{\text{max}}_{150-220} > 1.51$.  For $P\left(\alpha^{\text{max}}_{150-220} > 1.51\right) \ge 0.5$, the source is classified as dusty, for $P\left(\alpha^{\text{max}}_{150-220} > 1.51\right) < 0.5$, the source is classified as synchrotron.
\item External counterparts: Flag on sources with an associated detection in one of the external catalogs we cross-match.  See Section \ref{sec:xchecksection}.
\item Extendedness: Flag on sources that appear to be extended or are multiple members of the same source at physically offset locations due to being extended.  See Section~\ref{sec:extsrcsection}.
\item Redshift information: Measured redshift, if available.
\item Cut classification: Flag indicating a source is a member of the ``ext cut" (1), ``$z$ cut" (2), or SMG list (3).  See Section \ref{sec:cutsec}.
\end{enumerate}

\begin{deluxetable*}{l c c}
\centering
\tablecaption{Overview of Detections}
\tablehead{Criterion & N. sources $> 4.5\,\sigma$ & N. sources $>5\,\sigma$}
\startdata
95\,GHz detections & 2774 & 2416 \\
150\,GHz detections & 3909 & 3617 \\
220\,GHz detections & 1435 & 991 \\
Three-band detections & 722 & 645 \\
\hline
Sources classified as synchrotron-dominated & 3980 & 3506 \\
Sources classified as dust-dominated & 865 & 530 \\
Sources classified as SPT SMGs & 506 & 258 \\
Sources classified as low-$z$ LIRGs & 302 & 224 \\
Sources identified as stars & 10 & 10 \\
\enddata
\tablecomments{Sources are included if detected above 4.5 (or 5)$\,\sigma$ in at least one band except for three-band detections, which restricts to sources detected above the given threshold in all three bands.}
\label{tab:dettable}
\end{deluxetable*}

\subsection{Completeness}
\label{sec:completeness}
The completeness of the catalog for a given band is defined as the ratio of the number of sources we detect using the source-finding algorithm compared with the true number of sources in the map for a given flux.  
Due to the presence of noise in the maps, sources near the detection threshold may be missed by the source-finder if they happen to be coincident with a negative noise fluctuation which pulls their flux below the detection threshold.  
Completeness is important not only for the robustness of the catalog, but also for calculating number counts, discussed in the following section.  
The completeness is calculated in practice by performing the source-finding on a known population of sources at fixed flux values.  
We add a set of 100 simulated sources at a chosen flux level to random locations in the residual map (the optimally filtered map post-CLEANing, which is a good approximation to noise plus a background of sources below the detection threshold of the CLEANing).  
The source profile used is the real-space version of the transfer function (i.e. a beam with the timestream filtering applied) for the sector which contains the coordinates randomly chosen for the source, rotated to the proper angle.   
We then run the source-finder and cross match the returned detections with the known inputs.  
We repeat this process for a broad range of flux levels.  
The completeness as a function of flux is then given by $f_\text{compl} (S) = N_\text{recovered}/N_\text{input}$.  
Since the noise in our maps is to a good approximation Gaussian and sources are rare enough that the noise dominates the distribution of flux in the map, we would expect the completeness to follow an error function of the form
\begin{equation}
f_\text{compl} (S) = \frac{1}{\sqrt{2\pi\sigma^2}}\int_S^\infty e^{-(S' - S_0)^2/2\sigma^2}dS'
\end{equation}
where $S_0$ is the detection threshold, in this case 4.5 times the mean RMS noise in the map for each band.
Since this process is computationally expensive, we evaluate the completeness at a few discrete flux levels and fit the error function to those results and use it as a model of our completeness, and we estimate the errors on our completeness estimate using binomial statistics. 
We repeat this process for each band separately.

Galaxy clusters appear as compact negative signals at 95 and 150\,GHz via the thermal Sunyaev-Zel'dovich (SZ) effect (see \citealt{bleem15b} for a recent review and catalog release of clusters detected in 2500 square-degrees of the SPT-SZ survey; and \citealt{sunyaev72} for background on the SZ effect).  
Compact clusters with high significance can overlap and therefore cancel out emissive sources, which we do not account for in the completeness calculation.  
Using an assumed cosmological model and cluster mass function as well as SPT cluster selection functions, M13 calculated an expectation of one cluster large enough to cancel a 4.5-$\sigma$ emissive source per ten square degrees of SPT-SZ survey, which corresponds to roughly 10-20 clusters per field or roughly 250 total in the full SPT-SZ survey.  
Given the relatively low point source density in the SPT maps above the detection threshold, the likelihood of purely random overlap and cancellation is less than 1\% per field for 150 GHz, the band with the highest source density, and even though it is known that point sources and clusters have some preference for clustering, the effect on the completeness due to cluster overlap is expected to be relatively small.  

Flux levels averaged over all sectors per each field and band for 50\% and 95\% completeness are shown in Table~\ref{tab:completeness}.  
The median 95\% completeness across all fields is 12.89, 7.60, and 26.83\,mJy at 95, 150, and 220\,GHz, respectively.  

\begin{deluxetable*}{l | c c c c | c c c c | c c c c}
\centering
\tablecaption{Noise levels, completeness, and purity levels} \small

\tablehead{
& \multicolumn{4}{c|}{95\,GHz} & \multicolumn{4}{|c|}{150\,GHz} & \multicolumn{4}{|c}{220\,GHz}\\
Name & RMS & 50\% c. & 95\% c. &  \%p. & RMS & 50\% c. & 95\% c. &  \%p. & RMS & 50\% c. & 95\% c. & \%p.\\
& [mJy] & [mJy] & [mJy] & 4.5\,$\sigma$ & [mJy] & [mJy]  & [mJy] & 4.5\,$\sigma$ & [mJy] & [mJy]  & [mJy] & 4.5\,$\sigma$}
\startdata 
\fivefiffi 		& 2.25 	& 9.76	& 13.33	& 96.2 	& 1.01  	& 4.39 	& 6.00 	& 97.2 	& 2.97 	& 13.25 	& 18.09 	& 89.8\\
\twthreefiffi 	& 2.17 	& 9.43	& 12.87	& 93.1	& 0.939  	& 4.18 	& 5.71 	& 95.0	& 2.73 	& 11.86 	& 16.19 	& 83.9\\
\twonesix 		& 1.89  	& 8.20 	& 11.20  	& 96.7	& 1.11  	& 4.75 	& 6.49 	& 95.9	& 3.83	& 16.42 	& 22.42 	& 85.1\\
\threesix 		& 1.93  	& 8.30 	& 11.34  	& 93.8	& 1.13 	& 4.90 	& 6.69 	& 92.3	& 3.89 	& 16.58 	& 22.64 	& 82.0\\
\twonefif 		& 2.18 	& 9.44 	& 12.89   	& 94.1	& 1.28 	& 5.46	& 7.46 	& 95.3	& 4.36 	& 19.12 	& 26.12 	& 83.4\\
\fourten 		& 1.87 	& 8.15	& 11.13 	& 96.8	& 1.17 	& 5.22	& 7.13	& 97.3	& 4.14 	& 17.65	& 24.11	& 86.6\\
\zerofif 		& 2.24 	& 9.72	& 13.27 	& 96.2	& 1.30 	& 5.71	& 7.80	& 95.3	& 4.49 	& 19.65	& 26.83	& 80.8\\
\twothir 		& 2.15 	& 9.17	& 12.52	& 95.5	& 1.24 	& 5.44	& 7.43	& 93.9	& 4.16 	& 17.89	& 24.43	& 78.2\\
\onesix 		& 2.14 	& 9.25	& 12.64	& 94.3	& 1.27 	& 5.50	& 7.52	& 95.2	& 4.29 	& 18.77	& 25.64	& 79.0\\
\fivefourfi 		& 2.35 	& 10.35	& 14.13	& 92.5	& 1.34 	& 5.79	& 7.90	& 88.4	& 4.83 	& 20.96	& 28.62	& 80.8\\
\sixfiffi 		& 2.22 	& 9.55	& 13.04	& 93.0	& 1.30 	& 5.59	& 7.64	& 95.0	& 4.77 	& 20.50	& 28.00 	& 87.8\\
\twthreesix 	& 2.20 	& 9.54	& 13.03	& 94.7	& 1.29 	& 5.72	& 7.82	& 94.6	& 4.62 	& 20.47	& 27.96 	& 83.8\\
\twonefour 	& 2.25 	& 9.80	& 13.29	& 94.4	& 1.33 	& 5.90	& 8.05	& 93.0	& 4.84 	& 21.03	& 28.71 	& 84.9\\
\twtwo 		& 2.29 	& 10.16	& 13.88	& 94.7	& 1.33 	& 5.87	& 8.02	& 90.9	& 4.93 	& 21.75	& 29.70	& 68.4\\
\twthreefourfi 	& 2.18 	& 9.56	& 13.05 	& 94.2	& 1.29 	& 5.52	& 7.54	& 94.4	& 4.71 	& 20.65	& 28.19	& 78.9\\
\sixsix 		& 2.14 	& 9.08	& 12.40 	& 94.9	& 1.30 	& 5.61	& 7.66	& 93.1	& 4.91 	& 21.55	& 29.43 	& 88.1\\
\threefour 		& 2.11 	& 9.11	& 12.44	& 93.1	& 1.27 	& 5.57	& 7.60	& 95.3	& 4.54 	& 19.45	& 26.56 	& 87.1\\
\onefour 		& 2.21 	& 9.44	& 12.89 	& 96.3	& 1.28 	& 5.70	& 7.79 	& 92.8	& 4.62 	& 20.01	& 27.33	& 81.1\\
\sixfourfi 		& 2.17 	& 9.20	& 12.56 	& 94.2	& 1.30 	& 5.63	& 7.69 	& 94.8	& 4.85 	& 21.13	& 28.86	& 78.0
\enddata

\tablecomments{\label{tab:completeness} RMS noise for the matched-filtered maps, averaged across all sectors; 50\% and 95\% completeness levels; and purity levels at 4.5\,$\sigma$.}

\end{deluxetable*}

\subsection{Purity}
\label{sec:purity}
The purity of the catalog as a function of source signal-to-noise is defined as one minus the fraction of sources at that signal-to-noise or higher that are expected to be false detections due to noise in the map.  
To quantify the purity of the catalog, we estimate the number of detections above a given threshold in a simulated noise-only map and compare those with the number detected above the same significance in the real maps.  
We generate simulated noise maps from difference maps, which contain instrument noise and residual atmosphere.  
The method for generating difference maps is discussed in Section~\ref{sec:optfiltsection}.  
To the noise realizations, we add contributions from the power spectrum of primary anisotropies in the CMB, which is also a source of noise for our source detections.  
These noise fluctuations have a power spectrum determined from the best fit $\Lambda$CDM model to combined WMAP7 and SPT data~\citep{keisler11}.  
We also include an estimate of the thermal SZ effect, as well as contributions from the CIB in terms of a Poisson and clustered component.  
The component of the noise that we add to our simulations to account for the SZ effect is a Gaussian random field with power spectrum given by fitting measurements in \citet{shirokoff11}.  

Running the source-finder on these simulated maps, we calculate the purity as a function of signal-to-noise to be
\begin{equation}
f_\text{pure} = 1 - N_\text{false}/N_\text{total}
\end{equation}

Massive clusters in the real maps will contribute to impurity in the source-finding because the timestream filtering causes these objects to have positive wings, which can be detected as false sources.  
However, these false detections are easy to identify and quite rare in the real maps.  
We remove them from the catalog by hand, and a total of six sources are removed.  
Thus there is no need to include them in the purity simulations.

Table~\ref{tab:completeness} shows purity values averaged over all sectors per field and per band for detections $\geq 4.5\,\sigma$.  
The median purity for sources detected at $\geq 4.5\,\sigma$ across all fields for the full survey is 94.4\%, 94.8\%, and  83.4\% at 95, 150, and 220\,GHz, respectively.  
For sources detected at $\geq 5.0\,\sigma$, the median purity across all fields is 98.9\%, 97.6\%, and 95.1\% at 95, 150, and 220\,GHz, respectively.

\begin{deluxetable*}{l c c c c c c}
\centering
\tablecaption{Cross-matches with External Catalogs}
\tablehead{Survey Name & Band & Beam size & $\Sigma$ [1/deg$^2$] & $r_\text{assoc}$ [arcmin] & X-matches & P(random) [\%]}
\startdata
SPT & 95\,GHz (3.2\,mm) & 1.7\,arcmin & 1.10 & & & \\
 & 150\,GHz (2.0\,mm) & 1.2\,arcmin & 1.54 & & & \\
 & 220\,GHz (1.4\,mm) & 1.0\,arcmin & 0.57 & & & \\
SUMSS & 843\,MHz (36\,cm) & 45\,arcsec & 26.75 & 0.8 & 3427 & 1.49\\
PMN & 4850\,MHz (6\,cm) & 4.2\,arcmin & 1.75 & 2.5 & 1834 & 0.95\\
AT20G & 20\,GHz (1.5\,cm) & 4.6\,arcsec & 0.32 & 1.0 & 820 & 0.03\\
IRAS & 12, 25, 60, 100\,$\mu$m & 11-88\,arcsec & 4.72 & 1.5 & 318 & 0.92\\
AKARI-FIS & 65, 90, 140, 160\,$\mu$m & 24-59\,arcsec & 0.87 & 1.5 & 217 & 0.17\\
AKARI-IRC & 9, 18\,$\mu$m & 3.3-6.6\,arcsec & 5.18 & 0.5 & 56 & 0.11\\
WISE & 3.4, 4.6, 12, 22\,$\mu$m & 6.1-12\,arcsec & 45.32 & 0.7 & 734 & 1.94\\
RASS & 0.1-2.4\,keV &  & 3.53 & 1.5 & 447 & 0.69
\enddata
\tablecomments{Overview of bands, beam sizes, and source densities for the SPT-SZ catalog and external cross-match catalogs.  For each external catalog, the cross-match radius, the total number of cross-matches, and the probability of random association given the chosen cross-match radius and underlying source density of the external catalog are listed.}
\label{extcatstable}
\end{deluxetable*}

\subsection{External associations}
\label{sec:xchecksection}
To further characterize the nature of sources in the SPT catalog, we cross-match with seven external catalogs, ranging in wavelength from radio to X-ray.  
These include: 

\begin{itemize}
\item The Sydney University Molonglo Sky Survey (SUMSS, \citealt{mauch03}) at 843\,MHz
\item The Parkes-MIT-NRAO (PMN) Southern Survey \citep{wright94} at 4850\,MHz
\item The Australia Telescope 20-GHz Survey (AT20G, \citealt{murphy10})
\item The Infrared Astronomical Satellite Faint Source Catalog (IRAS-FSC, \citealt{moshir92}) at 12, 25, 60, and 100\,$\mu$m
\item The Infrared Astronomical Satellite AKARI, IRC Point Source Catalog \citep{yamamura10} at 9 and 18\,$\mu$m, and the FIS Bright Source Catalog \citep{ishihara10} at 65, 90, 140, and 160\,$\mu$m
\item The Wide-field Infrared Survey Explorer (WISE) AllWISE Source Catalog at 3.4, 4.6, 12, and 22\,$\mu$m 
\item The ROSAT All-Sky Survey (RASS) Bright Source Catalog~\citep{voges99} and Faint Source Catalog~\citep{voges00} at X-ray energies 0.1-2.4\,keV
\end{itemize}

Each external catalog is cross-matched with positions of SPT point sources in the catalog using a radial association criterion, as overviewed in Section~\ref{sec:radxmatchmethod}.  
An appropriate radius for association is determined for each external catalog by looking at the distributions of source separations, selecting a radius such that the probability of a random, false association is approximately 1\% and no greater than 2\%.  
The chosen radius for each catalog can be found in Table~\ref{extcatstable}.  
For most of the external catalogs, the density of sources is low enough that confusion within the SPT beam size is not an issue.  
For WISE, which has the highest source density, confusion becomes a problem for cross-matching with the detections in the shorter-wavelength WISE bands.  
Therefore, we restrict the source density in the WISE sources we cross-match with by applying a cut on the WISE catalog using the W4 22\,$\mu$m band, and restricting to only cross-matching with WISE sources that have W4 flux greater than 5\,mJy.  
We experimented with a more complex cross-matching scheme, incorporating source flux and number density, but found that a simple radial cross-match achieved comparable results.

\begin{figure}[t]
\begin{center}
\includegraphics[width=8.5cm]{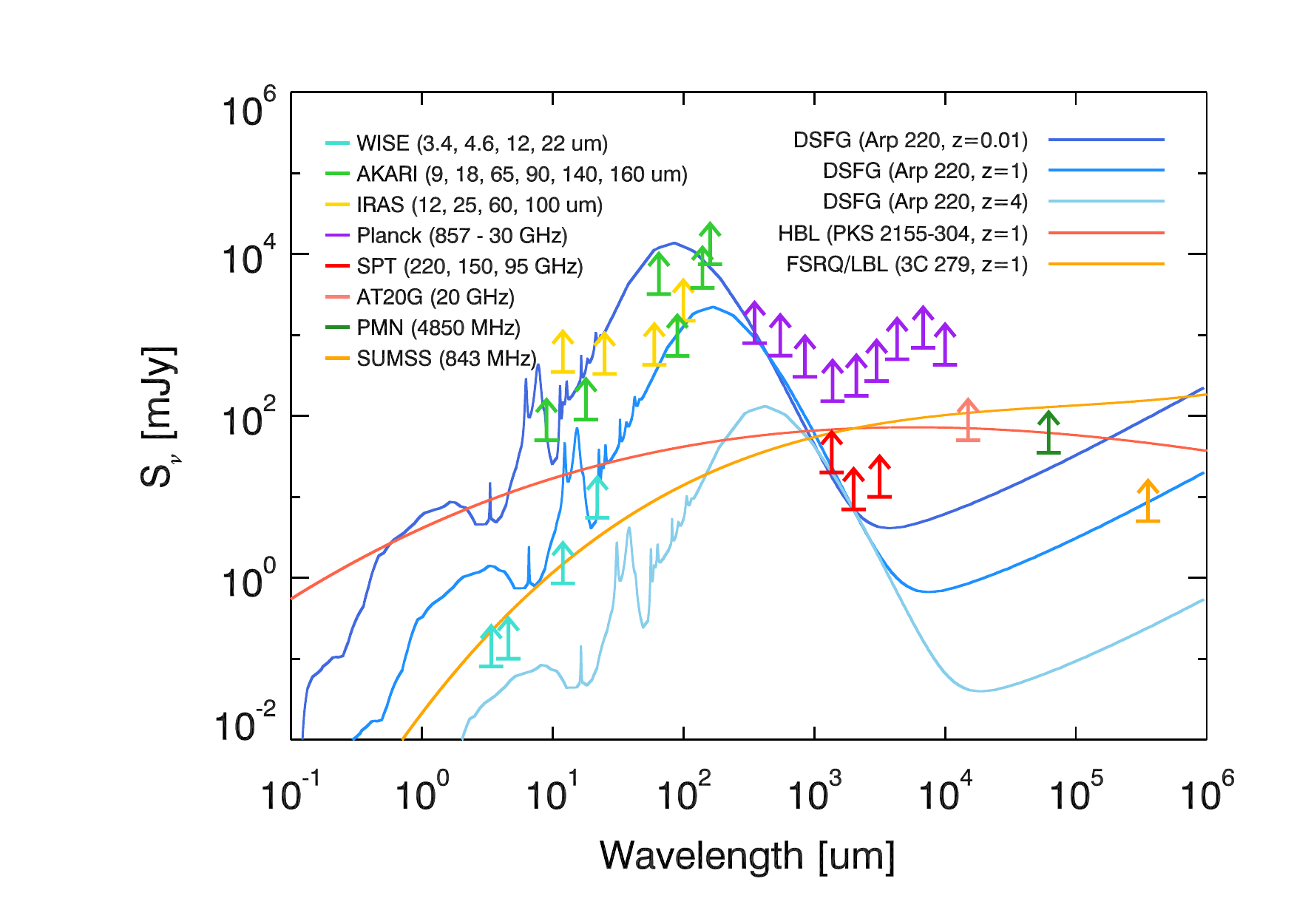}
\end{center}
\caption{A comparison of survey depths for SPT and wide-field surveys used to cross-match with the SPT source catalog.  Blue curves show example spectral energy distribution (SED) curves for dusty star-forming galaxies and their high-redshift component, SMGs, which are an Arp 220 SED shifted in redshift.  Red and orange curves show two example synchrotron SEDs for different types of flat-spectrum sources.
\label{fig:survdepth}}
\end{figure}

\begin{figure}[!ht]
\begin{center}
\includegraphics[width=8.5cm]{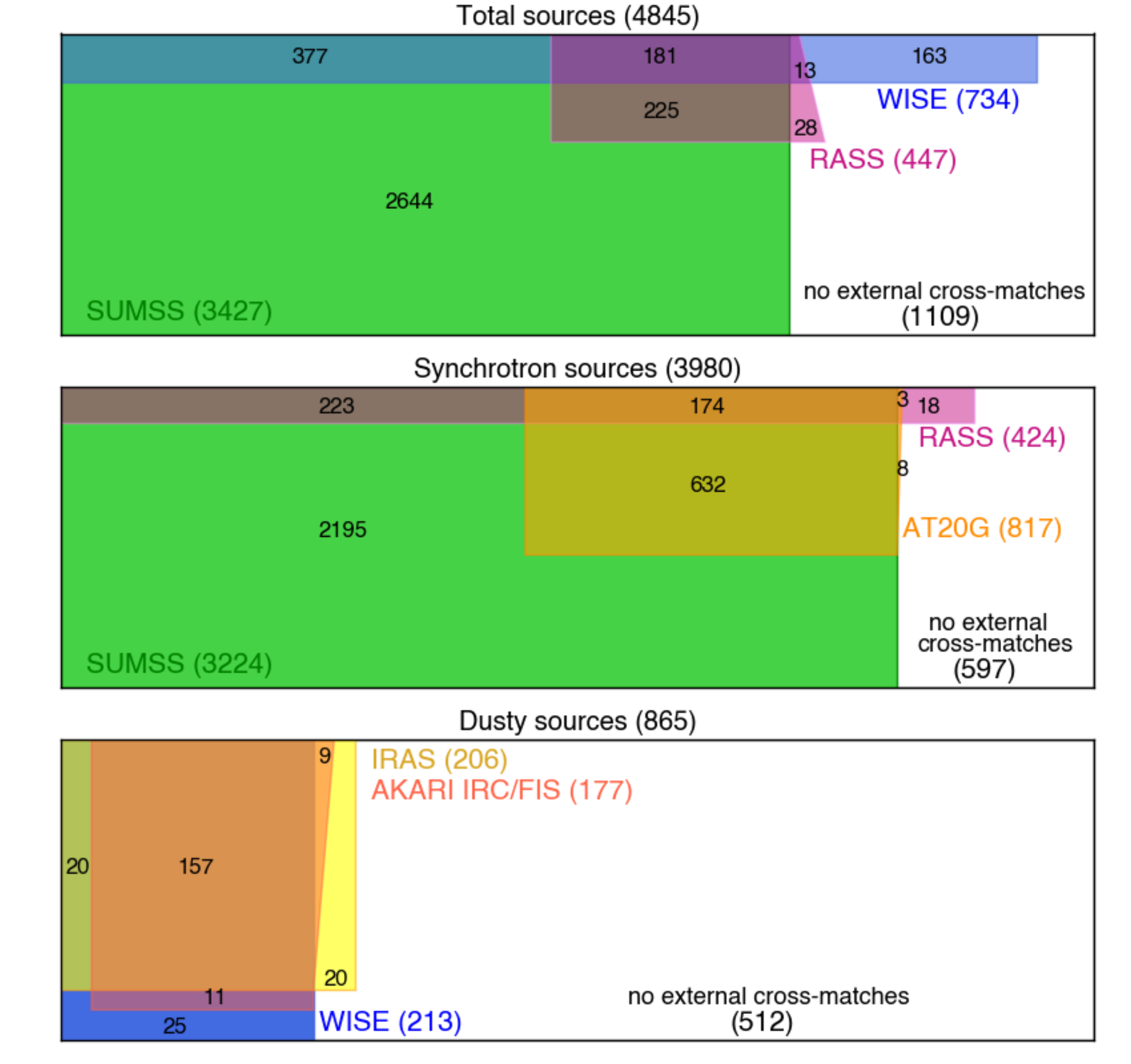}
\end{center}
\caption{Venn diagrams showing fractional cross-match overlap between the SPT-SZ catalog and external catalogs.  The top panel shows cross-matches with the full SPT-SZ catalog; the middle and bottom panels show cross-matches for synchrotron and dusty sources, respectively.  In each panel, colored regions indicate the proportion of SPT-SZ sources with cross-matches in each catalog, showing overlapping cross-matches between various catalogs, and white space indicates the fraction of SPT-SZ sources that have no cross-matches with catalogs displayed in that panel.  We find that of the 4845 total sources in the catalog, 1109 (23\%) have no cross-matches in external catalogs; 597 of these are classified as synchrotron (15\% of synchrotron sources), and 512 are classified as dusty (59\% of dusty sources). 
\label{fig:venndiag}}
\end{figure}

Figure~\ref{fig:survdepth} shows an overview of the wavelengths and detection thresholds of the external catalogs with which we cross-match the SPT-SZ catalog.
The figure also shows reference spectral energy distribution (SED) curves for DSFGs, modeled by an Arp 220 profile, and reference SEDs for two examples of flat-spectrum synchrotron sources.
Shifting the reference DSFG SED in redshift demonstrates how negative K-correction enables detection of high-redshift dusty sources in mm wavelengths.  
Surveys in the infrared observe dusty sources on the Wien side of their SED and therefore shift to a dimmer portion of the spectrum with increasing redshift, in addition to dimming from increasing source distance.
In contrast, mm/sub-mm wavelength surveys observe dusty sources on the Rayleigh-Jeans side of the spectrum and therefore shift to an intrinsically brighter part of the spectrum with increasing redshift, canceling the effect of dimming from increased distance.
Table~\ref{extcatstable} gives an overview of each survey and the number of cross-matches with the SPT catalog.  
A comparison of cross-matches per catalog, including cross-match overlap between surveys for the total SPT catalog as well as dusty and synchrotron sub-populations is illustrated in Figure~\ref{fig:venndiag}.  
The most ubiquitous cross-match for the SPT catalog is with the SUMSS survey in the radio, where 71\% of SPT sources have cross-matches in SUMSS.  
SUMSS is especially useful for cross-matches with synchrotron-dominated sources in the SPT-SZ survey since the wide-field radio survey has full coverage of the SPT-SZ area and is complete to a depth of 6\,mJy/beam at 5\,$\sigma$, 
The SUMSS beam is relatively large, making it not suitable for cross-matching with high-resolution optical and infrared catalogs, but confusion is not a significant issue when comparing with SPT, which has a similarly large beam and low source density.  
For dusty sources, IRAS in the infrared is particularly useful for identifying low-redshift dusty galaxies, but both WISE and AKARI overlap IRAS cross-matches considerably, as shown in Figure~\ref{fig:venndiag}.  

Of the 4845 sources in the catalog, 1109 have no cross-matches with external catalogs.  
84\% of these are detections in only one band, mostly in 150 GHz-only or 220 GHz-only; 10 sources with no cross-matches in external catalogs have detections in all three bands.

\subsection{Extended sources}
\label{sec:extsrcsection}

We expect that all extragalactic sources with redshifts greater than $ z \sim 0.05$ will be unresolved in the maps, given the instrumental beam size of roughly 1\,arcmin.  
There is a chance that very nearby sources or bright AGN with extended radio lobes may be resolved in the maps.  
We take a two-pronged approach to flagging extended sources: first, we fit a cutout around each detected source to a model constructed from the beam profile convolved with a non-symmetric 2D Gaussian and compare the $\Delta \chi^2$ of the fit to a model containing only the beam.  
Based on looking at the fields with the most obvious extended sources, we use a threshold of $\Delta \chi^2\geq7$ to flag sources as extended in the catalog.  
Second, to ensure that we are catching all sources that are detected as multiple detections of the same source in the CLEANing, we run a by-eye check of all sources within close proximity to other detections and flag sources that appear to be multiple detections at physically offset locations of the same, extended source.  
Each source flagged as possibly being a multiple detection of the same, extended object, is cross-checked with external catalogs to determine if the detections are indeed from the same object or from distinct objects that appear in our maps with close proximity.  
For the sake of completeness, we leave all detections in the catalog, but indicate the likelihood that a source is extended.  
In calculating the number counts, we calculate multiple versions of the counts, including using the extendedness information from both flagging methods.  
Fluxes for extended sources will be lower limits on the true flux, since the CLEANing is unable to accurately return flux for sources that do not look like our chosen source profile.    
Using the two methods discussed above, a total of 131 sources from the catalog are flagged as extended.

\subsection{Redshift associations}
\label{sec:zassoc}

Redshifts for SPT catalog sources are obtained from a combination of follow-up observations (e.g., \citealt{weiss13, strandet16}) and the literature.  
We obtain literature redshifts by querying the NASA/IPAC Extragalactic Database (NED) and using an association radius of 0.6\,arcmin.
743 sources in the catalog have identified redshifts; available redshift information is listed per source in the catalog and shown in Figure~\ref{fig:alphavsredshift}.  

\subsection{Star identification}
\label{sec:starident}

\begin{figure*}[!ht]
\begin{center}
\includegraphics[width=18.1cm]{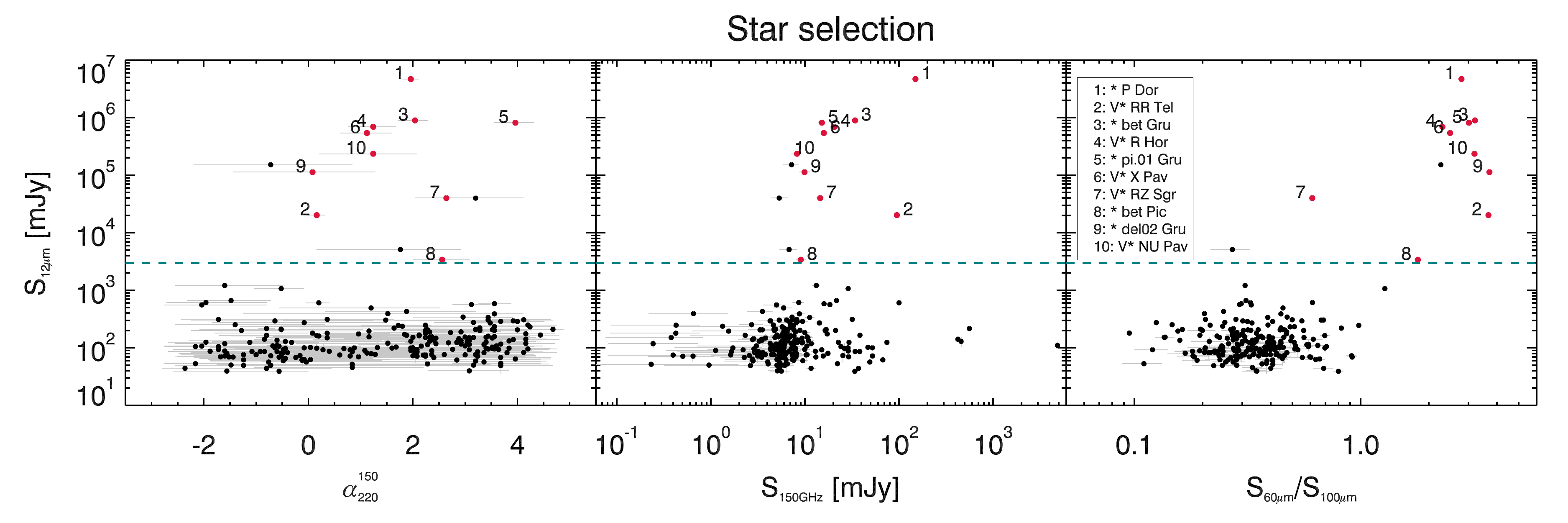}
\end{center}
\caption{Cross-matched IRAS 12\,$\mu$m flux versus SPT spectral index $\alpha_{150-220}$ ({\it left}), SPT flux at 150\,GHz ({\it middle}), and the ratio of cross-matched IRAS 60\,$\mu$m flux to 100\,$\mu$m flux ({\it right}).  Objects in the SPT-SZ catalog identified as stars are shown as red points, all other objects in the SPT-SZ catalog are shown as black points, with a rough separation threshold in 12\,$\mu$m flux shown as a teal dashed line.  We note that using SPT data alone, stars are not distinguishable from non-stellar SPT sources (left two panels), and that while the ratio of 60\,$\mu$m to 100\,$\mu$m IRAS flux can be used to identify stars (right panel), high flux at 12\,$\mu$m is sufficient for star identification.  The method for identifying stars is discussed in more detail in Section~\ref{sec:starident}, and further details regarding stars identified in the SPT-SZ catalog and other sources that appear with high flux at 12\,$\mu$m but are not identified as stars can be found in Section~\ref{sec:stardiscuss}.
\label{fig:starid}}
\end{figure*}

A small but interesting sub-population in the catalog are ten stars, identified primarily using their cross-matched IRAS flux at 12\,$\mu$m.  
In this section, we overview the method for separating stars from other objects in the SPT-SZ catalog; Section~\ref{sec:stardiscuss} discusses characteristics of the stars we see in the SPT-SZ catalog.
As shown in Figures \ref{fig:alpha1v2} and \ref{fig:starid}, the stars are not clearly identifiable using SPT data alone: their flux in SPT bands does not set them apart from other SPT sources and their spectral indices in SPT wavelengths span both synchrotron and dusty populations.  
However, looking at cross-matched flux in IRAS at 12\,$\mu$m, these sources have considerably higher flux than other sources in the SPT catalog. 

The primary selection effect for detecting stars at millimeter wavelengths is a bias toward either large-surface-area and highly-luminous stars or stars with excess emission from dust such that they will have sufficient flux at mm wavelengths to be detectable.  
Nine of the ten stars identified in the SPT catalog are red giants on the Asymptotic Giant Branch (AGB), most of which are late-type M stars.  
The remaining star is * $\beta$~Pic, which has a well-known dusty circumstellar disk~\citep{sheret04,riviere-marichalar14}.  
Red giants are luminous, large, and relatively cool, with surface temperatures of order a few thousand Kelvin.  
Therefore their stellar flux follows a blackbody distribution peaking around 1 to a few\,$\mu$m~\citep{bedding97, whitelock97} with a spectral index of 2 at longer wavelengths, and they are large and bright enough to be detectable at millimeter wavelengths, far into the Rayleigh-Jeans tail of their blackbody spectrum.
Dusty galaxies as well as flat- and steep-spectrum synchrotron sources have spectra that rise as a function of wavelength for wavelengths shorter than $\sim$ 100~$\mu$m, as shown in Figure~\ref{fig:survdepth}, whereas stars have spectra that are falling between $\lambda =$ a few\,$\mu$m to $\sim$ 100\,$\mu$m.  
Therefore, the ratio of IRAS 60\,$\mu$m/100\,$\mu$m flux should be less than one for non-stellar objects and greater than one for stars.  
This ratio can be used with relative success to identify stars, as shown in the right panel of Figure~\ref{fig:starid}, but we find that high IRAS 12\,$\mu$m flux on its own is a more effective criterion.  
In theory, it should be possible to use cross-matches with WISE at wavelengths shorter than the IRAS bands to more clearly identify stars, since stars should be even brighter at shorter IR wavelengths than IRAS, closer to the peak of the stellar SED; however, most of the stars observed in the SPT sample are so bright that the WISE flux measurements are saturated and unreliable.

\subsection{Cut selection criteria}
\label{sec:cutsec}

To assist with comparing number counts with models and to further characterize source populations within the catalog, we develop three source cuts using extendedness and external cross-match information.  
Because source fluxes are measured in maps that have been optimally filtered assuming sources are unresolved by the SPT beam, sources that are flagged as extended or measured in the SPT maps as multiple detections will have fluxes that are systematically underestimated and therefore may bias the number counts.  
We therefore develop two cuts to flag them for removal when calculating the number counts.  

First, in the extended cut, or ``ext cut," we flag all objects flagged as extended or detected as multiple detections but confirmed to be a single object, using the methods described in Section~\ref{sec:extsrcsection}.  
Sources identified as stars are also removed in the counts for this cut, since they are not included in the models with which we compare the counts.  
The extended cut removes 131 sources from the catalog as a whole, 36 of these are classified as synchrotron-dominated and 95 as dust-dominated.  

Second, we develop a cut to flag all low-redshift objects, using the redshift cross-match information discussed in Section~\ref{sec:zassoc}.  
Because the extended source flag used in the ``ext cut" involves in part a by-eye inspection of individual sources, a method was sought to remove extended objects more systematically.  
All extended sources appear large enough in the SPT maps to be resolved by the SPT beam, and therefore should all be relatively local and removable by cutting all objects with low measured redshift.  
However, cutting below a redshift threshold will remove additional sources as well.  The ``$z$ cut" trims all sources flagged as stars and all sources with cross-matched redshifts $z \le 0.1$, resulting in flagging 461 sources from the full catalog, of which 248 have a synchrotron classification and 213 have a dusty classification.  
Looking at the distributions of source angular sizes for SPT sources with NED identifications, we expect that the cut threshold of $z < 0.1$ will correspond roughly to cutting objects with angular sizes $\gtrsim 1$\,arcmin, roughly the size of the SPT beam.  
We verify that all sources flagged by ``ext cut" are included in those sources flagged by the ``$z$ cut." 

\begin{figure}
\begin{center}
	\includegraphics[width=8.5cm]{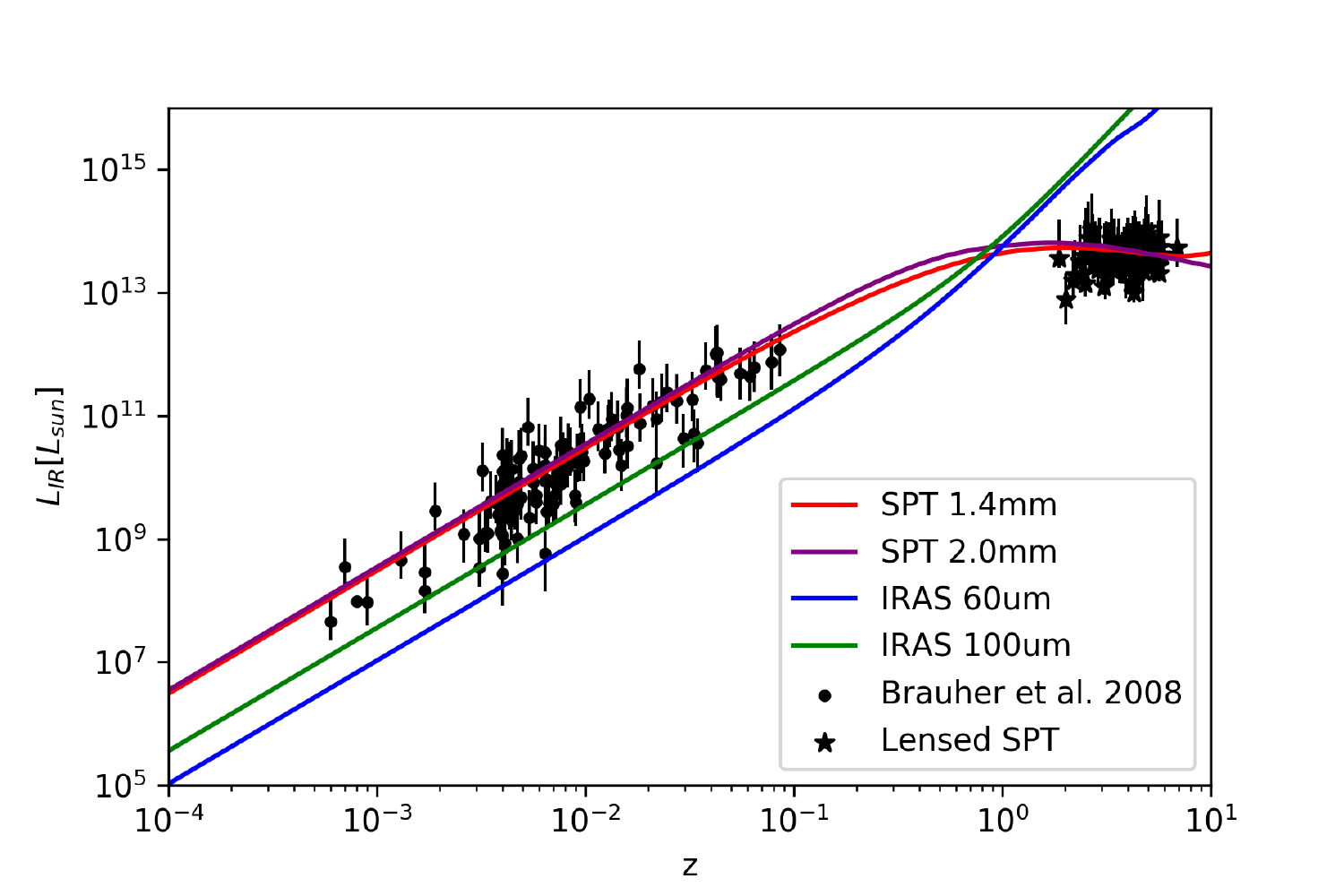}
\end{center}
\caption{Observed infrared luminosity versus redshift for low-redshift infrared galaxies~\citep{brauher08} and SPT SMGs with detection thresholds for an Arp-220-like dusty galaxy spectrum shown in solid lines.
\label{fig:iras_select}}
\end{figure}

To more cleanly select sources in the catalog that are likely to be high-redshift SMGs, which in the relatively high flux range probed by the catalog are likely to be gravitationally lensed, we develop a list of ``SPT SMGs" using more strict criteria than the ``$z$ cut" and ``ext cut" source lists.  
For this cut, we include only dust-dominated sources, we apply the same redshift criterion as the ``$z$ cut," and we also exclude any remaining detections with IRAS cross-matches.  
Although a few IRAS detections have been confirmed to be at relatively high redshift (e.g. APM 0827~\citealt{irwin98b}), these sources are few in the literature.  
Furthermore, IRAS detections are unlikely to be high-redshift objects, since, as shown in Figure~\ref{fig:survdepth}, dusty objects will be observed on the Wien side of the spectrum in the IRAS bands, which will shift to an intrinsically dimmer portion of the spectrum with increasing redshift, in addition to reduced flux from greater distance.  
In contrast, the negative K-correction of dusty sources in sub-/millimeter bands enables the detection of the same luminosity source out to high redshifts, as shown in Figure~\ref{fig:iras_select}.
Additionally, a cut on IRAS objects has been used successfully in previous SPT analyses as a proxy for trimming low-redshift objects \citep{vieira10,mocanu13}.
We also trim SPT detections that are measured as ``dipping" in the three SPT bands, meaning that they are sources with a dusty spectral index between 150 and 220\,GHz, but a synchrotron spectral index between 95 and 150\,GHz.  
In a reanalysis of SPT number counts from M13, \citet{mancuso15} identified a set of sources in the SPT-SZ survey that are relatively bright at 95 GHz and were classified as dusty galaxies using the SPT pipeline in M13.
When considering fluxes for each source across a wider range of frequencies than just the SPT data, \citet{mancuso15} note that these sources do appear to have significant emission in radio bands indicating synchrotron emission and don't have spectra that would clearly indicate that these sources are DSFGs.
The presence of synchrotron emission causes the spectrum in SPT bands to appear ``dipping," and although we note that the exact classification of these sources remains somewhat unclear, it provides evidence that ``dipping" sources in SPT data are not clearly DSFGs. 
Furthermore, a set of SPT ``dipping" sources have been observed in preliminary follow-up observations with LABOCA at 870\,$\mu$m.
While a few sources had measured fluxes at 870\,$\mu$m that would be consistent with the presence of dust, most sources had measured fluxes at 870\,$\mu$m that were too low to be consistent with a DSFG spectrum, indicating that they may be synchrotron sources with complicated spectra or may be blends of unrelated objects.
Follow-up spectroscopy with the VLT to obtain redshifts for a set of SPT ``dipping" sources have measured redshifts in the range $z = 0.85 - 2.32$, also indicating that these sources are less likely to be high-redshift SMGs.
We note that there are a couple dipping sources from V10 and M13 with follow-up observations that confirmed that they are high-redshift lensed objects.  
However, these objects either appear to have ``dipping" spectral behavior in the SPT bands due to superposition of the high-redshift object with its foreground lens or are a superposition of unrelated objects along the line of sight.  
Therefore, although we note that high-redshift dusty galaxies may possess significant synchrotron emission, and therefore may manifest in the SPT-SZ catalog with a ``dipping" spectrum, follow-up information on known SPT ``dipping" sources so far has indicated they are less likely to be high-redshift dusty galaxies, and therefore to be conservative, we exclude them from the SMG list.  
We find a total of 506 sources in the SMG list, of which 73 have detections above 4.5\,$\sigma$ at both 150 and 220\,GHz.  

\begin{figure*}[ht!]
\begin{center}
\includegraphics[width=18.1cm]{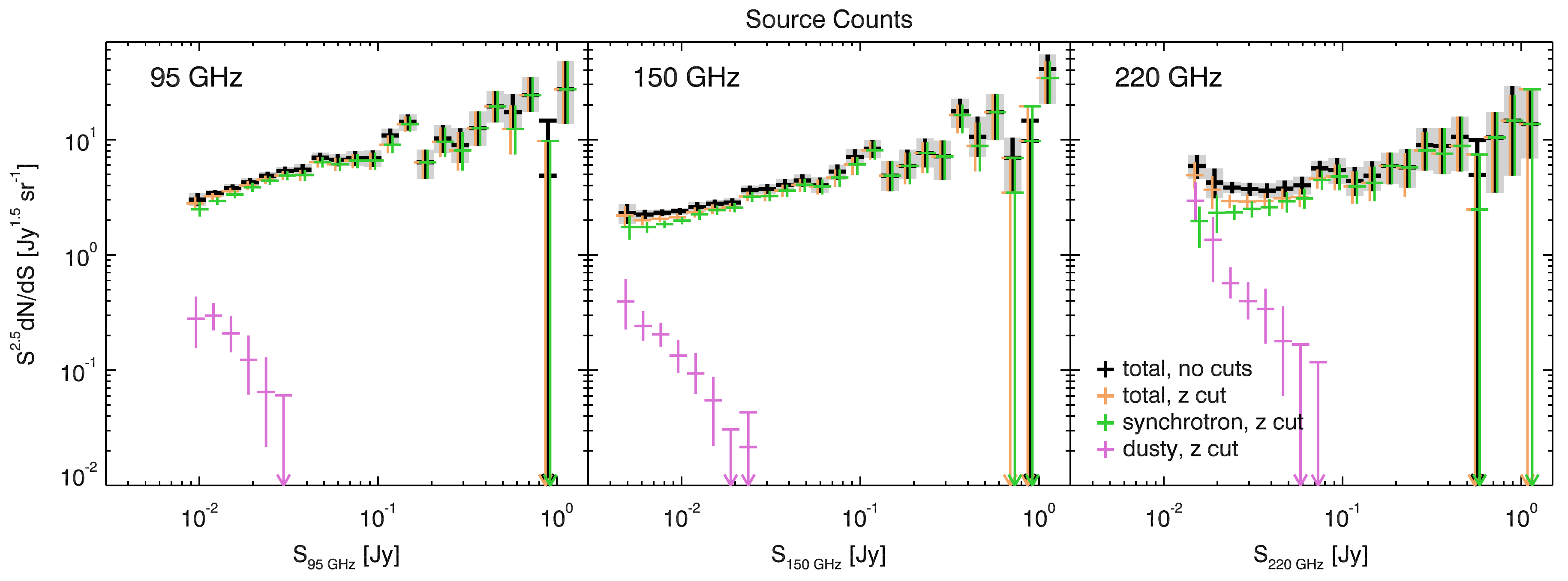}
\end{center}
\caption{Differential number counts of emissive sources in 2530 square degrees of the SPT-SZ survey.  Ten sources identified as stars have been removed.  Two versions of the total counts are shown: total counts with no cuts applied, other than stars, and total counts with the ``$z$ cut" applied (cutting all objects with measured redshifts below z $= 0.1$).  Counts for synchrotron-dominated sources (with ``$z$ cut" applied) are shown in green, and counts for dust-dominated sources (also with ``$z$ cut" applied) are shown in purple.  Details of the cuts can be found in Section~\ref{sec:cutsec} and plots comparing cut versions for dusty and synchrotron counts can be found in the appendix.
\label{fig:numcounts}}
\end{figure*}

\section{Number counts}
\label{sec:number_counts}

In addition to supplying a catalog of detected sources, we seek to calculate the expected number counts in each of our bands as a function of flux.  
The number counts provide a characterization of mm-wave source populations at different wavelengths, and can be used to constrain models of galaxy evolution.

To characterize the number counts at each of our three frequencies, we employ a bootstrap method developed in \citet{austermann09}.  
For each band, we select only the sources in the catalog that are detected above 4.5\,$\sigma$ in that particular band.  
For each source, using our chosen band as the flux prior band for deboosting, we select 50,000 triplets of source fluxes from the 3-dimensional flux posterior probability distribution for that source.  
Effectively this creates 50,000 mock catalogs.  
We resample each catalog by drawing fluxes with replacement for a number of sources that is a Poisson deviate of the true catalog size.  
We then calculate for each catalog the number of sources in each flux bin to find the differential number counts, and determine 16th, 50th, and 84th percentiles of the distribution of $dN/dS$ within each flux bin.  
The number counts are corrected for completeness in each bin using the simulations in Section~\ref{sec:completeness}.  
We plot our calculated number counts in Figure \ref{fig:numcounts}.  
We do not explicitly correct for purity in the number counts, since that will be accounted for by the deboosting which has generated the posteriors we draw from.  
The posteriors include fluxes below the detection threshold, and when drawing fluxes at random, there is a chance that fluxes below the detection threshold will be chosen.  
When this occurs, we remove them from the number counts calculation.

Figure~\ref{fig:numcounts} shows differential source counts per band for the full catalog population (excluding stars), as well as synchrotron and dusty population counts.  
The counts for synchrotron and dusty populations are generated using a probabilistic classification, where we calculate the corresponding $\alpha^{\text{max}}_{150-220}$ for each of the 50,000 flux resamplings of each source.  
We then calculate the probability that each resampling will be classified as dusty or synchrotron using the same cut as for the catalog sources, and associate it with the counts for its assigned population.  
Therefore, for a single source in the catalog, if it has a probability $p$ of having $\alpha^{\text{max}}_{150-220} \ge 1.51$, it will fall into the dusty source counts $p$ fraction of resamplings and will fall into the synchrotron source counts $1-p$ fraction of resamplings.  
Looking at Figure~\ref{fig:numcounts}, as we might expect, the synchrotron counts dominate at all frequencies, but dusty sources are much more prominent at 220\,GHz than in the other two bands, and exceed the synchrotron counts at the very lowest flux levels.  
The total counts are shown with two cut versions: no cuts applied (other than removing stars), and applying the ``$z$ cut," where we remove all sources with measured redshifts z $< 0.1$. 
Synchrotron-dominated and dust-dominated counts are also shown, with the ``$z$ cut" applied.
Tables~\ref{tab:diffcountstable90}, \ref{tab:diffcountstable150}, and \ref{tab:diffcountstable220} give the calculated $dN/dS$ number counts for 95, 150, and 220\,GHz, respectively, for the total catalog population, synchrotron and dusty populations with the ``$z$ cut" applied, and SMGs.

\begin{deluxetable*}{llllc}
\tabletypesize{\footnotesize}
\tablecaption{$95$~GHz differential counts
\label{tab:diffcountstable90}}
\tablehead{
\colhead{Flux range} & \colhead{$dN/dS$ total, no cuts} & \colhead{$dN/dS$ sync, $z$ cut} & \colhead{$dN/dS$ dust, $z$ cut} & \colhead{completeness} \\
\colhead{[Jy]} & \colhead{[${\rm Jy}^{-1} {\rm deg}^{-2}$]} & \colhead{[${\rm Jy}^{-1} {\rm deg}^{-2}$]} & \colhead{[${\rm Jy}^{-1} {\rm deg}^{-2}$]} & 
}
\startdata
$8.7\times10^{-3}-1.1\times10^{-2}$ & $(9.71_{-1.2}^{+1.3})\times 10^{1}$ & $(8.00_{-1.1}^{+1.3})\times 10^{1}$ & $9.00_{-4.0}^{+5.0}$ & $0.80$\\
$1.1\times10^{-2}-1.4\times10^{-2}$ & $(6.28_{-0.4}^{+0.4})\times 10^{1}$ & $(5.37_{-0.4}^{+0.4})\times 10^{1}$ & $5.44_{-1.4}^{+1.6}$ & $0.92$\\
$1.4\times10^{-2}-1.7\times10^{-2}$ & $(3.95_{-0.3}^{+0.3})\times 10^{1}$ & $(3.48_{-0.3}^{+0.3})\times 10^{1}$ & $2.17_{-0.7}^{+0.9}$ & $1.00$\\
$1.7\times10^{-2}-2.2\times10^{-2}$ & $(2.52_{-0.2}^{+0.2})\times 10^{1}$ & $(2.29_{-0.2}^{+0.2})\times 10^{1}$ & $(7.28_{-3.6}^{+4.6})\times 10^{-1}$ & $1.00$\\
$2.2\times10^{-2}-2.7\times10^{-2}$ & $(1.63_{-0.1}^{+0.1})\times 10^{1}$ & $(1.48_{-0.1}^{+0.1})\times 10^{1}$ & $(2.18_{-1.5}^{+2.2})\times 10^{-1}$ & $1.00$\\
$2.7\times10^{-2}-3.4\times10^{-2}$ & $(1.03_{-8.7\times10^{-2}}^{+9.3\times10^{-2}})\times 10^{1}$ & $9.33_{-0.9}^{+0.9}$ & $ 0 _{- 0 }^{+0.1}$ & $1.00$\\
$3.4\times10^{-2}-4.2\times10^{-2}$ & $5.97_{-0.6}^{+0.7}$ & $5.37_{-0.6}^{+0.6}$ &   & $1.00$\\
$4.2\times10^{-2}-5.3\times10^{-2}$ & $4.32_{-0.4}^{+0.4}$ & $3.95_{-0.4}^{+0.4}$ &   & $1.00$\\
$5.3\times10^{-2}-6.7\times10^{-2}$ & $2.36_{-0.3}^{+0.3}$ & $2.15_{-0.3}^{+0.3}$ &   & $1.00$\\
$6.7\times10^{-2}-8.3\times10^{-2}$ & $1.39_{-0.2}^{+0.2}$ & $1.32_{-0.2}^{+0.2}$ &   & $1.00$\\
$8.3\times10^{-2}-1.0\times10^{-1}$ & $(7.88_{-1.3}^{+1.3})\times 10^{-1}$ & $(7.50_{-1.3}^{+1.3})\times 10^{-1}$ &   & $1.00$\\
$1.0\times10^{-1}-1.3\times10^{-1}$ & $(7.03_{-1.0}^{+1.0})\times 10^{-1}$ & $(5.84_{-0.9}^{+1.2})\times 10^{-1}$ &   & $1.00$\\
$1.3\times10^{-1}-1.6\times10^{-1}$ & $(5.25_{-0.8}^{+0.8})\times 10^{-1}$ & $(5.01_{-0.7}^{+0.8})\times 10^{-1}$ &   & $1.00$\\
$1.6\times10^{-1}-2.1\times10^{-1}$ & $(1.33_{-0.4}^{+0.4})\times 10^{-1}$ & $(1.33_{-0.4}^{+0.4})\times 10^{-1}$ &   & $1.00$\\
$2.1\times10^{-1}-2.6\times10^{-1}$ & $(1.22_{-0.3}^{+0.4})\times 10^{-1}$ & $(1.14_{-0.3}^{+0.4})\times 10^{-1}$ &   & $1.00$\\
$2.6\times10^{-1}-3.2\times10^{-1}$ & $(6.07_{-1.8}^{+2.4})\times 10^{-2}$ & $(5.46_{-1.8}^{+2.4})\times 10^{-2}$ &   & $1.00$\\
$3.2\times10^{-1}-4.1\times10^{-1}$ & $(4.84_{-1.5}^{+1.9})\times 10^{-2}$ & $(4.84_{-1.5}^{+1.9})\times 10^{-2}$ &   & $1.00$\\
$4.1\times10^{-1}-5.1\times10^{-1}$ & $(4.25_{-1.2}^{+1.5})\times 10^{-2}$ & $(4.25_{-1.2}^{+1.5})\times 10^{-2}$ &   & $1.00$\\
$5.1\times10^{-1}-6.4\times10^{-1}$ & $(2.16_{-0.6}^{+0.9})\times 10^{-2}$ & $(1.54_{-0.6}^{+0.9})\times 10^{-2}$ &   & $1.00$\\
$6.4\times10^{-1}-8.0\times10^{-1}$ & $(1.72_{-0.5}^{+0.7})\times 10^{-2}$ & $(1.72_{-0.5}^{+0.7})\times 10^{-2}$ &   & $1.00$\\
$8.0\times10^{-1}-1.0$ & $(1.96_{-2.0}^{+3.9})\times 10^{-3}$ & $ 0 _{- 0 }^{+3.9\times10^{-3}}$ &   & $1.00$\\
$1.0-1.3$ & $(6.26_{-3.1}^{+4.7})\times 10^{-3}$ & $(6.26_{-3.1}^{+4.7})\times 10^{-3}$ &   & $1.00$\\
\enddata
\end{deluxetable*}
 
\begin{deluxetable*}{lllllc}
\tabletypesize{\footnotesize}
\tablecaption{$150$~GHz differential counts
\label{tab:diffcountstable150}}
\tablehead{
\colhead{Flux range} & \colhead{$dN/dS$ total, no cuts} & \colhead{$dN/dS$ sync, $z$ cut} & \colhead{$dN/dS$ dust, $z$ cut} & \colhead{$dN/dS$ SMGs} & \colhead{completeness} \\
\colhead{[Jy]} & \colhead{[${\rm Jy}^{-1} {\rm deg}^{-2}$]} & \colhead{[${\rm Jy}^{-1} {\rm deg}^{-2}$]} & \colhead{[${\rm Jy}^{-1} {\rm deg}^{-2}$]} & \colhead{[${\rm Jy}^{-1} {\rm deg}^{-2}$]} & 
}
\startdata
$4.4\times10^{-3}-5.6\times10^{-3}$ & $(4.04_{-0.8}^{+0.8})\times 10^{2}$ & $(3.06_{-0.7}^{+0.8})\times 10^{2}$ & $(6.90_{-3.0}^{+3.9})\times 10^{1}$ & $(5.91_{-3.0}^{+3.0})\times 10^{1}$ & $0.80$\\
$5.6\times10^{-3}-7.0\times10^{-3}$ & $(2.23_{-0.2}^{+0.2})\times 10^{2}$ & $(1.74_{-0.2}^{+0.2})\times 10^{2}$ & $(2.40_{-0.6}^{+0.8})\times 10^{1}$ & $(1.36_{-0.4}^{+0.5})\times 10^{1}$ & $0.90$\\
$7.0\times10^{-3}-8.7\times10^{-3}$ & $(1.31_{-8.8\times10^{-2}}^{+9.0\times10^{-2}})\times 10^{2}$ & $(1.04_{-7.9\times10^{-2}}^{+7.9\times10^{-2}})\times 10^{2}$ & $(1.16_{-0.3}^{+0.3})\times 10^{1}$ & $5.10_{-1.4}^{+1.6}$ & $0.97$\\
$8.7\times10^{-3}-1.1\times10^{-2}$ & $(7.69_{-0.5}^{+0.5})\times 10^{1}$ & $(6.38_{-0.5}^{+0.5})\times 10^{1}$ & $4.30_{-1.3}^{+1.6}$ & $1.79_{-0.5}^{+0.9}$ & $1.00$\\
$1.1\times10^{-2}-1.4\times10^{-2}$ & $(4.76_{-0.5}^{+0.5})\times 10^{1}$ & $(4.12_{-0.4}^{+0.4})\times 10^{1}$ & $1.72_{-0.6}^{+0.9}$ & $(8.58_{-4.3}^{+5.7})\times 10^{-1}$ & $1.00$\\
$1.4\times10^{-2}-1.7\times10^{-2}$ & $(2.89_{-0.3}^{+0.3})\times 10^{1}$ & $(2.54_{-0.2}^{+0.3})\times 10^{1}$ & $(5.71_{-3.4}^{+3.4})\times 10^{-1}$ & $(2.28_{-2.3}^{+2.3})\times 10^{-1}$ & $1.00$\\
$1.7\times10^{-2}-2.2\times10^{-2}$ & $(1.68_{-0.1}^{+0.1})\times 10^{1}$ & $(1.52_{-0.1}^{+0.1})\times 10^{1}$ & $ 0 _{- 0 }^{+0.2}$ &   & $1.00$\\
$2.2\times10^{-2}-2.7\times10^{-2}$ & $(1.23_{-0.1}^{+0.1})\times 10^{1}$ & $(1.08_{-0.1}^{+0.1})\times 10^{1}$ & $(7.27_{-7.3}^{+7.3})\times 10^{-2}$ & $(7.27_{-7.3}^{+7.3})\times 10^{-2}$ & $1.00$\\
$2.7\times10^{-2}-3.4\times10^{-2}$ & $7.13_{-0.7}^{+0.8}$ & $6.20_{-0.6}^{+0.7}$ &   &   & $1.00$\\
$3.4\times10^{-2}-4.2\times10^{-2}$ & $4.39_{-0.5}^{+0.6}$ & $3.93_{-0.5}^{+0.5}$ &   &   & $1.00$\\
$4.2\times10^{-2}-5.3\times10^{-2}$ & $2.73_{-0.3}^{+0.4}$ & $2.51_{-0.3}^{+0.3}$ &   &   & $1.00$\\
$5.3\times10^{-2}-6.7\times10^{-2}$ & $1.41_{-0.2}^{+0.2}$ & $1.38_{-0.2}^{+0.2}$ &   &   & $1.00$\\
$6.7\times10^{-2}-8.3\times10^{-2}$ & $1.06_{-0.2}^{+0.2}$ & $(9.40_{-1.6}^{+1.6})\times 10^{-1}$ &   &   & $1.00$\\
$8.3\times10^{-2}-1.0\times10^{-1}$ & $(8.06_{-1.3}^{+1.3})\times 10^{-1}$ & $(6.94_{-1.3}^{+1.1})\times 10^{-1}$ &   &   & $1.00$\\
$1.0\times10^{-1}-1.3\times10^{-1}$ & $(5.39_{-0.9}^{+1.0})\times 10^{-1}$ & $(5.24_{-0.9}^{+1.0})\times 10^{-1}$ &   &   & $1.00$\\
$1.3\times10^{-1}-1.6\times10^{-1}$ & $(1.79_{-0.5}^{+0.6})\times 10^{-1}$ & $(1.79_{-0.5}^{+0.6})\times 10^{-1}$ &   &   & $1.00$\\
$1.6\times10^{-1}-2.1\times10^{-1}$ & $(1.24_{-0.3}^{+0.5})\times 10^{-1}$ & $(1.24_{-0.4}^{+0.4})\times 10^{-1}$ &   &   & $1.00$\\
$2.1\times10^{-1}-2.6\times10^{-1}$ & $(9.12_{-2.3}^{+3.0})\times 10^{-2}$ & $(9.12_{-3.0}^{+2.3})\times 10^{-2}$ &   &   & $1.00$\\
$2.6\times10^{-1}-3.2\times10^{-1}$ & $(4.85_{-1.8}^{+1.8})\times 10^{-2}$ & $(4.85_{-1.8}^{+1.8})\times 10^{-2}$ &   &   & $1.00$\\
$3.2\times10^{-1}-4.1\times10^{-1}$ & $(6.78_{-1.9}^{+1.9})\times 10^{-2}$ & $(6.29_{-1.9}^{+1.5})\times 10^{-2}$ &   &   & $1.00$\\
$4.1\times10^{-1}-5.1\times10^{-1}$ & $(2.32_{-0.8}^{+1.2})\times 10^{-2}$ & $(1.93_{-0.8}^{+1.2})\times 10^{-2}$ &   &   & $1.00$\\
$5.1\times10^{-1}-6.4\times10^{-1}$ & $(2.16_{-0.9}^{+0.9})\times 10^{-2}$ & $(2.16_{-0.9}^{+0.9})\times 10^{-2}$ &   &   & $1.00$\\
$6.4\times10^{-1}-8.0\times10^{-1}$ & $(4.92_{-2.5}^{+2.5})\times 10^{-3}$ & $(2.46_{-2.5}^{+2.5})\times 10^{-3}$ &   &   & $1.00$\\
$8.0\times10^{-1}-1.0$ & $(3.92_{-3.9}^{+2.0})\times 10^{-3}$ & $(3.92_{-3.9}^{+3.9})\times 10^{-3}$ &   &   & $1.00$\\
$1.0-1.3$ & $(9.39_{-4.7}^{+3.1})\times 10^{-3}$ & $(7.83_{-3.1}^{+3.1})\times 10^{-3}$ &   &   & $1.00$\\
\enddata
\end{deluxetable*}
 
\begin{deluxetable*}{lllllc}
\tabletypesize{\footnotesize}
\tablecaption{$220$~GHz differential counts
\label{tab:diffcountstable220}}
\tablehead{
\colhead{Flux range} & \colhead{$dN/dS$ total, no cuts} & \colhead{$dN/dS$ sync, $z$ cut} & \colhead{$dN/dS$ dust, $z$ cut} & \colhead{$dN/dS$ SMGs}  & \colhead{completeness} \\
\colhead{[Jy]} & \colhead{[${\rm Jy}^{-1} {\rm deg}^{-2}$]} & \colhead{[${\rm Jy}^{-1} {\rm deg}^{-2}$]} & \colhead{[${\rm Jy}^{-1} {\rm deg}^{-2}$]} & \colhead{[${\rm Jy}^{-1} {\rm deg}^{-2}$]} & 
}
\startdata
$1.4\times10^{-2}-1.7\times10^{-2}$ & $(6.14_{-1.4}^{+1.5})\times 10^{1}$ & $(2.05_{-0.9}^{+0.7})\times 10^{1}$ & $(3.07_{-1.0}^{+1.4})\times 10^{1}$ & $(2.73_{-1.0}^{+1.0})\times 10^{1}$ & $0.83$\\
$1.7\times10^{-2}-2.2\times10^{-2}$ & $(2.52_{-0.7}^{+0.8})\times 10^{1}$ & $(1.37_{-0.5}^{+0.6})\times 10^{1}$ & $8.00_{-4.6}^{+4.6}$ & $6.86_{-3.4}^{+3.4}$ & $0.99$\\
$2.2\times10^{-2}-2.7\times10^{-2}$ & $(1.29_{-0.1}^{+0.2})\times 10^{1}$ & $7.87_{-1.1}^{+1.2}$ & $1.92_{-0.5}^{+0.7}$ & $1.51_{-0.5}^{+0.5}$ & $0.90$\\
$2.7\times10^{-2}-3.4\times10^{-2}$ & $7.15_{-0.9}^{+0.9}$ & $4.80_{-0.8}^{+0.8}$ & $(7.61_{-2.3}^{+3.5})\times 10^{-1}$ & $(5.86_{-2.3}^{+2.9})\times 10^{-1}$ & $0.99$\\
$3.4\times10^{-2}-4.2\times10^{-2}$ & $3.93_{-0.6}^{+0.6}$ & $2.82_{-0.5}^{+0.5}$ & $(3.70_{-1.9}^{+1.9})\times 10^{-1}$ & $(3.24_{-1.4}^{+1.9})\times 10^{-1}$ & $1.00$\\
$4.2\times10^{-2}-5.3\times10^{-2}$ & $2.33_{-0.4}^{+0.4}$ & $1.81_{-0.4}^{+0.4}$ & $(1.11_{-0.7}^{+1.1})\times 10^{-1}$ & $(1.11_{-0.7}^{+1.1})\times 10^{-1}$ & $1.00$\\
$5.3\times10^{-2}-6.7\times10^{-2}$ & $1.41_{-0.2}^{+0.3}$ & $1.09_{-0.2}^{+0.2}$ & $ 0 _{- 0 }^{+5.9\times10^{-2}}$ & $ 0 _{- 0 }^{+5.9\times10^{-2}}$ & $1.00$\\
$6.7\times10^{-2}-8.3\times10^{-2}$ & $1.13_{-0.2}^{+0.2}$ & $(8.93_{-1.6}^{+1.9})\times 10^{-1}$ & $ 0 _{- 0 }^{+2.4\times10^{-2}}$ & $ 0 _{- 0 }^{+2.4\times10^{-2}}$ & $1.00$\\
$8.3\times10^{-2}-1.0\times10^{-1}$ & $(6.19_{-1.5}^{+1.7})\times 10^{-1}$ & $(5.44_{-1.3}^{+1.5})\times 10^{-1}$ &   &   & $1.00$\\
$1.0\times10^{-1}-1.3\times10^{-1}$ & $(2.84_{-0.7}^{+0.9})\times 10^{-1}$ & $(2.54_{-0.7}^{+0.9})\times 10^{-1}$ &   &   & $1.00$\\
$1.3\times10^{-1}-1.6\times10^{-1}$ & $(1.79_{-0.5}^{+0.6})\times 10^{-1}$ & $(1.55_{-0.5}^{+0.6})\times 10^{-1}$ &   &   & $1.00$\\
$1.6\times10^{-1}-2.1\times10^{-1}$ & $(1.24_{-0.4}^{+0.4})\times 10^{-1}$ & $(1.24_{-0.4}^{+0.4})\times 10^{-1}$ &   &   & $1.00$\\
$2.1\times10^{-1}-2.6\times10^{-1}$ & $(6.84_{-2.3}^{+3.0})\times 10^{-2}$ & $(6.84_{-2.3}^{+3.0})\times 10^{-2}$ &   &   & $1.00$\\
$2.6\times10^{-1}-3.2\times10^{-1}$ & $(6.07_{-2.4}^{+2.4})\times 10^{-2}$ & $(5.46_{-1.8}^{+2.4})\times 10^{-2}$ &   &   & $1.00$\\
$3.2\times10^{-1}-4.1\times10^{-1}$ & $(3.39_{-1.5}^{+1.5})\times 10^{-2}$ & $(2.90_{-1.0}^{+1.9})\times 10^{-2}$ &   &   & $1.00$\\
$4.1\times10^{-1}-5.1\times10^{-1}$ & $(2.32_{-1.2}^{+1.2})\times 10^{-2}$ & $(1.93_{-0.8}^{+1.5})\times 10^{-2}$ &   &   & $1.00$\\
$5.1\times10^{-1}-6.4\times10^{-1}$ & $(6.16_{-6.2}^{+6.2})\times 10^{-3}$ & $(3.08_{-3.1}^{+6.2})\times 10^{-3}$ &   &   & $1.00$\\
$6.4\times10^{-1}-8.0\times10^{-1}$ & $(7.38_{-4.9}^{+4.9})\times 10^{-3}$ & $(7.38_{-4.9}^{+4.9})\times 10^{-3}$ &   &   & $1.00$\\
$8.0\times10^{-1}-1.0$ & $(5.89_{-3.9}^{+5.9})\times 10^{-3}$ & $(5.89_{-3.9}^{+3.9})\times 10^{-3}$ &   &   & $1.00$\\
$1.0-1.3$ & $(3.13_{-1.6}^{+3.1})\times 10^{-3}$ & $(3.13_{-3.1}^{+3.1})\times 10^{-3}$ &   &   & $1.00$\\
\enddata
\end{deluxetable*}
 
\section{Discussion of results}
\label{sec:discussion}

\subsection{Source catalog characteristics}

Of the 4845 sources in the catalog, 3980 (82.1\%) are classified as synchrotron sources, and 865 (17.9\%) as dusty sources, based on the probability that their $\alpha^{\text{max}}_{150-220}$ from deboosting is less or greater, respectively, than 1.51, the minimum of the summed posterior distribution of $\alpha^{\text{max}}_{150-220}$, as discussed in Section~\ref{sec:popsep}.   
1109 sources in the catalog, or about 23\%, have no cross-matches in external catalogs, and of those, 597 are classified as synchrotron and 512 as dusty.  937 or 84\% of the sources in the catalog with no external cross-matches are detected in only one band by SPT, and 172 (16\%) are detected in at least two bands.

Looking at Figure \ref{fig:alpha1v2}, we see that while a majority of sources in the catalog fit into the paradigm of two populations, dusty and synchrotron, with similar spectral indices between $95-150$~GHz and $150-220$~GHz, we also see some sources with a spectral break.  
To categorize different types of behavior, we look at the distributions of $\alpha^{\text{max}}_{95-150}$ for dusty and synchrotron sources, and see that $\alpha^{\text{max}}_{95-150}$ = 0.5 forms a relatively natural population separation, although this is a somewhat soft threshold.  
Using $\alpha^{\text{max}}_{95-150}$ = 0.5 and $\alpha^{\text{max}}_{150-220}$ = 1.51 as population thresholds, we divide the plots in Figure~\ref{fig:alpha1v2} into four quadrants: ``rising," ``falling," ``dipping," and ``peaking," though we stress that since the population break lines do not fall along $\alpha = 0$, the behavior of a source in one of the quadrants may not be as simple as the name suggests.  
For example, a source in the ``peaking" quadrant may have flux that rises with band between all three frequencies but with a spectral index shallow enough that the source was characterized as synchrotron.

\subsubsection{Synchrotron sources}

Using this categorization, we find that of the 3980 sources categorized as synchrotron, 3266 sources fall into the ``falling" category, sources that we expect to have their flux dominated by synchrotron emission and likely are characteristic synchrotron sources: steep-spectrum sources and flat-spectrum sources (blazars), including Flat-spectrum Radio Quasars (FSRQs) and BL Lacs (which can be further categorized as LBL (low-frequency peaked BL Lacs) and HBL (high-frequency peaked BL Lacs))~\citep{dezotti10,urry98}.

Considering all sources classified as synchrotron, we find $\alpha^{\text{max}}_{95-150}$ (as shown in Figure~\ref{fig:alpha1v2}) has a median value of $-0.6$ with a wide standard deviation of $1.2$,  and $\alpha^{\text{max}}_{150-220}$ has a median of $-0.7$ with standard deviation of $0.9$.  
Restricting to synchrotron sources detected at greater than 5.0\,$\sigma$ at 150 and 220\,GHz, these median spectral indices flatten and tighten slightly to median $\alpha^{\text{max}}_{95-150} = -0.6$ with a standard deviation of $0.4$ and median $\alpha^{\text{max}}_{150-220} = -0.6$ with a standard deviation of $0.5$.  
These numbers are the same if we restrict to only synchrotron sources in the ``falling" quadrant.

From models of synchrotron number counts, we expect that in our observing bands, synchrotron sources for the flux ranges spanned by the SPT catalog should be dominated by flat-spectrum sources, either FSRQs for sources with fluxes $\gtrsim$15\,mJy, or BL Lacs for sources with fluxes $\ltsim$15\,mJy, although steep-spectrum sources are expected to assume a larger portion of the synchrotron population at lower flux ranges as well~\citep{tucci11}.  
Flat-spectrum sources are expected to have spectral indices $\alpha > -0.5$, but according to \cite{tucci11}, the spectra of FSRQs will feature a spectral break which becomes more prominent at higher observing frequencies.  
For the ``C2Ex" model version from \citet{tucci11}, which is expected to be the model version in \citet{tucci11} that best predicts synchrotron number counts at our observing frequencies, the frequency at which the spectral break is predicted to occur is below our observing bands for all but the few very highest-flux sources in our catalog.  
Therefore, in the SPT bands, it is likely that FSRQs will appear as steep-spectrum sources, post-spectral break.  
In contrast, according to the \citet{tucci11} model, BL Lacs are expected to feature a spectral break at observing frequencies higher than the SPT bands, and therefore, BL Lacs should appear as flat-spectrum sources, but their population will be balanced out somewhat in the lower flux ranges by steep-spectrum sources.  
Therefore, we might expect that relatively high-flux synchrotron sources in the SPT catalog will appear with moderately steep spectral indices in our bands, and lower fluxes are likely to have a wider distribution of spectral indices, which may peak between flat- and steep- depending on the balance between FSRQs, BL Lacs, and steep-spectrum sources.  
Looking at the SPT catalog, we find this to be generally true: synchrotron sources with fluxes greater than 50\,mJy in at least two bands have a moderately steep median spectral index $\alpha^{\text{max}}_{95-150} = -0.6$, which is the same regardless of if we restrict to sources in the ``falling" quadrant or include all synchrotron-classified sources.  
Looking at synchrotron sources in the lower range of flux probed by the SPT-SZ catalog, $S_{150} < 20\,$mJy, but still detected above 4.5\,$\sigma$ at 150 and 220\,GHz such that they will have well-measured spectral indices, we find the same median $\alpha^{\text{max}}_{95-150}$ but with a wider distribution.  
We also note, however, that the width of the distribution in $\alpha^{\text{max}}_{95-150}$ will necessarily be wider for lower-flux sources just due to larger scatter from noise.

Sources in the ``peaking" quadrant are classified as synchrotron and have a flat, or falling index between 150\,GHz and 220\,GHz, but a rising spectral index between 95\,GHz and 150\,GHz.  
In the SPT-SZ catalog, there are 714 sources in this quadrant.  
88\% of the sources in the ``peaking" quadrant are single-band detections at 150\,GHz only, indicating that many have relatively low flux, given the noise threshold is lowest for 150\,GHz and sources just barely detected at 150\,GHz may be below the noise threshold at 90 and 220\,GHz.
Because they are low-significance detections and we apply relatively unrestrictive priors on spectral index in the deboosting, the flux deboosting for these sources is quite uncertain.
We expect that visible clustering of sources in the ``peaking" quadrant as shown in the upper right panel of Figure~\ref{fig:alpha1v2}, therefore, is likely due to the influence of edges of the applied spectral index priors in the deboosting, particularly because this clustering is not visible in the distributions of the raw spectral indices, shown in the upper left panel of Figure~\ref{fig:alpha1v2}.

44\% of the ``peaking" sources have cross-matches in SUMSS; only about 6\% have cross-matches in IRAS. 
As a check on the expected nature of the sources ``peaking" in the SPT bands, we consider all sources with cross-matches in SUMSS, and find that a large majority of the sources show flat or falling spectral behavior between the measured SUMSS flux at 843\,MHz and flux at 150\,GHz in SPT, indicating that they likely have spectra consistent with being flat- or steep-spectrum synchrotron sources.
A total of 20 sources have detected fluxes above $4.5\,\sigma$ in both 95 and 150\,GHz; of these, 19 have cross-matches in radio bands, including SUMSS.  
Of the sources with radio cross-matches, all but one have a spectral index relative to SUMSS consistent with being a flat-spectrum source, despite having $\alpha^{\text{max}}_{95-150} \gtrsim 0.5$.  
The remaining source with a radio cross-match has a spectral index relative to SUMSS that would categorize it as a steep-spectrum source.
For these sources ``peaking" in the SPT data that have relatively well-measured spectral indices in the SPT bands and cross-matches in radio catalogs, we expect these sources are likely AGN with significant self-absorption and disagreements in spectral behavior between the SPT bands and fluxes from radio cross-matches may be due to source variability over time.
There is a population of sources, known as Gigahertz peaked-spectrum (GPS) sources, that peak generally in the range 500\,MHz -- 10\,GHz~\citep{odea98} due to either self-absorption of synchrotron or to free-free absorption in the ionized outskirts of the source, with a subpopulation peaking at frequencies above 5\,GHz, known as High Frequency Peakers (HFPs)~\citep{dallacasa00}.
However, multi-frequency follow-up observations of both the original ``bright sample" of HFPs from \citet{dallacasa00} and ``faint sample" from \citet{stanghellini09} indicated that a large fraction of each sample, including all faint sources with the highest turn-over frequencies, were identified as flat-spectrum blazars, often with large variability between epochs~\citep{tinti05,orienti10}.
Combining this information with spectral index information relative to radio cross-matches for the SPT sources with relatively well-measured ``peaking" spectral indices, we expect these sources are not HFPs, and instead are likely flat- or steep-spectrum synchrotron sources.

\begin{figure*}[htp]
\begin{center}
\includegraphics[width=18.1cm]{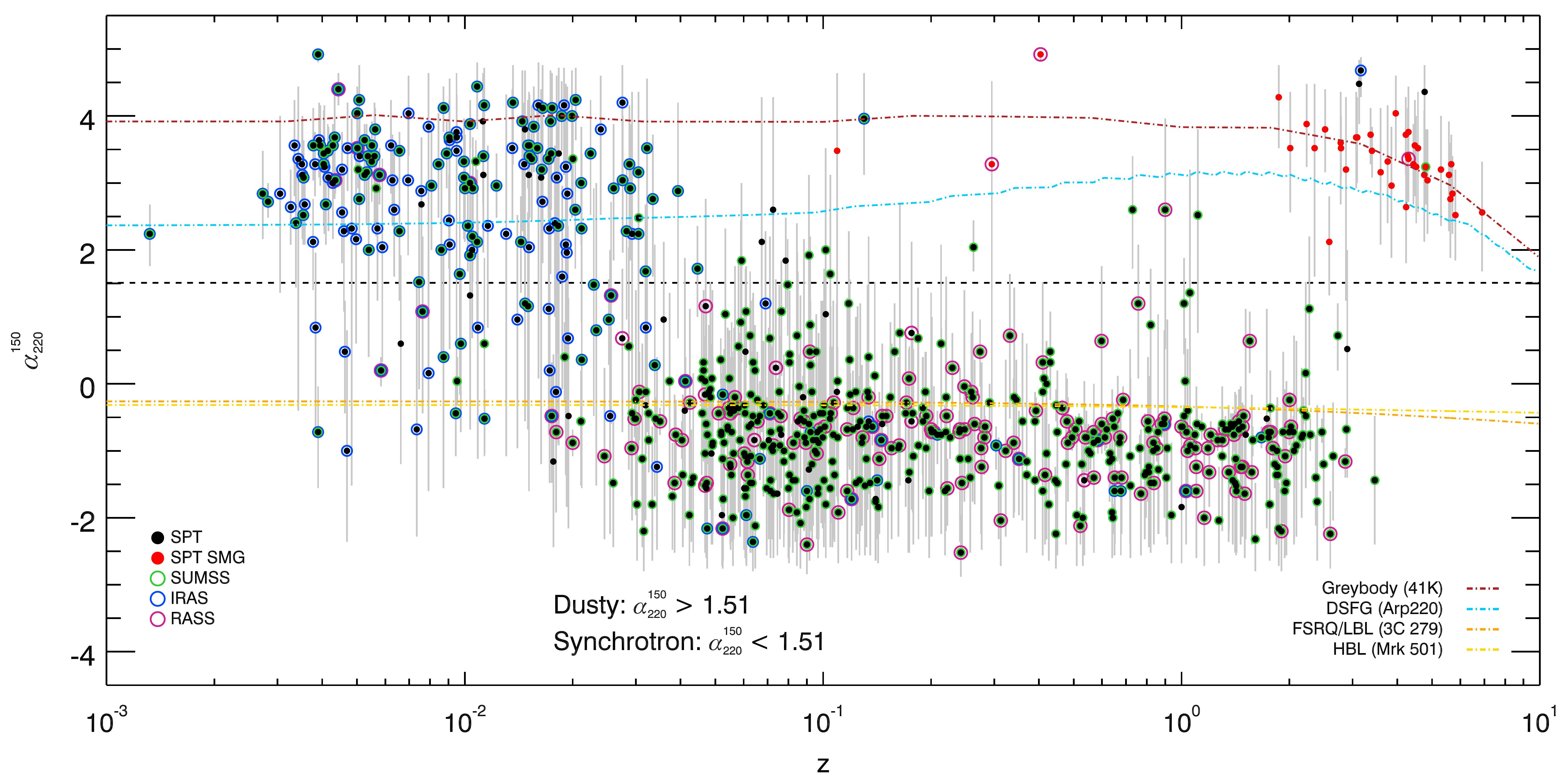}
\end{center}
\caption{Deboosted $\alpha_{150-220}$ versus redshift for sources in the SPT-SZ catalog with measured redshifts either from cross-matches in NED or follow-up observations with ALMA~\citep{weiss13,strandet16}.  Sources in the SMG list with measured redshifts are shown as red points, all other SPT-SZ catalog sources are shown as black points.  The black dashed line shows the separation between dusty and synchrotron sources.  Dot-dashed lines show two models of dusty sources: a 41\,K greybody (red) and an Arp 220 model (blue), and two models of flat-spectrum synchrotron sources: a model for flat-spectrum sources dominated by radio emission (FSRQs and LBLs) based on the spectrum of the blazar 3C 279, and a model for flat-spectrum sources dominated by X-ray emission (HBLs) based on the spectrum of the blazar Markerian 501.  Because redshifts are drawn from multiple sources, the selection function for SPT-SZ sources with measured redshifts is complicated.  We don't attempt to quantify the redshift completeness of the catalog, and this figure therefore is meant to illustrate the known redshift information for the catalog.
\label{fig:alphavsredshift}}
\end{figure*}

We note that from looking at proxy SED profiles for redshifted dusty galaxies as shown in Figure~\ref{fig:survdepth}, and shown in Figure~\ref{fig:alphavsredshift} in the following section, we would expect that very high-redshift dusty galaxies may also appear in the SPT data with a ``peaking" profile, with a rising spectral index between 95 and 150\,GHz, consistent with dust, but a spectral index between 150 and 220\,GHz that begins to flatten as the peak of the blackbody SED is redshifted into the SPT bands.
In the currently-employed classification scheme, these sources would be categorized as synchrotron-dominated.
A source with an Arp 220 spectrum would have a spectral index between 150 and 220\,GHz that flattens to below 1.51, the threshold for population separation used in the current catalog, at roughly a redshift of $z\sim10$.
We note that the dusty source in the SPT-SZ catalog with the highest measured redshift, $z = 6.9$~\citep{strandet17}, has spectral indices in the SPT-SZ bands that place it in the ``rising" quadrant, correctly categorizing it as a dusty source.

\subsubsection{Dusty sources}

Looking at all 865 dusty-classified sources, we find median spectral indices with relatively wide distributions of $\alpha^{\text{max}}_{95-150} = 1.7$ with a standard deviation of $1.5$ and $\alpha^{\text{max}}_{150-220} = 2.7$ with a standard deviation of $0.8$, which steepen to $\alpha^{\text{max}}_{95-150} = 2.3$ with a standard deviation of $1.3$ and $\alpha^{\text{max}}_{150-220} = 3.3$ with a standard deviation of $0.5$ when considering only sources detected above 5.0\,$\sigma$ at both 150 and 220\,GHz.  
They also steepen to $\alpha^{\text{max}}_{95-150} = 2.1$ and $\alpha^{\text{max}}_{150-220} = 2.8$ with a standard deviations of $0.9$ and $0.8$, respectively, when considering only dusty sources in the ``rising" quadrant (695 sources), and $\alpha^{\text{max}}_{95-150} = 2.5$ and $\alpha^{\text{max}}_{150-220} = 3.3$, with standard deviations of 1.0 and 0.5, respectively, for sources in the ``rising" quadrant detected above 5.0\,$\sigma$ at both 150 and 220\,GHz.  

We expect dusty galaxies observed in the frequency bands of SPT-SZ, where we are probing the Rayleigh-Jeans side of the spectrum, to follow a modified blackbody spectrum, $S_\nu \propto \nu^\beta B_{\nu}(\nu,T_d) \propto \nu^{\beta+2}$, where $\beta$, the dust emissivity spectral index is often assumed to be 1.5 and measured to be in the range 1 -- 2 for starburst galaxies~\citep{dunne01,magnelli12}.  
Thus, we expect to find measured spectral indices for dusty sources in the range $\alpha = 3-4$, and we find the SPT catalog to be relatively consistent with this, especially for dusty sources with well-measured spectral indices (detected at both 150 and 220 GHz).  

We find 170 sources in the ``dipping" quadrant, which are dusty-classified sources with typically greater flux at 95 and 220\,GHz relative to 150\,GHz.  
We expect some sources in this category to be nearby spiral galaxies or ULIRGs, sources with spectra containing both dust and synchrotron components.
For example, from Figure~\ref{fig:survdepth}, an Arp 220 SED with slightly more synchrotron emission would appear as a dipping source in the SPT-SZ bands.
In addition, multi-wavelength follow-up observations of powerful radio galaxies and steep radio-spectrum quasars have shown these sources can have spectra that dip at roughly 1\,mm~\citep{haas06}, with contributions to the emission at far-infrared and submm wavelengths expected to be caused by dust heating from the AGN and star formation, and observations with {\it Herschel} have shown that a substantial fraction of these sources may be radio-loud ULIRGs~\citep{barthel18,podigachoski16}.
As mentioned in Section~\ref{sec:cutsec}, follow-up observations at 870\,$\mu$m with LABOCA of a set of 21 sources with dipping spectra in SPT bands have indicated that while a few have fluxes at 870\,$\mu$m consistent with the presence of dust, most of these sources have fluxes at 870\,$\mu$m that are too low to be consistent with a DSFG spectrum.  
Five SPT ``dipping" sources with LABOCA follow-up have fluxes at 870\,$\mu$m that indicate, along with cross-matches in radio catalogs, their spectra are likely a combination of dust and synchrotron: SPT0420-55, SPT0427-47, SPT2117-58, SPT2147-55, and SPT2014-56.
The rest of the sources have non-detections or measured fluxes at 870\,$\mu$m that are too low to be consistent with the significant presence of dust.
Follow-up spectroscopy with the VLT for SPT ``dipping" sources have yielded measured redshifts in the range $z = 0.85 - 2.32$, indicating that they are unlikely to be high-redshift SMGs, though none of the five sources mentioned above with bright 870\,$\mu$m fluxes have measured redshifts.
Therefore, it's likely that this group of sources contains a combination of different types of objects, including low-redshift galaxies, moderate-redshift synchrotron sources or sources with spectra that include both dust and synchrotron components, and possibly sources that are blends of unrelated objects along the line of sight.
68\% of the sources in the ``dipping" quadrant have cross-matches in external catalogs, especially SUMSS, and most of the brightest ``dipping" sources have cross-matches in SUMSS, IRAS, and WISE, as expected for nearby galaxies.  

Most sources in the ``dipping" quadrant do not have a strong preference for falling in that quadrant: many are detections in only 95\,GHz or only 220\,GHz, meaning they have high uncertainties on their deboosted spectral indices, or they are relatively close to the threshold of a different quadrant.  
Using the posterior distributions for  $\alpha^{\text{max}}_{95-150}$ and  $\alpha^{\text{max}}_{150-220}$ to calculate a probability for each source to be deboosted into the dipping quadrant, of the four sources with greater than 90\% likelihood of being in the dipping quadrant, two are part of objects detected as multiple components in the source-finding, indicating that they are extended, and therefore likely low-redshift and possibly have greater uncertainty on their measured flux from not optimally matching the source profile used to extract them.   

We would also expect that galactic HII regions should fall in the ``dipping" quadrant.  
These sources are expected to be quite extended in the SPT-SZ maps, often appearing in the catalog as multiple source detections; therefore their measured fluxes are quite inaccurate.
Extended sources in the catalog that cross-match with galactic HII regions are mostly divided between the ``rising" and ``dipping" quadrants.  
These sources are flagged as extended in the catalog and are cut by both the ``ext cut" and ``$z$ cut."

For the list of 506 SMGs, we find median spectral indices of $\alpha^{\text{max}}_{95-150} = 2.1$ with a standard deviation of $0.9$ and $\alpha^{\text{max}}_{150-220} = 2.6$ with a standard deviation of $0.8$, which steepen to $\alpha^{\text{max}}_{95-150} = \alpha^{\text{max}}_{150-220} = 3.3$, with standard deviations of 0.7 and 0.4, respectively, for sources in the SPT SMG list detected above 5.0\,$\sigma$ at 150 and 220\,GHz (52 sources).

Follow-up observations of SMG candidates from the 2500-square-degree SPT-SZ survey using ALMA have measured redshifts for 39 sources, yielding a median redshift $z \sim 4$~\citep{weiss13,strandet16}, and all available measured redshifts for SMGs in the survey are shown in Figure~\ref{fig:alphavsredshift}.  
The source with the highest-measured redshift, SPT0311--58, with a redshift of $z = 6.9$~\citep{strandet17} has been observed to have massive rates of star formation likely triggered by the merging of two component galaxies, making it a very rare object observed well into the epoch of reionization~\citep{marrone18}.
While follow-up observations of SPT SMGs have indicated that most are subject to strong gravitational lensing, a number of sources show no evidence of gravitational lensing, indicating that they may be intrinsically extremely luminous or may be groups of dusty star-forming galaxies potentially in the early stages of forming a galaxy cluster, referred to as a `protocluster'~\citep{miller18}.  
SPT2349-56, a discovered protocluster in the 2500-square-degree SPT-SZ survey area, has been shown using deep ALMA spectral imaging to consist of at least 14 galaxies all at a redshift of 4.31 and undergoing massive star-formation in a relatively compact region~\citep{miller18}.
From a follow-up sample of roughly 90 SPT sources, a total of about 9 protocluster candidates have been discovered so far using detailed ALMA and LABOCA observations. 
These sources show similar characteristics to SPT2349-56, with typical measured redshifts $z \gtrsim 4$, demonstrating that discovered protoclusters in the SPT-SZ area can inform the study of structure formation in the very early universe.

\subsubsection{Redshift distribution of SPT-detected sources}

A total of 743 sources in the SPT-SZ catalog have measured redshifts from cross-matches in NED or follow-up observations.  
Of these, 531 are synchrotron-dominated sources and 212 are dust-dominated sources.  234 (163) synchrotron (dusty) sources with measured redshifts are at $z < 0.1$.
Figure~\ref{fig:alphavsredshift} shows the distribution of spectral index, $\alpha^{\text{max}}_{150-220}$, for all sources in the SPT-SZ catalog with measured redshifts, showing both dusty and synchrotron populations.  
Because redshifts are drawn from a variety of different sources, we don't attempt to quantify the completeness function of the redshift cross-matching with the SPT catalog; rather, this figure is to give an illustration of the known redshift information for the catalog.
The measured redshifts for high-redshift SPT SMGs are drawn primarily from follow-up observations with ALMA of sources discovered in the SPT-SZ survey area~\citep{weiss13,strandet16}.

A few individual sources in Figure~\ref{fig:alphavsredshift} warrant additional comment.  
Two sources with redshift associations where the measured redshift is  $> 3$ that have no cross-matches in external catalogs are not included in the SPT SMG list because both appear with a dipping spectrum in the SPT bands, due to blending of the lensed object with either its foreground lens or an unrelated source along the line of sight.  
An additional source with measured redshift $>3$ appears in the plot above with a cross-match with an IRAS detection, this cross-match is most likely a false association of two unrelated objects that fall just within the association radius.  
Finally, one source with a measured high redshift in the SPT SMG list has a cross-match with an X-ray detection in RASS; the lens for this high-redshift source is a galaxy cluster, and the X-ray detection is of the cluster, which falls within the RASS association radius of the background source.

\begin{figure*}[th!]
\begin{center}
\includegraphics[width=18.1cm]{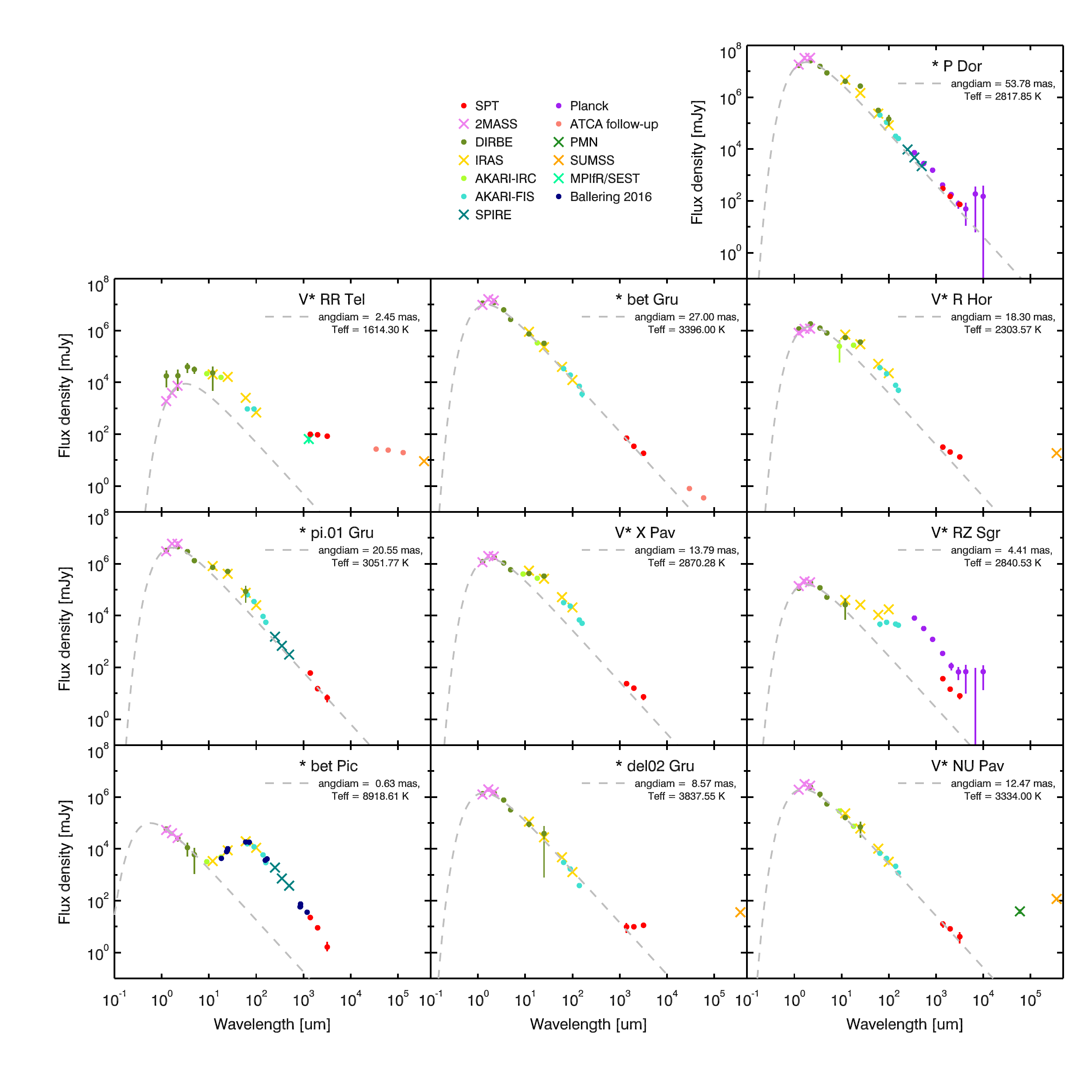}
\end{center}
\caption{Spectral energy distributions for the ten stars detected in the SPT-SZ catalog, with flux measurements drawn from the literature.  Grey dashed lines show blackbody model fits, where only fluxes from wavelengths in the range $1.25 - 5\,\mu$m are used in the fit.  In addition to available statistical errors from the literature, 10\% absolute calibration uncertainties have been applied to all fluxes from the literature.
\label{fig:starsed}}
\end{figure*}

\subsubsection{Millimeter-wavelength star characterization}
\label{sec:stardiscuss}

Millimeter wavelength observations of cool stars can provide interesting insight into the nature of these objects.  
The baseline expected flux measured at mm-wavelengths is due just to observing the Rayleigh-Jeans tail of the stellar blackbody radiation, where the stars detectable by SPT are those that are large and bright enough that despite observing the SED in a wavelength range where the flux is many orders of magnitude below the star's peak output, it is still detectable above the SPT noise level.  
As mentioned in Section~\ref{sec:starident}, nine of the ten stars in the SPT-SZ catalog are AGB stars, most of which are M type, and many of which are Mira variables or closely related to Miras, which are known to be very large and luminous, bright enough for the baseline flux in the Rayleigh-Jeans tail to be detectable.
However, excess mm-wavelength emission above the stellar SED or a spectral break from the expected $\nu^2$ of the stellar blackbody may indicate the presence of dust, including spinning dust, or potentially stellar winds, though the effects of winds are likely to be subdominant to dust because the stellar atmosphere will be optically thin at millimeter wavelengths~\citep{ogorman17,tram19}.

Gathering flux measurements from the literature, Figure~\ref{fig:starsed} shows SEDs for each of the ten stars detected in the SPT-SZ catalog. 
Other than * $\beta$~Pic, which is an A-type star and therefore much hotter than the other stars detected in the catalog (but which has a well-known dusty debris disc causing excess emission in longer wavelengths), the stars detected in the SPT-SZ catalog are expected to have effective temperatures of roughly a few thousand Kelvin, and therefore should have blackbody spectra that peak at $\sim 1$ -- a few microns.
Therefore, to explore the possibility of excess flux at millimeter wavelengths, we fit a simple blackbody model where the model is constrained using only data in the wavelength range of the blackbody peak, $1.25 - 5\,\mu$m, to fit for the blackbody effective temperature and the star's angular diameter.

Six stars in the catalog show flux in the SPT bands relatively consistent with blackbody fits to just the baseline stellar SED: *~P~Dor, *~bet~Gru, *~pi.01~Gru, V*~R~Hor, V*~X~Pav, and V*~NU~Pav.  
The other four stars identified in the SPT data have somewhat different spectral behavior from following the tail of the stellar blackbody.  
Two stars show excess emission but similar spectral indices: * $\beta$~Pic and V*~RZ~Sgr.  
As mentioned above, * $\beta$~Pic has a well-documented debris disc with a median dust temperature of 79\,K~\citep{riviere-marichalar14}. 
This star shows strong excess flux in the SPT bands relative to the expected stellar SED and a spectral index slightly steeper than the expected $\nu^2$ of the stellar blackbody, likely due to dust modification of the blackbody spectrum.  
V*~RZ~Sgr shows an excess of flux in mm-wavelengths but with a typical blackbody spectral index.  
It is known to have an optical nebula~\citep{whitelock94} and a circumstellar shell large enough to be resolved by IRAS at 60\,$\mu$m with a measured radius of 4.3\,arcmin~\citep{young93}, and therefore the extra emission observed in the SPT bands is consistent with the significant presence of dust.  
Correspondingly, V*~RZ~Sgr is observed to be extended in the SPT-SZ catalog, and is flagged accordingly.
Therefore, the measured SPT fluxes are likely to underestimate the true flux, and this is the likely cause of the disagreement between the measured SPT fluxes and those from Planck~\citep{planck18-54}, as shown in Figure~\ref{fig:starsed}, given the larger Planck beam.

Two stars show spectral indices distinctly different from the stellar blackbody: V*~RR~Tel and del02~Gruis.  
V*~RR~Tel shows quite flat spectral indices as well as excess flux in SPT bands, although we also note that the simple blackbody model is a poor fit to the data.  
V*~RR~Tel is known to be a symbiotic nova, with a red giant in mutual orbit with a white dwarf~\citep{ivison95}, and its distinct spectrum may indicate the presence of significant stellar winds~\citep{gudel02} or the effect of ionization from the white dwarf.  
del02~Gru, a red giant, also has a relatively flat measured spectral index in SPT bands.
We note that both V*~RR~Tel and del02~Gru have spectral indices measured between the SPT bands that are relatively consistent with measured fluxes in radio catalogs, where the measured flux in radio frequencies has clearly departed from the blackbody spectrum.
Similarly, V*~R~Hor, V*~NU~Pav, and *~bet~Gru have detections in radio catalogs that also show a break from the blackbody spectrum.
The flux measurements in SPT bands for these three are relatively consistent with a blackbody spectrum, but show some departure, potentially consistent with their radio fluxes.

As can be seen in Figure~\ref{fig:starid}, three sources in the SPT catalog have cross-matches with sources in IRAS with fluxes at 12\,$\mu$m comparable to the stars but are not identified as stars. 
Looking at each of these objects individually by hand and comparing with data in external surveys, two appear to be likely false cross-matches due to blends of multiple objects superimposed in the SPT maps along the line of sight, making accurate cross-match identification difficult.  
A third SPT object appears in the catalog as a repeat cross-match with the IRAS source identified as V*~RZ~Sgr, due to either being a blend of unrelated objects along the line of sight near V*~RZ~Sgr or possibly a multiple detection of V*~RZ~Sgr itself, which is known to be extended, as noted above.

\begin{figure*}[th!]
\begin{center}
\includegraphics[width=18.1cm]{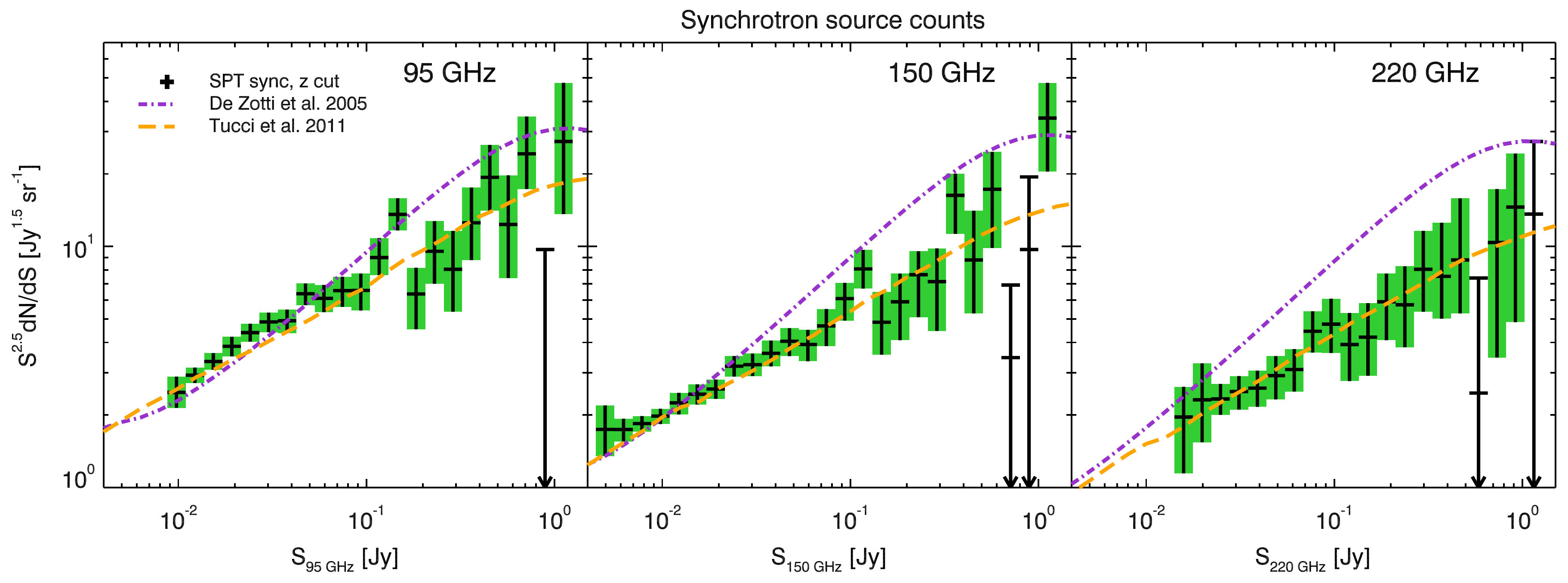}
\end{center}
\caption{Differential number counts of SPT synchrotron-dominated sources.  All sources in the catalog that are identified as stars as well as all sources flagged with the ``$z$ cut" (i.e. sources with measured redshifts $z < 0.1$) have been removed from the catalog prior to the counts calculation.  A comparison of the effects of different cut versions on the number counts can be found in the appendix.  Overplotted are the \citet{dezotti05} and the \citet{tucci11} models, which have not been fit to the SPT data.
\label{fig:sync_counts}}
\end{figure*}

\subsection{Number counts characterization}

\subsubsection{Synchrotron source population}

Differential number counts per band for synchrotron-dominated sources are shown in Figure~\ref{fig:sync_counts} along with comparison to two models: \citet{dezotti05} and \citet{tucci11}, neither of which have been fit to the SPT counts.  
The SPT counts shown are calculated using the method described in Section~\ref{sec:number_counts} on the SPT source population with the ``$z$ cut" flagged sources removed.  
Plots comparing non-cut and various cut versions of the synchrotron counts are shown in Figure~\ref{fig:sync_counts_appendix} in the Appendix.

The \citet{dezotti05} cosmological evolution model includes separate components for multiple synchrotron populations, including primarily steep-spectrum radio sources and two populations of flat-spectrum sources (blazars): flat-spectrum radio quasars (FSRQs) and BL Lacs.
It describes each population with a comoving luminosity function extrapolated to higher frequencies using a simple power law ($\alpha =-0.1$) for flat-spectrum sources and some spectral steepening for steep-spectrum sources. 

Similar to \citet{dezotti05}, the \citet{tucci11} model extrapolates source counts using spectral behavior measured at low radio frequencies (5 GHz).  
But the extrapolation is developed using characteristics of the physical mechanisms of emission for different populations, focusing specifically on flat-spectrum sources which dominate the number counts at cm- to mm-wavelengths and fluxes brighter than $\sim$ 10\,mJy.  
The spectrum of emission from flat-spectrum sources is expected to break at some frequency in the range of 10-1000\,GHz and steepen at higher frequencies due to both electron cooling as electrons are injected from the AGN core into the jets and from a reduction of the apparent size of the optically thick core with increasing observing frequency, such that high-frequency observations become dominated by the optically thin jets, which have a steeper spectrum~\citep{tucci11}.
This effect is most prominent for higher-flux sources, because these are more likely to be flat-spectrum sources.  
The SPT-SZ counts are compared with the ``C2Ex" version of the \citet{tucci11} model, which is the version that best fits data at frequencies $\gtrsim$ 100\,GHz, as confirmed mainly with comparison to Planck ERCSC counts, which has strong constraining power at the highest flux ranges due to full-sky coverage.

While historically the \citet{dezotti05} model has been broadly successful in extrapolating to higher frequencies~\citep{dezotti10}, because the \citet{dezotti05} model does not include a spectral break for flat-spectrum sources, we expect that it will become less of a good fit to the counts relative to the \citet{tucci11} model with increasing observing frequency and increasing source flux.  
Looking at Figure~\ref{fig:sync_counts}, while the \citet{dezotti05} model is in moderate agreement with the SPT-SZ counts at 95\,GHz, it becomes an increasingly poor fit to the counts at higher frequencies, particularly at high fluxes, where FSRQs will dominate.  
The \citet{tucci11} model is a reasonably good fit to the data at all three SPT-SZ frequency bands across the flux ranges probed by the SPT-SZ catalog.  

\begin{figure*}[ht!]
\begin{center}
\includegraphics[width=12cm]{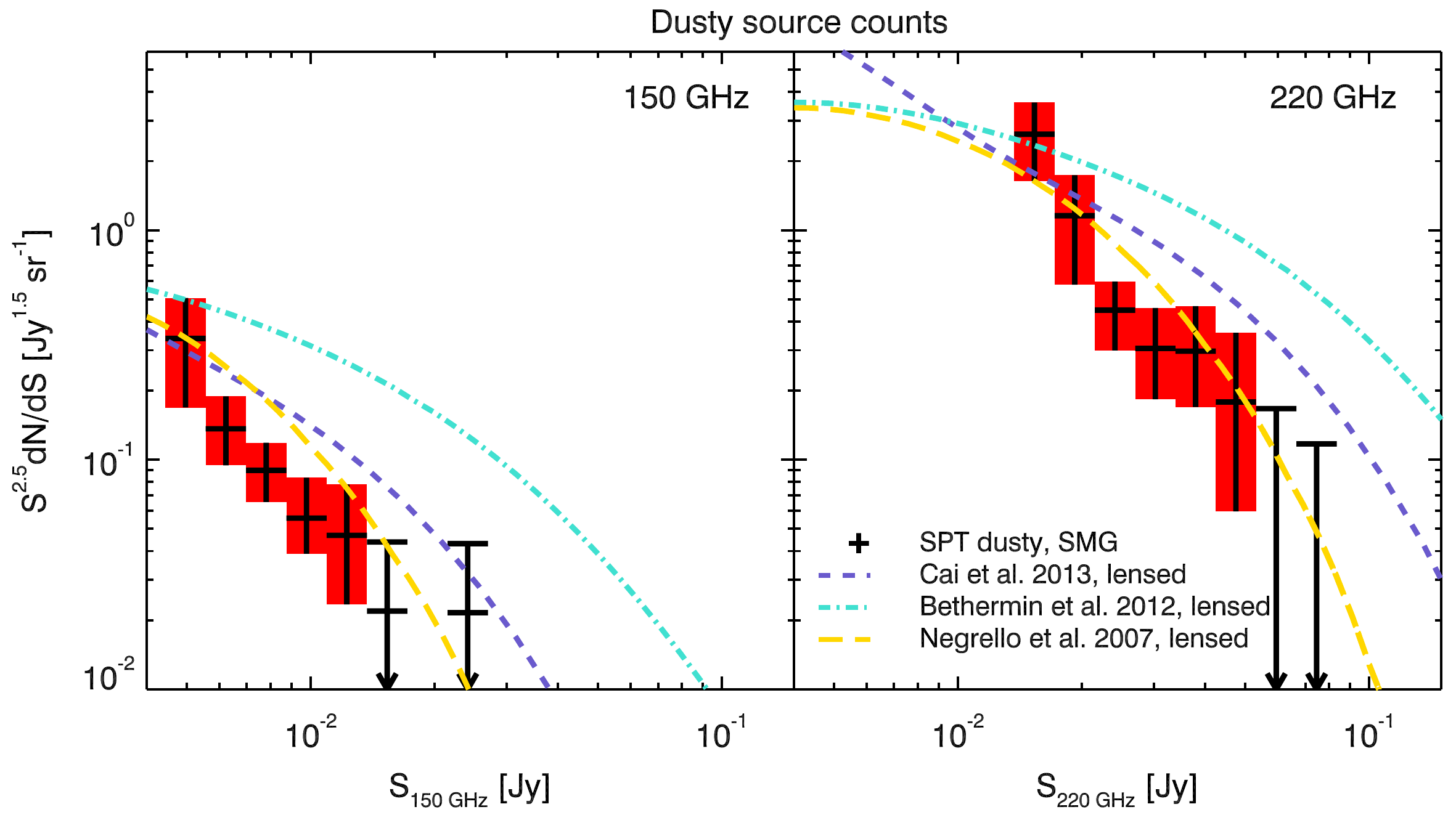}
\end{center}
\caption{Differential number counts for sources in the SPT SMG list.  Overplotted are lensed components of \citet{bethermin12}, \citet{cai13}, and \citet{negrello07} models, none of which have been fit to the SPT data.
\label{fig:dust_counts}}
\end{figure*}

\subsubsection{Dusty source population} 

Differential number counts per band for sources in the the SMG list are shown in Figure~\ref{fig:dust_counts} with comparisons to the lensed components of three representative models: \citet{bethermin12}, \citet{cai13}, and \citet{negrello07}.  
Because measured fluxes in the SPT-SZ catalog for low-redshift sources that are extended will underestimate the true source flux, we restrict our comparisons of the number counts for dusty populations with models to the SPT SMG list, for which comparison with models is the most straightforward.  
The SPT SMG list comprises dusty sources where all low-redshift sources, IRAS cross-matches, and dipping sources, which may be contaminated by blending with source lenses or unrelated objects along the line of sight, have been removed.
We expect this population to be dominated by sources that are magnified by gravitational lensing.
None of the models considered here have been fit to the SPT number counts.
Figure~\ref{fig:dust_counts_appendix} in the Appendix shows a comparison of different cut versions for the dusty number counts, including no cuts (only stars), ``ext cut," ``$z$ cut," and the SMG list.
All three models we compare with combine forward-physical and backward-phenomenological components to describe different populations of the observable galaxy population, including late-type warm (starburst) and cold (normal) galaxies, as well as lensed and unlensed spheroidals and protospheroidals.

The \citet{negrello07} model includes counts for protospheroidals as modeled using the physical model by \citet{granato04}.  
The \citet{granato04} counts have been rescaled at 850\,$\mu$m to agree with counts from the SCUBA SHADES survey \citep{coppin06}.  
Protospheroidals virialized generally at $z \gtrsim 1.5$, and the $z \ltsim 1.5$ contribution is considered to be dominantly from starburst and disc galaxies, which are modeled from local luminosity functions at 60\,$\mu$m.  
The strongly lensed component to the \citet{negrello07} model was calculated similarly to \citet{perrotta02, perrotta03} but using the \citet{granato04} model for protospheroidals.  
When comparing with the SMG list, as in Figure~\ref{fig:dust_counts}, we have trimmed sources with fluxes greater than 200\,mJy at 60\,$\mu$m from the \citet{negrello07} model.
Of the dusty source counts models considered in this work, the \citet{negrello07} lensed model is the closest to reproducing the SPT SMG counts, but we note that it is the model that is the most fine-tuned to reproduce counts at (sub-)mm-wavelengths.

The \citet{bethermin12} model includes main sequence and starburst galaxies as the two main components of the model, using one SED per component from libraries from {\it Herschel}.  
Because phenomenological or hybrid models are limited by lacking physical underpinnings describing the evolution of the luminosity function, instead the \citet{bethermin12} model is based on two distinct star-formation mechanisms and their evolution, one for each galaxy component, based on the work in \citet{sargent12}.  
The contribution from strong gravitational lensing is accounted for by applying a magnification factor to the luminosity function.  
The lensed component of the \citet{bethermin12} model overestimates the SPT SMG number counts for all but the very lowest fluxes at both 150 and 220\,GHz.

The \citet{cai13} model is a hybrid model, combining a physical, forward model for spheroidal galaxies and backward-evolution model for late-type galaxies, based on observations that early-type galaxies are dominated by older stellar populations, while late-type galaxies have younger stellar populations.  
They improve on previous models by considering components of the flux for protospheroidal galaxies from star formation and central AGN in a unified way, rather than being considered separately.  
Protospheroidal galaxies are modeled using \citet{granato04}, and low-$z$ galaxy populations are considered in two populations: ``warm" starburst galaxies and ``cold" late-type galaxies.  
A magnification factor is applied to account for strong lensing of high-redshift protospheroidals.
While the \citet{cai13} model counts are lower than \citet{bethermin12}, this model also over-predicts the SPT SMG counts for generally all but the lowest flux bins probed by the SPT data at both 150 and 220\,GHz.

Disagreement of the data with models can help place constraints on the maximum magnification factor for the strong gravitational lensing of protospheroidals.  
Both the \cite{cai13} and \cite{bethermin12} models assume no upper bound on strong lensing magnification factor (assuming sources were point-like).  
According to \cite{bonato14} (see Figure 2), applying a maximum magnification factor of $20-30$ (corresponding to a physical source size of slightly less than $\simeq3$\,kpc~\citep{lapi12}), provides good agreement with SPT source counts from M13 for dusty sources with no IRAS cross-matches.
The SMG list reported in the current work is more conservative in defining SMG candidates relative to M13, restricting by IRAS cross-match but also measured redshift and removing ``dipping" sources.
The number counts for the SMG list in the current work are consistent with the dusty counts with no IRAS cross-matches in M13 at 220\,GHz and are slightly lower at 150\,GHz.
A maximum magnification factor of $20-30$ is in good agreement with follow-up observations with ALMA of a set of roughly 50 lensed SMGs from the 2500-square-degree SPT-SZ survey, where the measured median magnification factor is $\sim 6$, with a maximum of 30~\citep{spilker16}.  
The implied source physical sizes from the \citet{lapi12} model for a maximum magnification factor of 30 are also consistent with follow-up observations, where the measured size distribution of strongly lensed sources is found to be consistent with that of unlensed sources~\citep{spilker16}.

\begin{figure*}[ht!]
\begin{center}
\subfigure{\includegraphics[width=18.1cm]{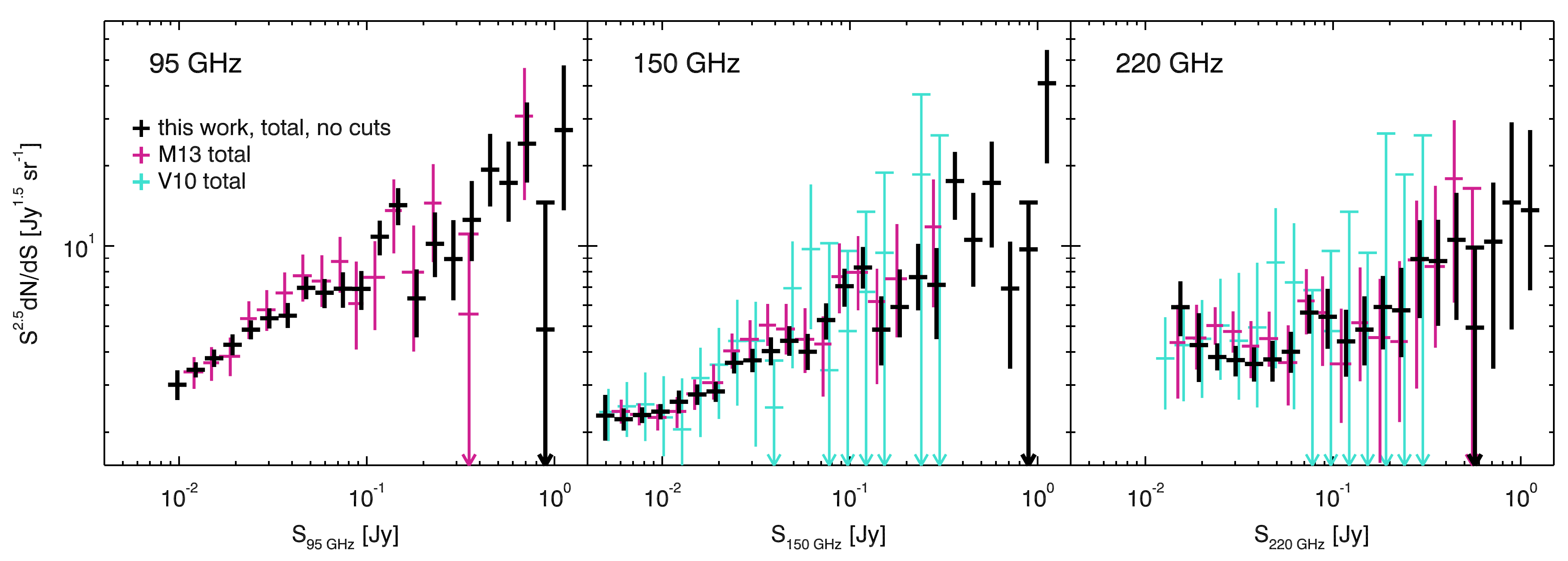}}
\end{center}
\caption{Total source counts (with no cuts applied other than removing sources identified as stars) for the current work, using 2530 square-degrees of the SPT-SZ survey, compared with number counts from M13 (calculated using 771 square degrees of sky area) and V10 (calculated using 87 square degrees of sky area in 150 and 220\,GHz only).  V10 used a detection threshold of $3\,\sigma$, whereas M13 and the current work use $4.5\,\sigma$ to increase the purity of the sample, which resulted in V10 probing a slightly lower flux range than M13 and the current work at 220\,GHz.
\label{fig:SPT_compare_total_counts}}
\end{figure*}

\subsection{Comparison with previous SPT-SZ point source results}

As the third and final compact source data release from SPT-SZ, the full 2530 square-degree analysis covers a factor of 3.3 times the area of the previous release, M13, including 1759 square-degrees of previously unanalyzed data.  
Due to alterations of the source-finding pipeline and differences in mask areas to aid full survey coverage, the five sky fields covered by the previous two analyses, V10 and M13, have been reanalyzed in the current analysis.
For \fivefiffi~and \twthreefiffi~which were originally observed in 2008 and then reobserved in 2010 and 2011 to add 95\,GHz coverage and greater depth at 150 and 220\,GHz, we have incorporated the previously unanalyzed 2010 and 2011 data.  
Although a large majority of the sources extracted in the five reanalyzed fields are consistent with past reported catalogs, there are slight differences in sources extracted between M13, V10, and the current analysis.  
These differences are due mainly to the lower noise in 150\,GHz for \fivefiffi~and \twthreefiffi, slight differences in the masks used for each field, and the slightly different treatment of map noise used for source detection, as discussed in Section~\ref{sec:clean}.  
We confirm that the sources that differ generally have signal-to-noise values very close to the detection threshold or are located at the edges of the survey area, which are affected by slight differences in masking.

Figure~\ref{fig:SPT_compare_total_counts} shows a comparison of total number counts between the three generations of compact source catalog releases with data from SPT-SZ: V10, M13, and the current work.
As expected, the increase in sky area with generally comparable noise level also reduces the error bars on the calculated number counts and adds a few flux bins of counts that were either upper limits or missing from M13.  
The error bars on the uncut version of the counts, which are most directly comparable to the M13 counts, reduce by roughly 50\%, consistent with the amount of area increase.  

The smaller error bars allow the SPT number counts to be more constraining of the parameters of galaxy evolution models.  
For the synchrotron counts, the \citet{tucci11} model more clearly agrees with the number counts than the older \citet{dezotti05} model, as shown in Figure~\ref{fig:sync_counts}, whereas the M13 counts showed a weaker preference between models, particularly at 95 and 150\,GHz.  
Since the main difference between the two models is the inclusion of a spectral break for FSRQs, the greater constraint shows a clear preference for the presence of a spectral break, although the counts are not constraining enough to provide much further information on the models, such as the break frequency, which might further constrain the AGN core size.  
Similarly, the smaller error bars for the dusty counts also provide clearer constraints, particularly on parameters governing lensed sources, such as the expected lensing magnification factor, as discussed in the previous section.

\subsection{Comparison of SPT-SZ with other mm-wavelength surveys}

\begin{figure*}[ht!]
\begin{center}
\subfigure{\includegraphics[width=18.1cm]{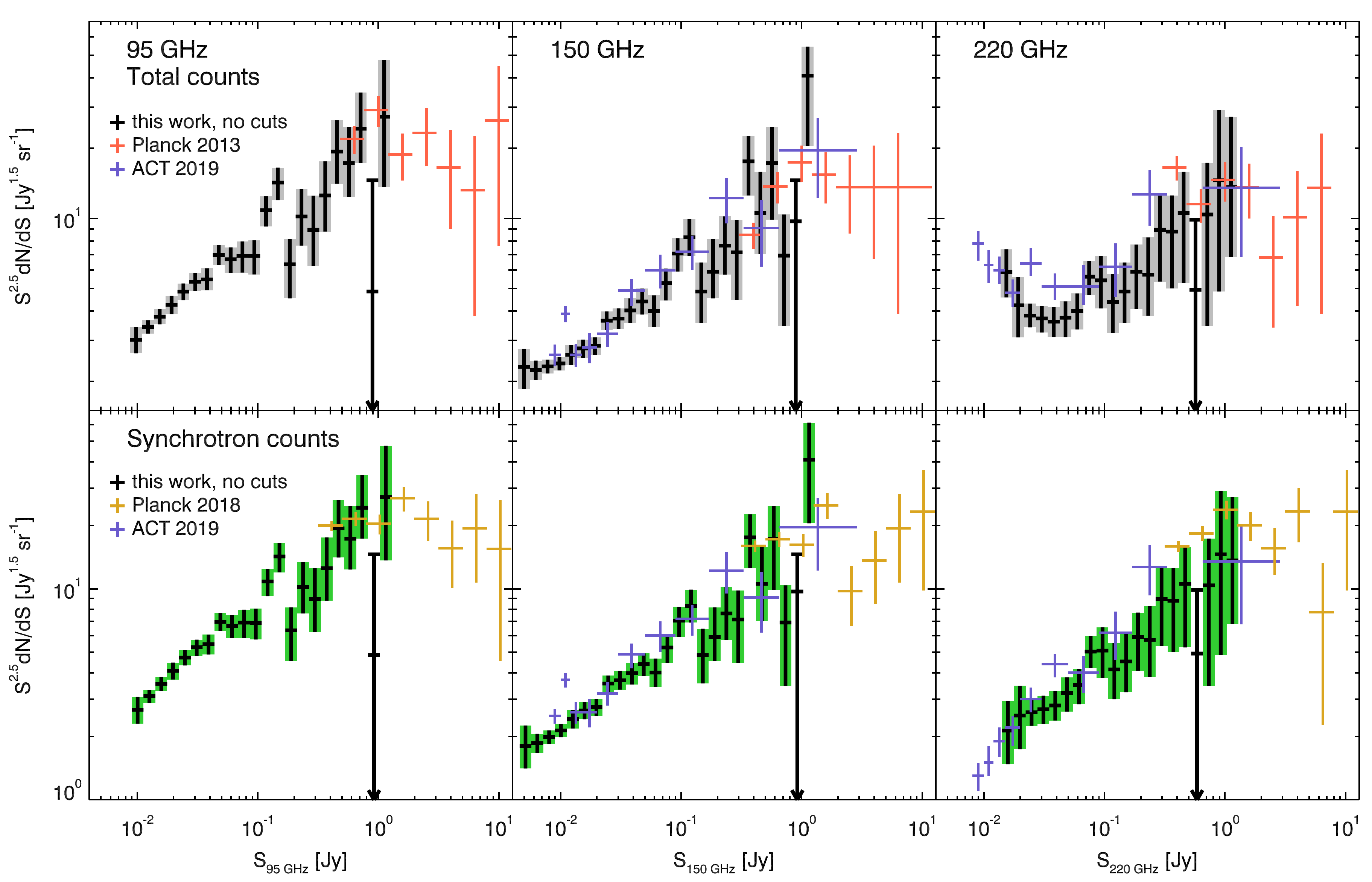}}
\end{center}
\caption{{\it Upper:} Total source counts (with no cuts applied other than removing sources identified as stars) for the current work, compared with total source counts from Planck (\citealt{planck12-7}) and the Atacama Cosmology Telescope (ACT, \citealt{gralla19}).  {\it Lower:} Synchrotron-dominated counts for the current work (shown with no cuts applied other than removing stars), compared with the Planck Multi-frequency Catalog of Non-thermal Sources~\citep{planck18-54}, where the counts were calculated applying the 70\% Galactic mask, GAL070, and ACT AGN counts from \citet{gralla19}.
\label{fig:SPT_compare_Planck_ACT}}
\end{figure*}

The upper panels of Figure~\ref{fig:SPT_compare_Planck_ACT} provide a comparison between the total counts from SPT-SZ from the current work (with only stars removed) with number counts from the all-sky survey from Planck at 100, 143, and 217\,GHz~\citep{planck12-7}, and number counts from the Atacama Cosmology Telescope (ACT, \citealt{gralla19}).  
The total counts from ACT presented here have been drawn from the sum of synchrotron- and dust-dominated sub-populations, as presented in Table 5 of \citet{gralla19}.  
The lower panels of Figure~\ref{fig:SPT_compare_Planck_ACT} compare synchrotron-dominated counts from SPT-SZ (similarly shown with only stars removed) compared with number counts from the Planck Multi-frequency Catalog of Non-thermal Sources~\citep{planck18-54}, excluding Galactic-plane sources using the GAL070 mask, and counts for AGN sources from ACT~\citep{gralla19}.
The total and synchrotron-dominated number counts from SPT-SZ are broadly consistent with both Planck and ACT across the full range of fluxes overlapping between the different surveys.

\begin{figure*}[ht!]
\begin{center}
\subfigure{\includegraphics[width=18.1cm]{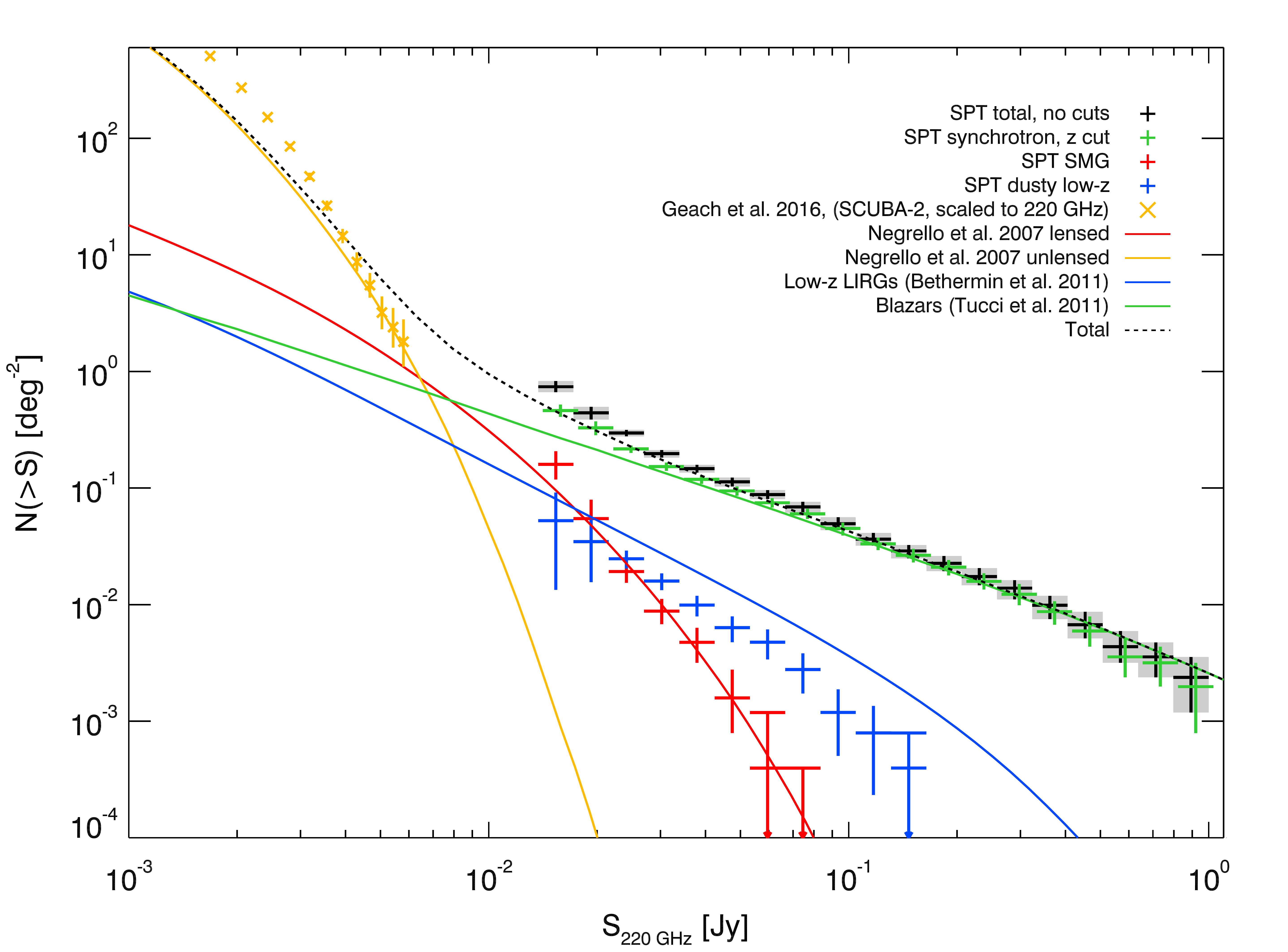}}
\end{center}
\caption{Cumulative number counts at 220\,GHz for total, synchrotron-dominated, and dust-dominated source populations inÊthe SPT-SZ catalog.
\label{fig:smg_comparison_counts}}
\vspace{0.2in}
\end{figure*}

As an overview of the different source populations present in the SPT-SZ catalog, Figure~\ref{fig:smg_comparison_counts} shows cumulative number counts at 220\,GHz, including total, synchrotron-, and dust-dominated source populations.
Contributions to dusty counts are considered in three components: low-$z$ LIRGs, SPT SMGs, and unlensed high-$z$ sources, with empirical counts from the SCUBA-2 instrument~\citep{geach17}, scaled from 850\,$\mu$m to 220\,GHz using an SED for Arp 220 shifted to $z\,\sim\,2.5$.  
SPT number counts for low-$z$ dusty sources have been calculated using sources that are trimmed by the ``$z$ cut." 
We know the SPT flux measurements for these sources will be biased low, and this is one of the causes of the slight discrepancy with the \citet{bethermin11} model counts for these sources.
The counts for this population are shown here primarily for illustration.
The \citet{negrello07} lensed-only model agrees well with cumulative number counts calculated from our SPT SMG list.

\section{Conclusion}
\label{sec:conclusion}

We have presented a catalog of 4845 compact sources extracted from 2530 square degrees of the SPT-SZ survey with fluxes measured in three bands centered at 95, 150, and 220\,GHz.  
Sources in the catalog are detected in at least one band with a significance of 4.5\,$\sigma$ or higher.  
Because the raw source fluxes will be subject to a positive bias due to the underlying source number counts being a steep function of flux, we apply a Bayesian deboosting method to report corrected fluxes and spectral indices.  
The deboosting method is also used to separate sources into synchrotron-dominated and dust-dominated populations using their deboosted spectral index between 150 and 220\,GHz.  
Synchrotron sources (with flat or falling spectral indices between 150 and 220\,GHz) are consistent with AGN, and dust-dominated sources (with rising spectral indices between 150 and 220\,GHz) are consistent with dusty star-forming galaxies, including a population of high-redshift dusty galaxies, which we refer to as SMGs.  
In the relatively bright flux ranges and moderate field depths probed by the SPT-SZ survey, we expect the high-redshift dusty sources we observe will be dominated by sources subject to strong gravitational lensing.  
We further categorize this population by developing an ``SPT SMG list," which contains 506 sources.  
With the largest currently available sky area with arcmin resolution, the SPT-SZ survey provides a powerful lever arm for finding the brightest and rarest high-redshift dusty sources.  
Sources in the SMG list with previous follow-up observations have measured redshifts up to $z = 6.9$, demonstrating that the survey has detected sources as far back as the epoch of reionization~\citep{strandet17}.
Similarly, protoclusters discovered in the 2500-square-degree area with redshifts $\gtrsim 4$ demonstrate that this dataset can probe high-redshift structure formation~\citep{miller18}.

Number counts for total, synchrotron-, and dust-dominated populations have also been calculated.  
The number counts probe flux ranges from $8.7\times 10^{-3} - 1.3$\,Jy at 95\,GHz, $4.4\times 10^{-3} - 1.3$\,Jy at 150\,GHz, and $1.4\times 10^{-2} - 1.3$\,Jy at 220\,GHz.  
We find that our synchrotron population number counts as well as catalog spectral indices are consistent with models from \cite{tucci11}, in which FSRQs are expected to dominate the brighter fluxes probed by the catalog, but featuring a spectral break at higher observing frequencies resulting in moderately steep spectral indices for FSRQs measured in the SPT-SZ bands.

As expected, number counts for the dusty source population are subdominant to those for the synchrotron population at all fluxes we probe, except at the lowest flux range at 220\,GHz.  
Focusing on number counts for the SMG list, we find that of all of the models we compare with, the \citet{negrello07} lensed model agrees the best with our measured counts at 150 and 220\,GHz; the \citet{cai13} and \citet{bethermin12} models lensed components generally overestimate our measured counts.  

As the third and final compact source catalog release from the SPT-SZ survey, our catalog and number counts are consistent with prior-released results from the SPT-SZ survey: \citet{vieira10} and \citet{mocanu13}.  
The current work features a roughly 50\% reduction in the error bars on our measured number counts, consistent with the increase in sky area utilized.  
We find our measured number counts are also consistent with all-sky results from Planck~\citep{planck12-7,planck18-54} and recently-released results from ACT~\citep{gralla19}.  

Looking to the future, SPT-3G, the camera most-recently installed on the South Pole Telescope, is undertaking observations currently that will push the flux detection threshold for compact sources by roughly an order of magnitude relative to SPT-SZ over a 1500 square-degree area, thus probing populations of both lensed and unlensed dusty sources and overlapping with flux ranges probed by the original deep but narrow surveys from instruments such as SCUBA.
Anticipated noise levels in the completed SPT-3G survey are roughly 140, 130, and 760\,$\mu$Jy at 95, 150, and 220\,GHz.
This survey will allow for unprecedented study of the highest-redshift dusty, star-forming galaxies, as well as provide powerful constraints on the development and evolution of extragalactic radio source populations.
SPT-3G will thus push the SPT source detection threshold into the population of unlensed dusty sources, enabling consistency checks with source counts measured in small field areas observed with ALMA~(e.g. \citealt{hatsukade18,zavala18}), while also covering large field areas, further enabling the discovery of extremely rare sources, such as protoclusters, advancing our understanding of structure formation in the early universe.

\begin{acknowledgments}

The SPT is supported by the NSF through grant PLR-1248097, with partial support through PHY-1125897, the Kavli Foundation and the Gordon and Betty Moore Foundation grant GBMF 947.
D.P.M. and J.D.V. acknowledge support from the US NSF under grants AST-1715213 and AST-1716127.
J.D.V. acknowledges support from an A. P. Sloan Foundation Fellowship.
C. Reichardt acknowledges the support from an Australian Research CouncilÕs Future Fellowship (FT150100074). 
B.B. is supported by the Fermi Research Alliance LLC under contract no. De-AC02-07CH11359 with the U.S. Department of Energy.
Work at Argonne National Lab is supported by UChicago Argonne LLC, Operator of Argonne National Laboratory (Argonne). 
Argonne, a U.S. Department of Energy Office of Science Laboratory, is operated under contract no. DE-AC02-06CH11357.
This material is based upon work supported by the U.S. Department of Energy, Office of Science, Office of High Energy Physics under Award Number DE-SC-0015640.
Thanks to Dr. Graham Harper for conversations regarding cool stars and their emission mechanisms, and thanks to Diego Herranz Mu\~noz from the Planck Collaboration and Dr. Megan Gralla from the ACT collaboration for supplying number counts.

This publication makes use of data products from the Wide-field Infrared Survey Explorer, which is a joint project of the University of California, Los Angeles, and the Jet Propulsion Laboratory/California Institute of Technology, funded by the National Aeronautics and Space Administration. 
This research uses observations from the AKARI mission, a JAXA project with the participation of ESA.  
This publication makes use of data products from the Two Micron All Sky Survey, which is a joint project of the University of Massachusetts and the Infrared Processing and Analysis Center/California Institute of Technology, funded by the National Aeronautics and Space Administration and the National Science Foundation.  
This publication makes use of data products from the {\it Herschel Observatory}; {\it Herschel} is an ESA space observatory with science instruments provided by European-led Principal Investigator consortia and with important participation from NASA.  
This research makes use of data products from the Planck Observatory (http://www.esa.int/Planck), an ESA science mission with instruments and contributions directly funded by ESA Member States, NASA, and Canada.
\end{acknowledgments}

\begin{figure*}[!ht]
\begin{center}
\includegraphics[width=18.1cm]{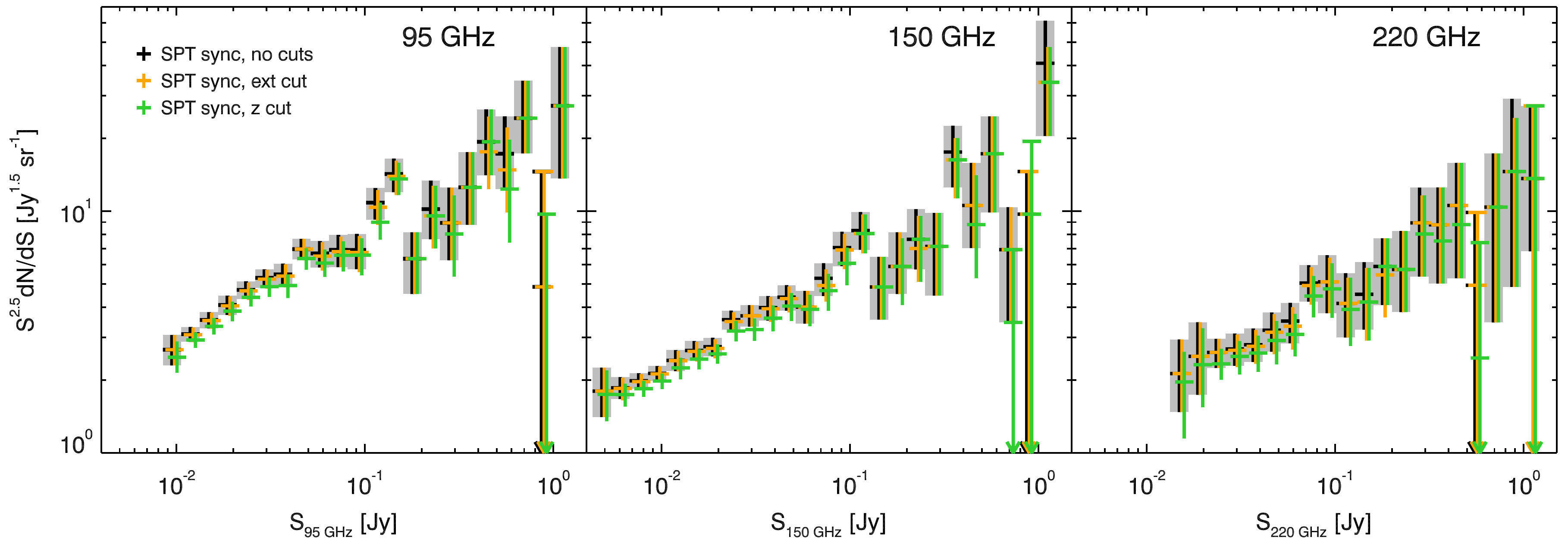}
\end{center}
\caption{Number counts of SPT synchrotron-dominated sources, showing a comparison of different cut versions.  All sources in the catalog that are identified as stars have been removed from the catalog prior to the counts calculation.  Three cut versions are shown: 1) only stars have been removed, ``no cuts."  2) Cutting all stars and all sources flagged as extended or detected as multiple detections and confirmed to be from a single object, ``ext cut,"  and 3) cutting all sources with measured redshifts $z < 0.1$, ``$z$ cut."
\label{fig:sync_counts_appendix}}
\end{figure*}

\begin{figure*}[!ht]
\begin{center}
\includegraphics[width=18.1cm]{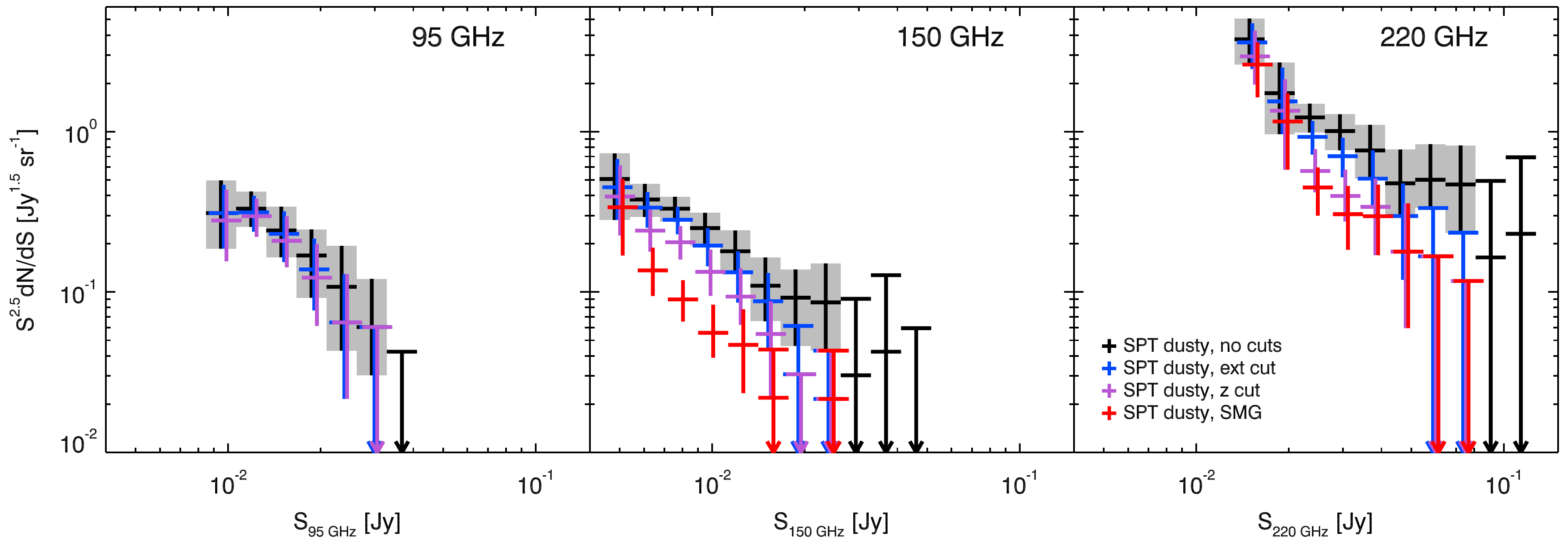}
\end{center}
\caption{Number counts of SPT dust-dominated sources, showing a comparison of different cut versions. All sources identified as stars have been removed from the catalog prior to the counts calculation.  Four cut versions are shown: 1) only stars have been removed, ``no cuts."  2) Cutting all stars and all sources flagged as extended or detected as multiple detections and confirmed to be from a single object, ``ext cut,"  3) cutting all sources with measured redshifts $z < 0.1$, ``$z$ cut," and 4) the SPT SMG list, cutting all sources with measured redshifts, IRAS cross-matches, or dipping spectral behavior in the SPT bands.
\label{fig:dust_counts_appendix}}
\end{figure*}

\appendix{Figures~\ref{fig:sync_counts_appendix} and \ref{fig:dust_counts_appendix} show comparisons between different versions of cuts applied to the SPT catalog when calculating number counts for synchrotron and dusty populations (where ten sources identified as stars have been removed from all versions of the counts): no cuts, ``ext cut," ``$z$ cut," and counts calculated from the SPT SMG list.
}

\newpage

\bibliography{../../../BIBTEX/spt}

\end{document}